\documentclass[useAMS,usenatbib]{mn2e}
\usepackage{natbib}
\usepackage{amssymb,amsmath,amsfonts}
\usepackage{epsfig}
\usepackage{mathptmx}
\usepackage{longtable}
\usepackage{graphicx}
\usepackage{comment}
\usepackage{color}
\usepackage[caption=false]{subfig}

\def\reff@jnl#1{{\rm#1\/}}

\def\aj{\reff@jnl{AJ}}                  % Astronomical Journal
\def\araa{\reff@jnl{ARA\&A}}            % Annual Review of Astron and Astrophys
\def\apj{\reff@jnl{ApJ}}                        % Astrophysical Journal
\def\apjl{\reff@jnl{ApJ}}               % Astrophysical Journal, Letters
\def\apjs{\reff@jnl{ApJS}}              % Astrophysical Journal, Supplement
\def\ao{\reff@jnl{Appl.Optics}}         % Applied Optics
\def\apss{\reff@jnl{Ap\&SS}}            % Astrophysics and Space Science
\def\aap{\reff@jnl{A\&A}}               % Astronomy and Astrophysics
\def\aapr{\reff@jnl{A\&A~Rev.}}         % Astronomy and Astrophysics Reviews
\def\aaps{\reff@jnl{A\&AS}}             % Astronomy and Astrophysics, Supplement
\def\azh{\reff@jnl{AZh}}                        % Astronomicheskii Zhurnal
\def\baas{\reff@jnl{BAAS}}              % Bulletin of the AAS
\def\jrasc{\reff@jnl{JRASC}}            % Journal of the RAS of Canada
\def\memras{\reff@jnl{MmRAS}}           % Memoirs of the RAS
\def\mnras{\reff@jnl{MNRAS}}            % Monthly Notices of the RAS
\def\pra{\reff@jnl{Phys. Rev. A}}         % Physical Review A: General Physics
\def\prb{\reff@jnl{Phys. Rev. B}}         % Physical Review B: Solid State
\def\prc{\reff@jnl{Phys. Rev. C}}         % Physical Review C
\def\prd{\reff@jnl{Phys. Rev. D}}         % Physical Review D
\def\prl{\reff@jnl{Phys. Rev. Lett}}      % Physical Review Letter
\def\pasp{\reff@jnl{PASP}}              % Publications of the ASP
\def\pasj{\reff@jnl{PASJ}}              % Publications of the ASJ
\def\qjras{\reff@jnl{QJRAS}}            % Quarterly Journal of the RAS
\def\skytel{\reff@jnl{S\&T}}            % Sky and Telescope
\def\solphys{\reff@jnl{Solar~Phys.}}    % Solar Physics
\def\sovast{\reff@jnl{Soviet~Ast.}}     % Soviet Astronomy
\def\ssr{\reff@jnl{Space~Sci.Rev.}}     % Space Science Reviews
\def\zap{\reff@jnl{ZAp}}                        % Zeitschrift fuer Astrophysik
\def\nat{\reff@jnl{Nature}}             % Nature 

\def\p#1by#2{{\partial{#1} \over \partial{#2}}}
\def\pp#1by#2#3{{\partial^2{#1} \over \partial{#2}\partial{#3}}}
\def\d#1by#2{{{\rm d}{#1} \over {\rm d}{#2}}}
\def\dd#1by#2#3{{{\rm d}^2{#1} \over {\rm d}{#2}{\rm d}{#3}}}

\title[]{SZ observations with AMI of the hottest galaxy clusters detected in the XMM-Newton Cluster Survey} %with the Arcminute Microkelvin Imager}

\label{firstpage}

\author[Shimwell et~al.]{AMI Consortium:
 Timothy W. Shimwell$^1$\thanks{E-mail: Timothy.Shimwell@csiro.au},
 Carmen Rodr{\'i}guez-Gonz{\'a}lvez$^2$,
    \newauthor
  Farhan Feroz$^{3}$,
  Thomas M. O. Franzen$^{1}$,
  Keith J. B. Grainge$^{3,4,5}$,
  Michael P. Hobson$^{3}$, 
  \newauthor
  Natasha Hurley-Walker$^6$,
  Anthony N. Lasenby$^{3,4}$,
  E. J. Lloyd-Davies$^7$,
  \newauthor
  Malak Olamaie$^3$,
 Yvette C. Perrott$^3$,
 Guy G. Pooley$^3$,
 Clare Rumsey$^3$,
 \newauthor
 A. Kathy Romer$^7$,
 Richard D. E. Saunders$^{3,4}$,
 Anna M. M. Scaife$^8$,
 \newauthor
  Michel P. Schammel$^3$,
 Paul F. Scott$^3$,
 David J. Titterington$^3$,
 Elizabeth M. Waldram$^3$\\
 $^1$ CSIRO Astronomy \& Space Science, Australia Telescope National Facility, PO Box 76, Epping, NSW 1710, Australia \\
 $^2$ Spitzer Science Center, MS 220-6, California Institute of Technology, Pasadena, CA 91125, USA \\
 $^3$ Astrophysics Group, Cavendish Laboratory, J J Thomson Avenue, Cambridge, CB3 0HE\\
 $^4$ Kavli Institute for Cosmology Cambridge, Madingley Road, Cambridge, CB3 0HA\\
 $^5$ University of Manchester, Jodrell Bank Centre for Astrophysics, Alan Turing Building, Oxford Road, Manchester, M13 9PL \\
 $^6$ International Centre for Radio Astronomy Research, Curtin Institute of Radio Astronomy, 1 Turner Avenue, Technology Park, Bentley, WA 6845, Australia \\
 $^7$ Astronomy Centre, University of Sussex, Falmer, Brighton, BN1 9QH  \\
 $^8$ School of Physics \& Astronomy, University of Southampton, Southampton, SO17 1BJ
}

\begin{document}

\maketitle

\begin{abstract}
\noindent
We have obtained deep SZ observations towards 15 of the apparently hottest XMM Cluster Survey (XCS) clusters that can be observed with the Arcminute Microkelvin Imager (AMI). We use a Bayesian analysis to quantify the significance of our SZ detections. We detect the SZ effect at high significance towards three of the clusters and at lower significance for a further two clusters. Towards the remaining ten clusters, no clear SZ signal was measured. We derive cluster parameters using the XCS mass estimates as a prior in our Bayesian analysis. For all AMI-detected clusters, we calculate large-scale mass and temperature estimates while for all undetected clusters we determine upper limits on these parameters. We find that the large-scale mean temperatures derived from our AMI SZ measurements (and the upper limits from null detections) are substantially lower than the XCS-based core-temperature estimates. For clusters detected in the SZ, the mean temperature is, on average, a factor of 1.4 lower than temperatures from the XCS. For clusters undetected in SZ, the average 68\% upper limit on the mean temperature is a factor of 1.9 below the XCS temperature. 

% This factor is 1.4 for SZ detections, 1.9 for the non-detection upper limit and 1.7 if you us both the detections and upper limits  of the null detections.

\end{abstract}

\begin{keywords}
cosmology: observations Ð cosmic microwave background Ð galaxies: clusters Ð Sunyaev ZelÕdovich Ð  X-rays: clusters 
\end{keywords}

\section{Introduction}

The Sunyaev-Zel'dovich (SZ; \citealt{Sunyaev_1972}; see e.g. \citealt{Birkinshaw_1999} and  \citealt{Carlstrom_2002}) effect is a secondary anisotropy in the cosmic microwave background (CMB) radiation, and it is caused by the inverse-Compton scattering of CMB photons off intracluster electrons. SZ observations of clusters complement those obtained with X-ray satellites -- the X-ray Bremsstrahlung emission from the hot intracluster electrons is dependent upon $n_{e}^2\Lambda(T_e)$, whereas the SZ signal follows $n_{e}T$, where $n_{e}$ is the electron density and $\Lambda(T_e)$ is the electron cooling function which is approximately given by $T_{e}^{1/2}$ (\citealt{Sarazin_1986}). Due to its weaker dependence on the electron density, the SZ effect is an excellent probe of large-scale cluster parameters such as mass, temperature and Compton Y-parameter. %and for imaging destroyed density peaks  arising from, for example, strong merger events. 
X-ray data, on the other hand, have the advantage over SZ of  having significantly better resolution. This allows for e.g., the inner regions of the clusters and their density profiles to be characterized more precisely, albeit to within a smaller radius for all but a few recent measurements taken with the {\sc{Suzaku}} satellite, see e.g., \cite{Kawaharada_2010}, \cite{Akamatsu_2011}, \cite{Walker_2012a} and \cite{Walker_2012b}, and with the {\sc{Chandra}} satellite, see e.g., \cite{Bonamente_2012}. A further distinction is that the X-ray emission is dependent upon the luminosity distance to the cluster; hence, to detect distant clusters, the sensitivity of the X-ray observations must be high or the cluster must be very luminous. %still, in the low signal-to-noise (SNR) regime, correct background subtraction becomes increasingly important. 
In contrast, the SZ surface brightness is wholly independent of redshift and therefore the integrated SZ flux density depends only on the angular diameter distance which is a weak function of redshift for z $\gtrapprox 0.5$.

Large SZ cluster surveys, such as those from ACT (e.g., \citealt{Hincks_2010} and \citealt{Marriage_2011}), SPT (e.g \citealt{Vanderlinde_2010},  \citealt{Carlstrom_2011}, \citealt{Williamson_2011}  and \citealt{Reichardt_2012})  and {\sc{Planck}} (e.g \citealt{Tauber_2010} and \citealt{ESZ}) have already discovered over a hundred new clusters and many more are expected to be found in the near future. Fulfilling the cosmological potential of these surveys relies, amongst other things, on accurate determination of cluster masses. Disentangling biases and other effects in the process of calculating cluster mass from observables is challenging and it may be that multi-frequency data are required.

The Arcminute Microkelvin Imager (AMI; see \citealt{Zwart_2008}) is a radio interferometer optimized for SZ observations at 16\,GHz. AMI has been used to observe several samples of well-known galaxy clusters (see e.g., \citealt{Hurley-Walker_2011}, \citealt{Zwart_2011}, \citealt{Carmen_2011}, \citealt{Hurley-Walker_2012}, \citealt{Carmen_2012} and \citealt{Schammel_2012}) and has also conducted a blind SZ survey (see \citealt{Shimwell_2012}). In this paper we present AMI observations of the hottest observable clusters in the  XMM Cluster Survey (XCS) catalogue (\citealt{Mehrtens_2011}). These optically confirmed clusters span a wide range of XCS-quoted 0.05 -- 100\,keV band rest-frame luminosities ($L_{\rm{X}},0.42-47.9\times 10^{37}$\,W), redshifts ($z$, 0.15-1.13) and temperature estimates ($T_{\rm{X}}$, 5.2-14.7\,keV). So they may: improve our understanding of a variety of cluster systems; test the numerous scaling relations between SZ and X-ray-derived parameters and observables; and help identify causes of possible discrepancies in derived cluster parameters such as mass.

\begin{comment}
Furthermore, deep 16\,GHz observations of galaxy clusters are useful to estimate the density excess of radio sources towards clusters. Such studies help reveal  the nature of radio sources in clusters and are also vital to assess the level of contamination caused by radio sources to present and future SZ instruments, as well as to other CMB measurements. Several large-scale studies have characterized this overdensity at lower frequencies (see e.g., \citealt{Sommer_2011}) but there have been very few such studies at similar frequencies (see e.g., \citealt{Coble_2007} and \citealt{Muchovej_2010}).

The outline of this paper is as follows: in Section 2 we present the sample; Sections 3, 4 and 5 describe the AMI, the observations and the data reduction procedure respectively; Section 6 outlines the SZ analysis; Section 7 describes the methods used to obtain X-ray mass estimates; Section 8 gives details on individual cluster observations; and Section 9 presents a more general discussion of our results. We stress that some readers will wish to skip Section 8 and jump to Section 9.

\end{comment}

We assume a concordance $\rm{\Lambda}$CDM cosmology with
$\rm{\Omega_{m}}$ = 0.3, $\rm{\Omega_\Lambda}$ = 0.7 and H$_{0}$ = 70 km\,s$^{-1}$Mpc$^{-1}$.
Coordinates are J2000.

\section{Cluster sample}
\label{Sec:cluster-sample}

The XCS survey focussed on analysing archival XMM-Newton data to detect galaxy clusters. In the XCS catalogue there are a total of 503 X-ray detected and optically confirmed galaxy clusters of which 255 are new to the literature. The methodology used for the X-ray analysis is described in \cite{Lloyd_Davies_2010} and the first data release, together with the optical analysis methodology, is presented in \cite{Mehrtens_2011}, both of which we now summarise. (Note that for the majority of the detected clusters the derived 0.05 -- 100\,keV band luminosity, radius, redshift and temperature are available in the XCS catalogue\footnote{http://xcs-home.org/}.)

In the X-ray analysis, extended X-ray sources are identified as candidate clusters and the location, ellipticity and shape of each is quantified. For several candidates, cluster spectroscopic redshifts have been determined or values are taken from the literature. % determined from dedicated follow-up observations or luminous red galaxies detected in the Sloan Digital Sky Survey (SDSS). 
When no spectroscopic redshift is determined, photometric redshifts have been obtained from single colour ($r-z$) images of the candidates from either dedicated follow-up observations or from public SDSS data using an algorithm based on \cite{Gladders_2000} that makes use of a colour-magnitude relation. The temperature and luminosity are derived by fitting models that describe the hot plasma (\citealt{Mewe_1986}) and photoelectric absorption (\citealt{Morrison_1983}) to the spectra, allowing for potential contamination from point sources and cool cores through spectral fitting. 
%The fractional temperature errors increase with increasing cluster temperature and decreasing X-ray counts; for simulated X-ray clusters at 5\,keV a fractional temperature error of 10\% is expected for 2000 counts and of 60\% for 200 counts. 
The X-ray surface brightness is then characterized by a spherically symmetric $\beta$-profile (\citealt{Cavaliere_1978}), allowing for contamination from cool cores and AGN. Under the assumption that the cluster gas is isothermal, the radius $r_{Y}$ can be determined, where $r_{Y}$ is defined as the radius inside which the mean total density is Y times the critical density of the Universe at the cluster redshift, $\rm{\rho_{crit,z}}$. Using the derived radius the luminosity can be aperture-corrected to obtain $L_{X,Y}$. The XCS derived luminosities have been compared with the 0.001 -- 50\,keV band luminosities and temperatures from \cite{Pacaud_2007} and are found to be in good agreement. 
For more details on the X-ray catalogue see \cite{Lloyd_Davies_2010}, \cite{Mehrtens_2011} and references within.

We have used AMI to conduct 16\,GHz observations towards those XCS clusters within AMI's easy declination range ($20^{\circ} < \delta < 80^{\circ}$) and with an X-ray mean temperature estimate greater than 5\,keV. This left 34 clusters (see Table \ref{XMM_CLUSTERS_LIST}) which we refer to as our full sample. A further 14 XCS clusters with $T_{\rm{X}}<5$\,keV lie close enough to these hot clusters to be within the AMI fields of view.

At 16\,GHz, contamination from radio point sources can lead to significant obscuration of an SZ effect and prevent its detection and/or the estimation of robust cluster parameters. Fifteen clusters from our full sample have been excluded from our analysis, either because of there being $\geq15$\,mJy of flux within 10$\arcmin$ of the XCS cluster position, or because of a source with a peak flux density greater than 15\,mJy on the primary beam-attenuated SA image (see Table \ref{XMM_CLUSTERS_LIST} for a summary of the radio source environment towards each of the clusters in our full sample). Further, we excluded XMJ1000+6839, XMJ1217+4729 and XMJ1217+4728 due to the presence of bright, extended emission away from the pointing centre, and we excluded XMJ1122+4659 due to the existence of faint, extended emission at the location of the cluster which we could not confidently remove. We emphasise that in our observations of clusters from the XCS sample, there is particularly high radio source contamination which is possibly a selection effect due to the nature of the XCS: it uses images from targeted X-ray observations that were made because they are \emph{interesting}, often because they contain bright X-ray sources -- and there is a correlation between X-ray bright and radio bright. Indeed, within our full sample, we find, for example, the galaxies NGC 4258 and NGC5548 and the calibrator sources J1110+4817 (0.13\,Jy) and J1407+2827 (1.9\,Jy).

\begin{figure}
\begin{center}
\includegraphics[width=7.9cm,clip=,angle=0.]{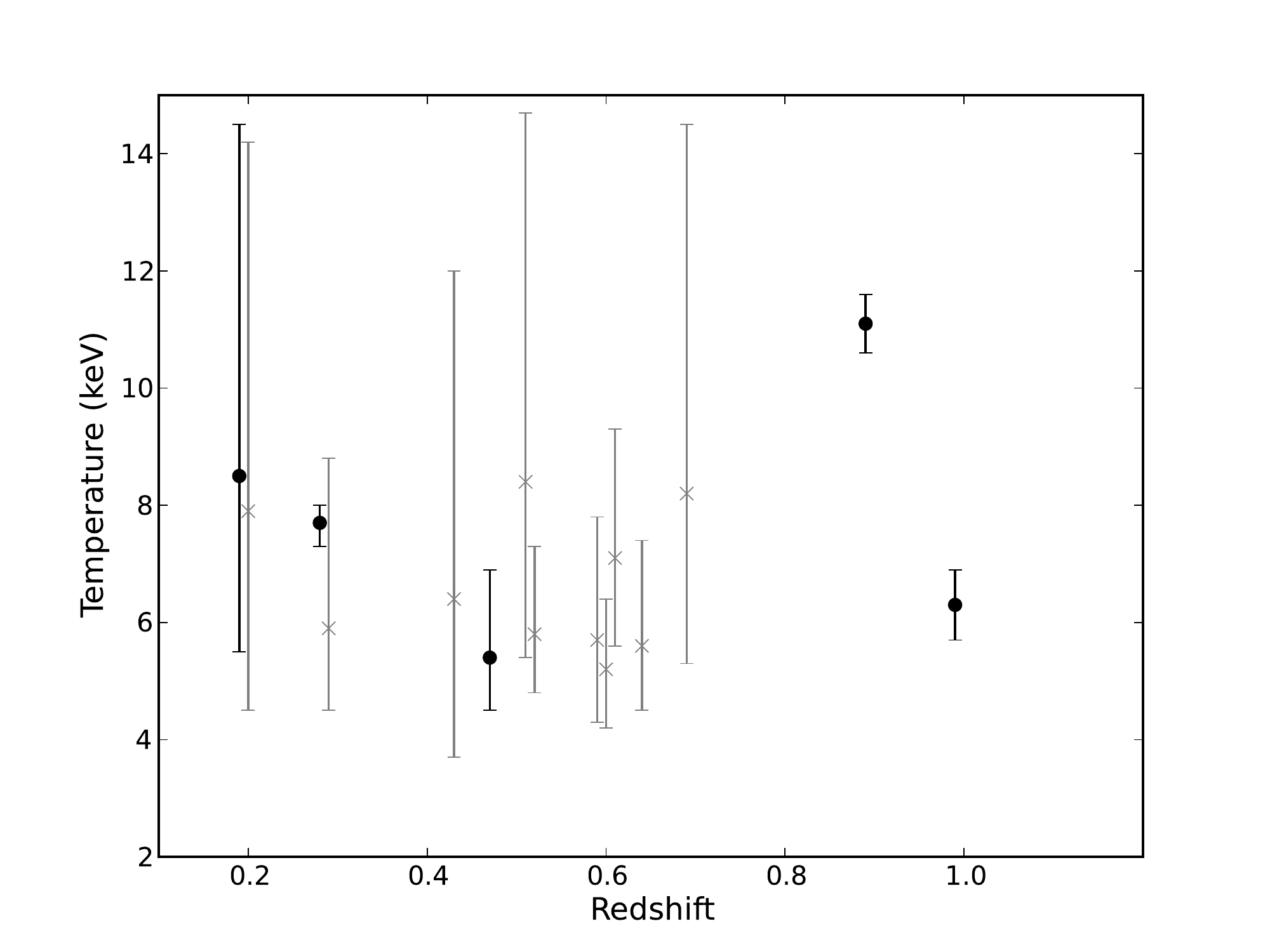} % ODD size to get them onto the right page if width=8.0cm they are all pushed to the end.
\includegraphics[width=7.9cm,clip=,angle=0.]{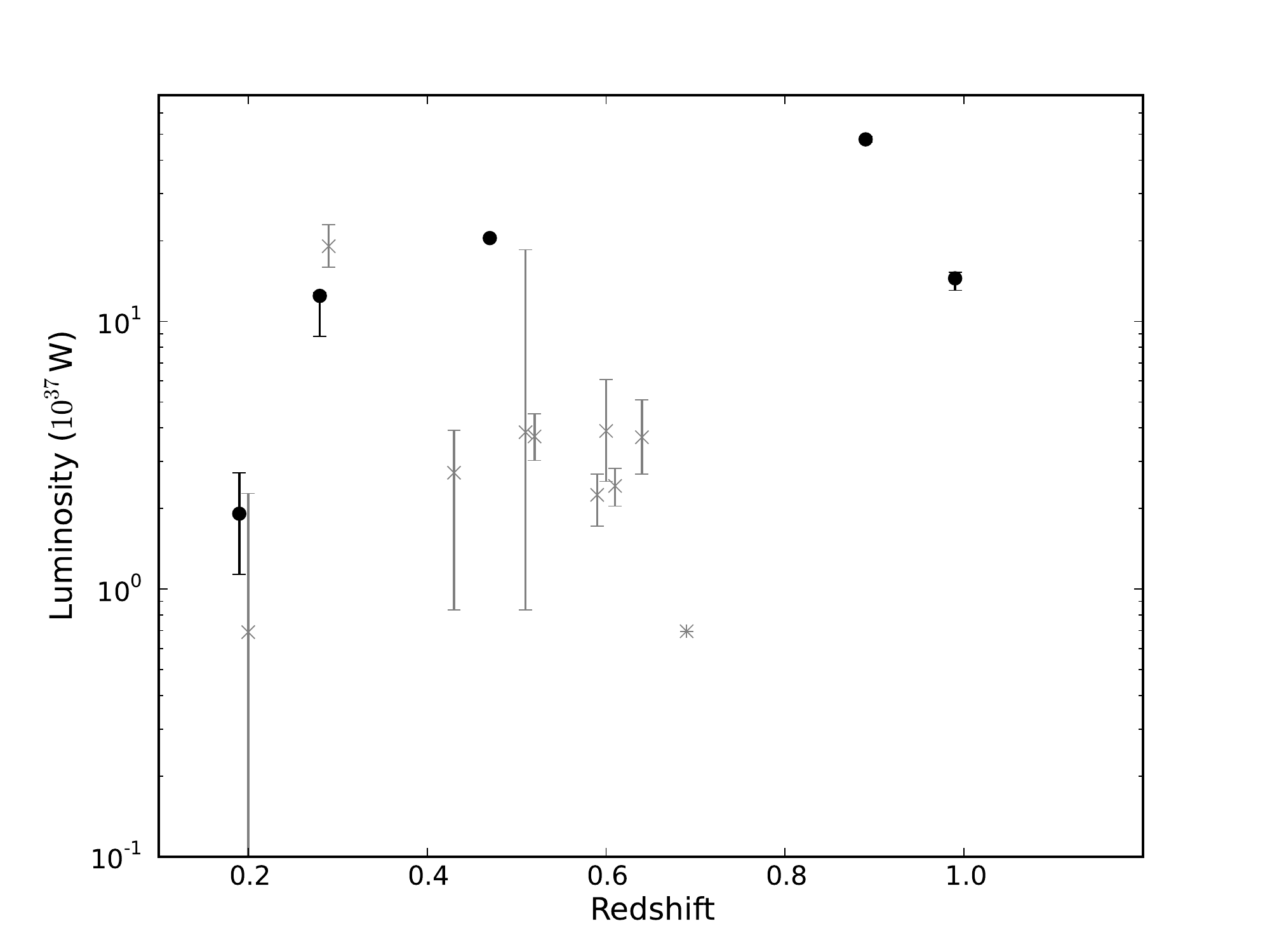}
\includegraphics[width=7.9cm,clip=,angle=0.]{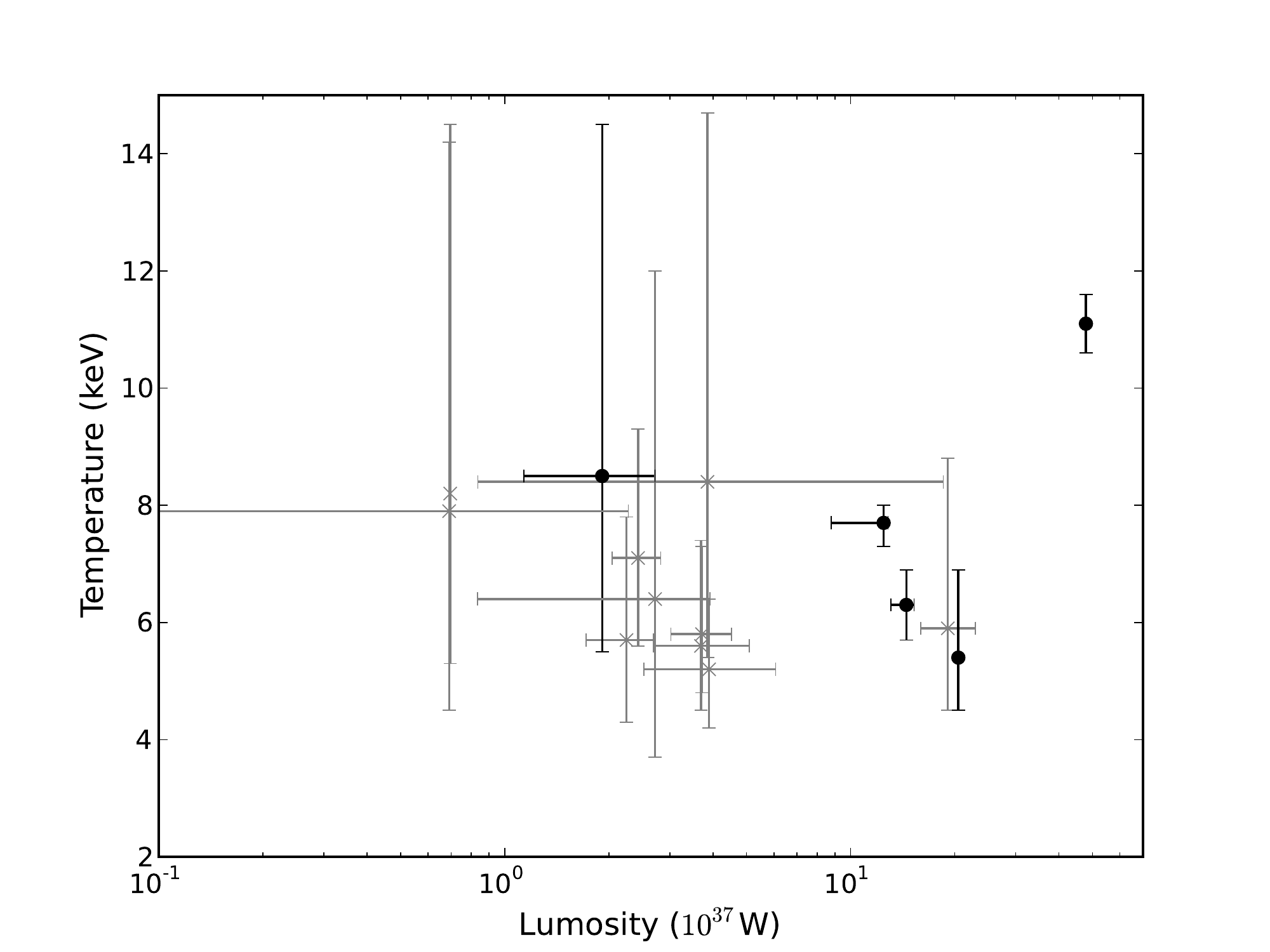}
\caption{The redshifts, temperatures and luminosities of the 15 XCS clusters %with $20^{\circ} < \delta < 80^{\circ}$  and a mean temperature greater than 5\,keV (\citealt{Mehrtens_2011}) 
that are in our SZ sample. 
%Clusters excluded from our sample due to significant point source contamination are represented with grey crosses and the remaining clusters are represented by black circles. Errors in the luminosities for XMJ0041+2526, XMJ0847+3448, XMJ1050+5737 and XMJ1115+5319 cannot be plotted as they are not provided in the XCS catalogue \citealt{Mehrtens_2011}. 
Errors in the luminosities for XMJ1050+5737 and XMJ1115+5319 cannot be plotted as they are not provided in the XCS catalogue. Clusters that are detected by AMI are indicated with black circles whereas those that are undetected by AMI are represented by grey $\times$ symbols. Our SZ sample spans $z$ (some are photometric estimates) of $0.19-0.99$, $L_{X}$ (0.05 -- 100\,keV band) of $0.7-47.9\times10^{37}$\,W and $T_{X}$ (XCS estimate) of $5.2-11$\,keV.
\label{fig:XMM-z-temp}}
\end{center}
\end{figure}

After the removal of the clusters a total of 15 hot clusters remain, to which we refer as our SZ sample.  The redshifts, temperatures, and luminosities for all 15 XCS clusters in the SZ sample {\textcolor{black} {are shown in Figure \ref{fig:XMM-z-temp} and listed with the other 19 clusters in our full sample in Table \ref{XMM_CLUSTERS_LIST}}}. There are nine lower-temperature XCS clusters within 15$\arcmin$ of these XCS clusters (see Table \ref{ALL_CLUSTERS_CLOSE}).  We have searched the literature for other known clusters within the 15$\arcmin$ of the SZ sample clusters and eight additional clusters -- not in the XCS catalogue -- were found. These additional clusters are shown in Table  \ref{ALL_CLUSTERS_CLOSE} and were found in the maxBCG, (\citealt{Koester_2007}), MCXC (\citealt{Piffaretti_2011}) and Zwicky (\citealt{Abell_1995}) catalogues. We have included Abell 1758A (13:32:43.10 +50:32:58.99) in Table \ref{ALL_CLUSTERS_CLOSE} even though this cluster is the same as XMJ1332+5031. (The \cite{Piffaretti_2011} position may be more accurate as some of the X-ray emission from XMJ1332+5031 extends off the edge of the XCS search region.)

% Have only tried MaxBCG, Abell + Zwicky, Planck, MaxBCG -- I'm sure to be missing a couple

\begin{table*}
\caption{A summary of the X-ray-derived cluster properties from \citealt{Mehrtens_2011} (also see Figure \ref{fig:XMM-z-temp}), the sensitivity of our 16-GHz AMI observations and the details of the observed radio source environments towards the 34 XCS clusters comprising our full cluster sample, which has been selected according to $20^{\circ} < \delta < 80^{\circ}$ and $T_{\rm{X}}>5$\,keV. Redshifts were obtained by different methods: ``spec" means optical spectroscopic redshift; ``phot" means photometric redshift derived from single colour  ($r-z$)  images by an algorithm based \citealt{Gladders_2000}; and ``X-ray" implies X-ray spectroscopic redshift. The X-ray luminosity is from the 0.05 -- 100\,keV band. Errors on the luminosities of XMJ0041+2526, XMJ0847+3448, XMJ1050+5737 and XMJ1115+5319 are not provided in the XCS catalogue. The X-ray counts are background subtracted. Given our LA rastering technique (Figure \ref{fig:XMJ1226+3332-LA-noise}), we have presented inner and outer thermal-noise estimates where the noise in the region of the hot XCS cluster is the inner estimate. {\textcolor{black} {In the final column we indicate which clusters are in the SZ sample.}}
} % Cluster list from Romer email 15/07/11, all above dec 20 with T >5keV
 \label{XMM_CLUSTERS_LIST}
\begin{tabular}{lcccccccccc}
\hline
Identity  & Redshift & X-Ray      & X-Ray       & X-ray & $\sigma_{SA}$          & Inner & Outer  & Total LA & Peak SA & {\textcolor{black} {SZ}} \\ 
          &          & luminosity & temperature & counts & (0.6k$\lambda$ taper) &     $\sigma_{LA}$                 &    $\sigma_{LA}$                 & flux density     & flux density & {\textcolor{black} {sample}} \\ 
           &          & inside $r_{200}$  & keV         & & mJy             &  mJy         &      mJy     & within 10'  & mJy \\ 
          &          & 1$\times10^{37}$\,W &             &        &                &                     &                     &    mJy            & \\ \hline 
 XMJ0041+2526 & 0.15 (spec) & $1.13$ & $14.7^{+8.4}_{-3.7}$ & 1410 & 0.29 (0.32) & 0.16 & 0.14 & 15.8 & 13.4 & {\textcolor{black} {$\times$}}\\ [0.2em]
 XMJ0046+4204 & 0.30 (X-ray) & $6.99^{+0.25}_{-0.27}$ & $6.9^{+0.6}_{-0.6}$ & 10443 & 0.20 (0.29) & 0.08 & 0.09 & 21.3 & 10.6 & {\textcolor{black} {$\times$}}\\ [0.2em]
 XMJ0110+3305 & 0.60 (phot) & $3.9^{+2.17}_{-1.37}$ & $5.2^{+1.2}_{-1.0}$ & 587 & 0.11 (0.12) & 0.10 & 0.11 & 6.8 & 5.5 & {\textcolor{black} {$\surd$}}\\ [0.2em]
 XMJ0116+3303 & 0.64 (phot) & $3.69^{+1.4}_{-1.0}$ & $5.6^{+1.8}_{-1.1}$ & 423 & 0.16 (0.22) & 0.14 & 0.15 & 11.5 & 6.6 & {\textcolor{black} {$\surd$}}\\ [0.2em]
 XMJ0515+7939 & 0.61 (phot) & $2.43^{+0.39}_{-0.39}$ & $7.1^{+2.2}_{-1.5}$ & 347 & 0.12 (0.15) & 0.08 & 0.09 & 10.4 & 3.6 & {\textcolor{black} {$\surd$}}\\ [0.2em]
 XMJ0830+5241 & 0.99 (X-ray) & $14.48^{+0.76}_{-1.43}$ & $6.3^{+0.6}_{-0.6}$ & 3674 & 0.10 (0.13) & 0.06 & 0.09 & 8.8 & 2.1 & {\textcolor{black} {$\surd$}}\\ [0.2em]
 XMJ0837+5532 & 0.28 (phot) & $0.65^{+0.55}_{-0.37}$ & $5.2^{+8.5}_{-2.1}$ & 136 & 1.09 (1.47) & 0.12 & 0.13 & 50.6 & 20.4 & {\textcolor{black} {$\times$}}\\ [0.2em]
 XMJ0847+3448 & 0.56 (spec) & $4.85$ & $5.6^{+0.7}_{-0.5}$ & 1494 & 0.45 (0.55) & 0.14 & 0.12 & 41.5 & 12.2 & {\textcolor{black} {$\times$}}\\ [0.2em]
 XMJ0901+6006 & 0.29 (phot) & $19.09^{+3.86}_{-3.16}$ & $5.9^{+2.9}_{-1.4}$ & 1379 & 0.14 (0.17) & 0.14 & 0.14 & 7.5 & 3.8 & {\textcolor{black} {$\surd$}}\\ [0.2em]
 XMJ0916+3027 & 0.59 (phot) & $2.25^{+0.44}_{-0.53}$ & $5.7^{+2.1}_{-1.4}$ & 295 & 0.15 (0.21) & 0.13 & 0.13 & 5.2 & 3.7 & {\textcolor{black} {$\times$}}\\ [0.2em]
 XMJ0918+2114 & 1.01 (spec) & $3.13^{+0.53}_{-0.63}$ & $8.3^{+5.3}_{-2.9}$ & 289 & 0.75 (1.03) & 0.13 & 0.13 & 2.0 & 25.7 & {\textcolor{black} {$\times$}}\\ [0.2em]
 XMJ0923+2256 & 0.19 (spec) & $1.91^{+0.81}_{-0.78}$ & $8.5^{+6.0}_{-3.0}$ & 713 & 0.17 (0.27) & 0.14 & 0.15 & 3.6 & 1.9 & {\textcolor{black} {$\surd$}}\\ [0.2em]
 XMJ0925+3059 & 0.52 (phot) & $3.72^{+0.80}_{-0.70}$ & $5.8^{+1.5}_{-1.0}$ & 1015 & 0.12 (0.17) & 0.09 & 0.14 & 7.1 & 6.3 & {\textcolor{black} {$\surd$}}\\ [0.2em]
 XMJ0953+6947 & 0.21 (spec) & $1.01^{+2.98}_{-0.65}$ & $5.7^{+1.1}_{-0.7}$ & 2291 & 2.85 (4.19) & 0.21 & 0.26 & 9.8 & 44.0 & {\textcolor{black} {$\times$}}\\ [0.2em]
 XMJ1000+6839 & 0.47 (phot) & $1.79^{+0.85}_{-1.12}$ & $5.4^{+1.3}_{-0.9}$ & 720 & 0.20 (0.26) & 0.12 & 0.12 & 12.0 & 6.2 & {\textcolor{black} {$\surd$}}\\ [0.2em]
 XMJ1031+3113 & 0.37 (spec) & $0.39^{+0.36}_{-0.25}$ & $5.8^{+8.7}_{-2.2}$ & 314 & 0.58 (0.82) & 0.11 & 0.12 & 6.0 & 29.4 & {\textcolor{black} {$\times$}}\\ [0.2em]
 XMJ1050+5737 & 0.69 (spec) & $0.7$ & $8.2^{+6.3}_{-2.9}$ & 377 & 0.13 (0.15) & 0.08 & 0.09 & 10.4 & 1.2 & {\textcolor{black} {$\surd$}}\\ [0.2em]
 XMJ1053+5735 & 1.13 (spec) & $3.27^{+0.46}_{-0.29}$ & $5.7^{+0.8}_{-0.6}$ & 1903 & 0.12 (0.15) & 0.09 & 0.08 & 29.2 & 19.4 & {\textcolor{black} {$\times$}}\\ [0.2em]
 XMJ1104+3544 & 0.56 (phot) & $4.32^{+6.01}_{-4.04}$ & $10.7^{+14.6}_{-4.4}$ & 178 & 0.14 (0.16) & 0.10 & 0.11 & 24.6 & 7.0 & {\textcolor{black} {$\times$}}\\ [0.2em]
 XMJ1109+4827 & 0.51 (phot) & $3.38^{+0.37}_{-0.40}$ & $6.2^{+1.6}_{-1.4}$ & 1164 & 0.18 (0.19) & 0.11 & 0.13 & 20.6 & 26.2 & {\textcolor{black} {$\times$}}\\ [0.2em]
 XMJ1115+5319 & 0.47 (spec) & $20.48$ & $5.4^{+1.5}_{-0.9}$ & 1359 & 0.21 (0.26) & 0.09 & 0.11 & 10.0 & 1.5 & {\textcolor{black} {$\surd$}}\\ [0.2em]
 XMJ1122+4659 & 0.44 (phot) & $0.68^{+0.09}_{-0.20}$ & $5.4^{+2.6}_{-1.8}$ & 575 & 0.11 (0.15) & 0.10 & 0.10 & 10.0 & 1.9 & {\textcolor{black} {$\times$}}\\ [0.2em]
 XMJ1217+4729 & 0.27 (spec) & $23.2^{+13.1}_{-10.8}$ & $9.8^{+6.6}_{-3.7}$ & 188 & 0.15 (0.21) & 0.11 & 0.11 & 4.1 & 8.3 & {\textcolor{black} {$\times$}}\\ [0.2em]
 XMJ1217+4728 & 0.27 (phot) & $0.42^{+0.15}_{-0.18}$ & $5.5^{+6.7}_{-2.5}$ & 192 & 0.17 (0.22) & 0.10 & 0.11 & 2.5 & 13.3 & {\textcolor{black} {$\times$}}\\ [0.2em]
 XMJ1226+3332 & 0.89 (phot) & $47.87^{+1.13}_{-1.10}$ & $11.1^{+0.5}_{-0.5}$ & 16801 & 0.12 (0.18) & 0.05 & 0.08 & 7.3 & 4.6 & {\textcolor{black} {$\surd$}}\\ [0.2em]
 XMJ1259+2830 & 0.52 (phot) & $3.13^{+0.18}_{-0.48}$ & $8.5^{+1.7}_{-1.6}$ & 1991 & 1.12 (1.24) & 0.12 & 0.15 & 2.3 & 105.8 & {\textcolor{black} {$\times$}}\\ [0.2em]
 XMJ1309+5739 & 0.2 (phot) & $0.69^{+1.59}_{-0.64}$ & $7.9^{+6.3}_{-3.4}$ & 526 & 0.12 (0.15) & 0.12 & 0.13 & 6.7 & 4.3 & {\textcolor{black} {$\surd$}}\\ [0.2em]
 XMJ1332+5031 & 0.28 (spec) & $12.46^{+0.37}_{-3.66}$ & $7.7^{+0.3}_{-0.4}$ & 2596 & 0.10 (0.13) & 0.09 & 0.09 & 11.4 & 5.6 & {\textcolor{black} {$\surd$}}\\ [0.2em]
 XMJ1406+2830 & 0.55 (spec) & $2.49^{+3.76}_{-0.97}$ & $5.6^{+1.8}_{-1.3}$ & 325 & 1.98 (2.46) & 0.19 & 0.25 & 13.0 & 185.1 & {\textcolor{black} {$\times$}}\\ [0.2em]
 XMJ1418+2511 & 0.29 (spec) & $6.26^{+0.46}_{-0.42}$ & $6.4^{+0.4}_{-0.4}$ & 5203 & 0.16 (0.19) & 0.12 & 0.13 & 28.5 & 7.7 & {\textcolor{black} {$\times$}}\\ [0.2em]
 XMJ1423+3828 & 0.43 (phot) & $2.72^{+1.20}_{-1.89}$ & $6.4^{+5.6}_{-2.7}$ & 197 & 0.14 (0.17) & 0.09 & 0.11 & 6.1 & 8.6 & {\textcolor{black} {$\surd$}}\\ [0.2em]
 XMJ1429+4241 & 0.92 (spec) & $10.84^{+11.7}_{-6.62}$ & $5.4^{+0.3}_{-0.3}$ & 1887. & 0.81 (0.86) & 0.09 & 0.10 & 28.8 & 13.5 & {\textcolor{black} {$\times$}}\\ [0.2em]
 XMJ1437+3415 & 0.51 (phot) & $3.86^{+14.68}_{-3.02}$ & $8.4^{+6.3}_{-3.0}$ & 376 & 0.12 (0.15) & 0.08 & 0.09 & 6.4 & 2.8 & {\textcolor{black} {$\surd$}}\\ [0.2em]
 XMJ1542+5359 & 0.64 (phot) & $4.74^{+2.36}_{-3.29}$ & $5.3^{+3.0}_{-1.6}$ & 338 & 0.47 (0.58) & 0.10 & 0.11 & 3.4 & 23.7 & {\textcolor{black} {$\times$}}\\ [0.2em]\hline
 \end{tabular}
\end{table*}

\begin{table*}
\caption{A summary of the additional 17 clusters catalogued in the literature that we found within 15$\arcmin$ of the 15 clusters in our SZ sample (see Section \ref{Sec:cluster-sample}). Nine of these are low temperature ($<5$keV) XCS clusters. For three of these low temperature XCS clusters there is no derived X-ray temperature. We have included the \citealt{Piffaretti_2011}  position for XMJ1332+5031 (Abell 1758A) because the XCS coordinates are affected by part of the X-ray cluster emission being beyond the edge of the XCS search area (see Figure \ref{Fig:SA-obs1}). The redshift and other parameters for the \citealt{Abell_1995} cluster close to XMJ0901+6006 are unknown. The \citealt{Piffaretti_2011} catalogue contains the $r_{500}$ rest-frame luminosity (0.1-2.4\,keV) rather than at $r_{200}$ but contains no estimate of the cluster temperature. References 1-4 correspond to \citealt{Piffaretti_2011}, \citealt{Mehrtens_2011}, \citealt{Abell_1995} and \citealt{Koester_2007} respectively.} %We have been unable to find a luminosity or temperature estimate for each of these additional clusters.}
 \label{ALL_CLUSTERS_CLOSE}
 \begin{tabular}{lcccccccc}
 \hline
Cluster Name & Right Ascension  & Declination & Redshift & Luminosity & Temperature & Close to & Separation  & Reference  \\ 
  &    &  &  & inside $r_{200}$  & keV &  & ($\arcmin$) &   \\ 
  &    &  &  & 1$\times10^{37}$\,W  &  &   &   &   \\ 
\hline 
 NGC 0410 & 01:10:58.10 & 33:08:58.00 &  0.018 & 0.025 ($r_{500}$)  & -- & XMJ0110+3305 & 7.9 & 1   \\ [0.2em]
 XMJ0116+3257 & 01:16:24.20 & 32:57:17.10 & 0.45 (phot) & $0.45^{+0.39}_{-0.19}$ & $1.4^{+0.3}_{-0.2}$ & XMJ0116+3303& 6.4 & 2 \\ [0.2em]
 XMJ0831+5234 & 08:31:15.00 & 52:34:53.89 & 0.52 (phot) & -- & -- & XMJ0830+5241& 9.9 &  2 \\ [0.2em]
  %XMJ0847+3451 & 8:47:1.40 & 34:51:12.80 & 0.43 (phot) & $0.8^{0.12}_{-0.11}$ & $2.6^{0.3}_{-0.5}$ & XMJ0847+3448& 3.21 & \cite{Mehrtens_2011}\\ excluded cluster
 % XMJ1052+5730 & 10:52:36.60 & 57:30:54.29 & 0.61 (phot) & $-99.0^{-99.0}_{--99.0}$ & $-99.0^{-99.0}_{--99.0}$ & XMJ1053+5735& 10.26 & \cite{Mehrtens_2011} \\ excluded cluster
 % XMJ1053+5720 & 10:53:18.50 & 57:20:43.69 & 0.34 (spec) & $0.48^{0.16}_{-0.29}$ & $2.2^{0.2}_{-0.1}$ & XMJ1053+5735& 14.97 & \cite{Mehrtens_2011} \\ excluded cluster
 % XMJ1217+4728 & 12:17:54.50 & 47:28:53.79 & 0.27 (phot) & $0.42^{0.15}_{-0.18}$ & $5.5^{6.7}_{-2.5}$ & XMJ1217+4729& 1.74  & \cite{Mehrtens_2011} \\ excluded cluster
 % XMJ1217+4729 & 12:17:44.60 & 47:29:21.50 & 0.27 (spec) & $23.2^{13.11}_{-10.81}$ & $9.8^{6.6}_{-3.7}$ & XMJ1217+4728& 1.74 & \cite{Mehrtens_2011} \\ excluded cluster
 % XMJ1406+2833 & 14:6:46.0 & 28:33:8.30 & 0.52 (phot) & $1.04^{-99.0}_{--99.0}$ & $2.3^{1.4}_{-0.6}$ & XMJ1406+2830& 7.03  & \cite{Mehrtens_2011} \\ excluded cluster
 % XMJ1417+2511 & 14:17:31.20 & 25:11:39.10 & 0.18 (spec) & $0.03^{-99.0}_{--99.0}$ & $1.8^{0.8}_{-0.4}$ & XMJ1418+2511& 13.83 & \cite{Mehrtens_2011} \\ excluded cluster
Zwicky 2094 & 09:02:26.40 &  60:16:12.00 & -- & --& --  &  XMJ0901+6006 & 14.6 &  3  \\ [0.2em]
maxBCG\\141.2857+31.0615  & 09:25:08.57 & 31:03:41.47 &  0.205 & --  & -- & XMJ0925+3059 & 9.3 &  4  \\ [0.2em]
 XMJ0925+3054 & 09:25:44.40 & 30:54:31.89 & 0.41 (phot) & $0.61^{+0.42}_{-0.4}$ & $1.3^{+0.3}_{-0.2}$ & XMJ0925+3059& 4.4 & 2 \\ [0.2em]
 XMJ0926+3103 & 09:26:41.40 & 31:03:08.80 & 0.67 (phot) & -- &-- & XMJ0925+3059& 12.7 &  2 \\ [0.2em]
 XMJ0926+3101 & 09:26:50.70 & 31:01:27.30 & 0.49 (phot) & $0.71^{+0.12}_{-0.27}$ & $3.5^{+1.9}_{-0.9}$ & XMJ0925+3059& 14.1 &  2  \\ [0.2em]
XMJ1226+3345 & 12:26:43.30 & 33:45:49.90 & 0.63 (phot) & -- & --& XMJ1226+3332& 13.3  &  2 \\ [0.2em]
XMJ1226+3343 & 12:26:56.40 & 33:43:29.40 & 0.50  & $3.83^{+3.26}_{-2.15}$ & $4.8^{+0.4}_{-0.3}$ & XMJ1226+3332& 10.7  &  2   \\ [0.2em]
maxBCG\\186.7603+33.3155 & 12:27:02.48 & 33:18:55.92 &  0.257 & --  &  --& XMJ1226+3332 & 14.0 &   2 \\ [0.2em]
A1758B & 13:32:30.20 & 50:24:32.00 &  0.280 &  10.99 ($r_{500}$)   & -- & XMJ1332+5031 & 8.3 & 1   \\  [0.2em]
A1758A & 13:32:43.10 & 50:32:58.99 &  0.280 &   4.18 ($r_{500}$)  &--  & XMJ1332+5031 & 2.2 & 1   \\ [0.2em] % COULD MOVE THIS TO XCS TABLE 
BVH2007 173 & 13:34:20.40 & 50:31:05.00 &  0.620 &   3.40 ($r_{500}$)  & -- & XMJ1332+5031 & 13.6 & 1   \\ [0.2em]
% A1758A & 13:32:42.38 & 50:31:39.36 &  0.2799 & XMJ1332+5031 & 1.99 & Planck \\  ALREADY THE MAIN XCS CLUSTER
%  maxBCG203.1601+50.5599& 13:32:38.42 & 50:33:35.70 &  0.284 & XMJ1332+5031 & 3.11 & maxBCG \\ ALREADY THE MAIN XCS CLUSTER
maxBCG \\ 215.6632+38.301 & 14:22:39.16 & 38:18:03.50 &  0.192 &  -- &  --& XMJ1423+3828 & 11.3 &  4  \\ [0.2em]
XMJ1437+3414 & 14:37:24.60 & 34:14:27.90 & 0.44 (phot) & $0.46^{+1.63}_{-0.31}$ & $2.2^{+1.0}_{-0.6}$ & XMJ1437+3415& 2.3  & 2  \\ [0.2em]
 XMJ1437+3408 & 14:37:43.20 & 34:08:07.80 & 0.49 (phot) & $2.23^{+1.06}_{-0.89}$ & $4.2^{+2.8}_{-1.6}$ & XMJ1437+3415& 9.3  & 2   \\[0.2em] \hline
%Abell 1758 & 13:32:32.8 &  50:31:36.87 & 0.0 & XMJ1332+5031 & 3.62 & Abell\\ \\ ALREADY THE MAIN XCS CLUSTER
%Zwicky 2358 & 9:25:0.0 &  31:2:59.99 & 0.0 & XMJ0925+3059 & 10.6 & Zwicky\\ \\ ALREADY THE MAIN XCS CLUSTER
%Zwicky 3942 & 11:15:4.80 &  53:22:48.0 & 0.0 & XMJ1115+5319 & 3.38 & Zwicky\\ \\ ALREADY THE MAIN XCS CLUSTER
%Zwicky 5744 & 13:9:21.60 &  57:40:12.0 & 0.0 & XMJ1309+5739 & 1.79 & Zwicky\\ \\ ALREADY THE MAIN XCS CLUSTER
%Zwicky 6054 & 13:32:26.40 &  50:31:47.99 & 0.0 & XMJ1332+5031 & 4.51 & Zwicky\\ \\ ALREADY THE MAIN XCS CLUSTER
 \end{tabular}
\end{table*}

\section{AMI}

AMI consists of two arrays: the Large Array (LA) -- the source subtractor -- is an eight-element array of 12.8-m dishes with a resolution $\approx$30$\arcsec$; the Small Array (SA) -- principally the SZ array -- is a ten-element array of 3.7-m dishes with a resolution $\approx$3$\arcmin$. The resolution of the SA is designed to match the typical angular size of massive galaxy clusters, whereas the LA operates at higher resolution for accurate radio-source detection at the expense of lower surface brightness sensitivity. Both arrays operate with a 4.5-GHz bandwidth centred on 16\,GHz. A technical summary of AMI is presented in Table \ref{AMI_tech} and a more complete description is given in \cite{Zwart_2008}.
 
 %At this frequency the SZ signal is not significantly contaminated by primary CMB anisotropies but the radio source density is larger than at higher frequencies. Given the LA-measured positions, flux-densities and spectral indices of radio sources, the effects of these contaminating sources can largely be removed from the SA data leaving signals from just the SZ decrement, CMB imprints, and faint radio sources below the LA detection limit. A technical summary of AMI is presented in Table \ref{AMI_tech} and a more complete description is given in \cite{Zwart_2008}.

\begin{table}
\caption{AMI summary.}
 \label{AMI_tech}
\begin{tabular}{lccc}
\hline 
   & SA  & LA   \\\hline 
Antenna diameter & 3.7\,m & 12.8\,m \\ 
Number of antennas & 10 & 8 \\ 
Number of baselines & 45 & 28 \\ 
Baseline length & 5--20\,m & 18--110\,m \\ 
16-GHz power primary beam FWHM  & 19.6$'$ & 5.6$'$ \\ 
Synthesized beam FWHM &  $\approx$ 3$'$ & $\approx$ 30$''$ \\ 
Array flux-density sensitivity & 30\,mJy\,$\rm{s^{-1/2}}$ & 3\,mJy\,$\rm{s^{-1/2}}$ \\ 
Frequency range & \multicolumn{2}{|c|}{13.5--18.0\,GHz} \\ 
Bandwidth & \multicolumn{2}{|c|}{4.5\,GHz} \\ 
Number of channels & \multicolumn{2}{|c|}{6} \\ 
Channel bandwidth & \multicolumn{2}{|c|}{0.75\,GHz} \\ 
Polarization measured & \multicolumn{2}{|c|}{I + Q} \\ \hline
 \end{tabular}
\end{table}

\section{Observations}

The  SA and LA were used to observe the 34 XCS clusters between 2009 November 6 and 2012 June 6. SA observations were centred on the clusters' X-ray coordinates and were interleaved every hour with 400 second phase-calibrator observations. For LA observations we typically used a 61+19pt hexagonal raster \footnote{A ``61+19 point raster" observation consists of 61 pointings separated by 4$\arcmin$ forming a hexagonal shape, of which the inner 19 pointings have lower noise levels. As a result, the area of interest in the vicinity of the cluster centroid is covered with deep observations and the shallower data on outer regions can be used to identify any relatively bright radio sources.} pointing grid centred on the cluster and these were interleaved every 10 min with 1 min phase-calibrator observations. For both arrays bi-daily observations of 3C48 or 3C286 are used for amplitude calibration (the assumed flux densities are consistent with the \citealt{Rudy_1987} model of Mars and are given in \citealt{Franzen_2011}). The bright phase calibrators are from \citealt{Patnaik_1992}, \citealt{Browne_1998} and \citealt{Wilkinson_1998}. %Catalogues produced from the Jodrell Bank VLA Survey. 

A summary of the AMI observations and the 16\,GHz source properties is given in Table \ref{XMM_CLUSTERS_LIST}.

%Lit search:

%XMJ0110+3305 -- 1 cluster, in MCXC sample (plus other samples but its the same cluster)
%XMJ0116+3303 -- no other large clusters
%XMJ0515+7939 -- no other large clusters
%XMJ0830+5241 -- no other large clusters
%XMJ0901+6006 -- no other large clusters (the cluster is a MAXBCG cluster -- there is a zwicky cluster towards the top left of our image, just north of the decrement)
%XMJ0916+3027 -- no other large clusters
%XMJ0923+2256 -- no there large clusters (the cluster is a MAXBCG cluster)

\section{Data reduction, mapping and source finding}

Initial data reduction was performed on the raw SA and LA data using the local software tool  \textsc{reduce}, following the techniques outlined in \cite{Davies_2009}, to produce calibrated data in $uv$ FITS format. In addition, the data quality was checked for systematic errors by using the two jack-knife tests that are described in \cite{Shimwell_2012}. We found no significant problem at the location of any of the 15 hot XCS clusters in our SZ sample.

\begin{figure}
\begin{center}
\includegraphics[width=8.0cm,clip=,angle=0.]{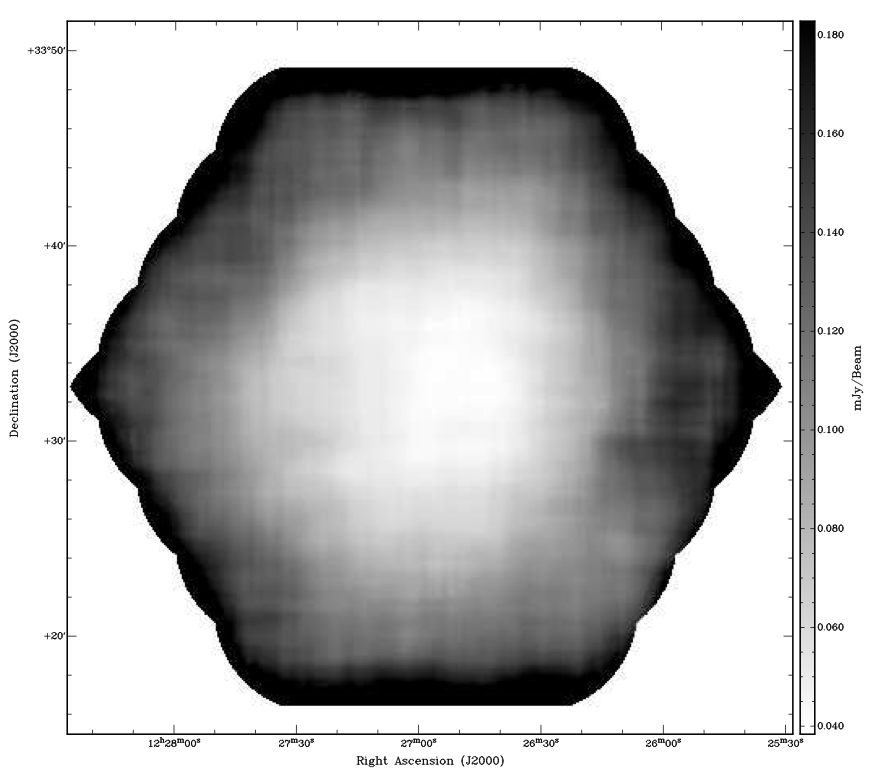}
\caption{The noise distribution for the LA primary beam corrected 61+19pt hexagonal grid of pointings centred on XMJ1226+3332  as calculated by \textsc{sourcefind}.}
\label{fig:XMJ1226+3332-LA-noise}
\end{center}
\end{figure}

The reduced LA data were mapped in \textsc{aips}\footnote{http://www.aips.nrao.edu} and \textsc{clean}ed to three times the thermal noise with a single \textsc{clean} box encompassing the primary beam. Images from individual LA pointings were then stitched together to form a raster image. Sources with flux densities greater than  4$\sigma_{LA}$ were identified on the LA maps by the \textsc{sourcefind} software package, which is described in \cite{Franzen_2011}. Unlike  \cite{Franzen_2011} we do not use the theoretical noise estimates and instead rely on \textsc{sourcefind} to determine the noise across the map. This technique has the advantages of identifying noisy regions around bright sources and of being more sensitive to confusion noise; it thus provides a more accurate estimate of the noise at a particular point on the image. An example of the noise variation across an LA image is given in Figure \ref{fig:XMJ1226+3332-LA-noise}. 

% It would be nice to have a description of the flood filling algorithm for reference to tom.

The reduced SA data were also mapped in \textsc{aips}. Images were \textsc{clean}ed to three times the thermal noise with a single \textsc{clean} box encompassing the primary beam. Using  \textsc{sourcefind}, the noise was determined across the SA maps; the noise on these maps typically varies by 10\% across the image but, in the regions of bright sources, the noise is higher. 

\begin{figure*}
\centerline{\includegraphics[width=6.0cm,,clip=,angle=0.]{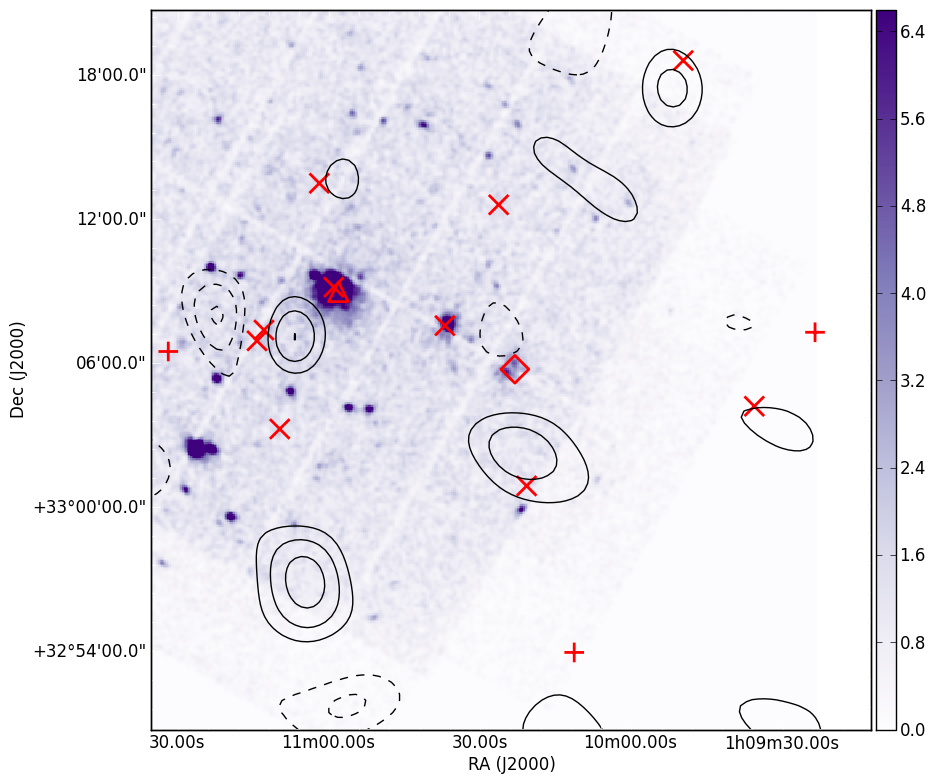}\qquad\includegraphics[width=6.0cm,,clip=,angle=0.]{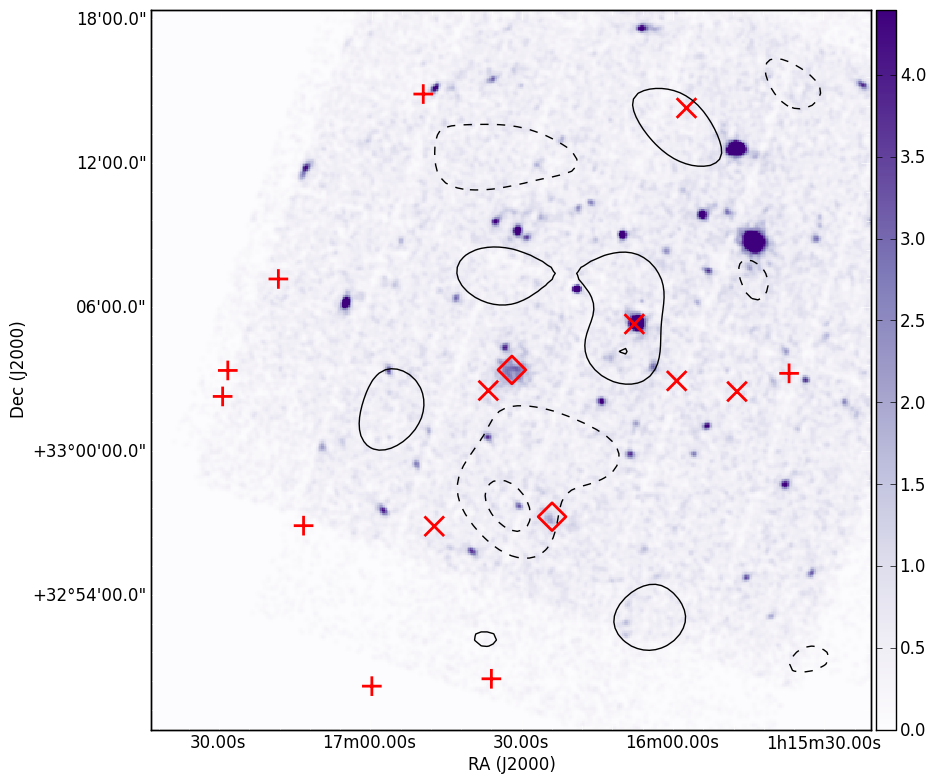}\includegraphics[width=6.0cm,,clip=,angle=0.]{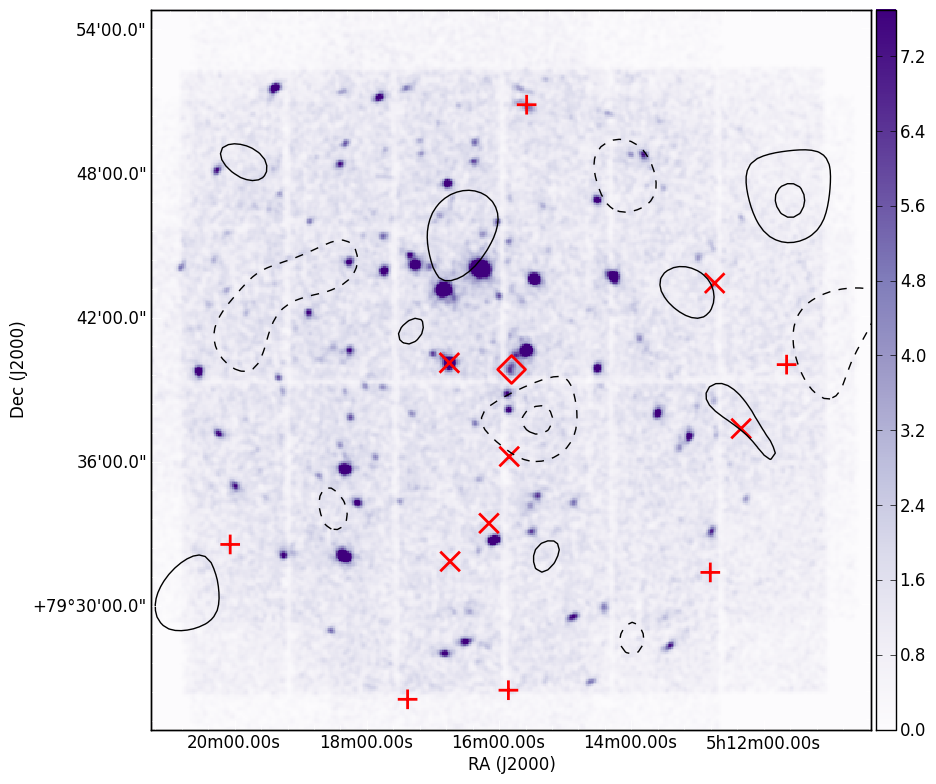}}

Left: XMJ0110+3305. Centre: XMJ0116+3303. Right: XMJ0515+7939.

\centerline{\includegraphics[width=6.0cm,,clip=,angle=0.]{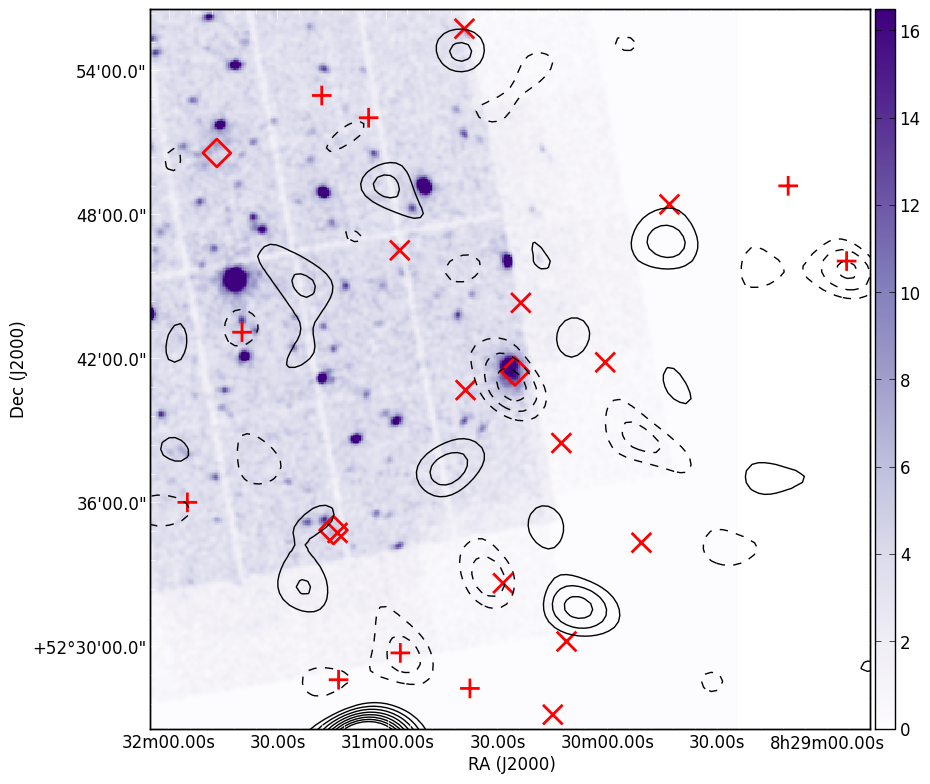}\qquad\includegraphics[width=6.0cm,,clip=,angle=0.]{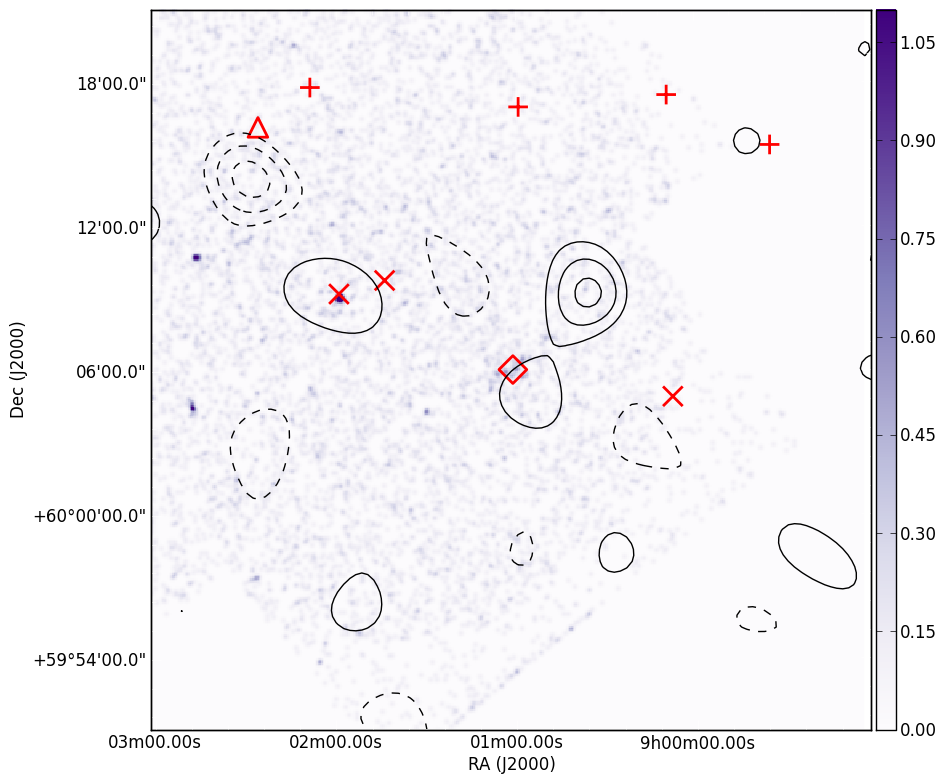}\includegraphics[width=6.0cm,,clip=,angle=0.]{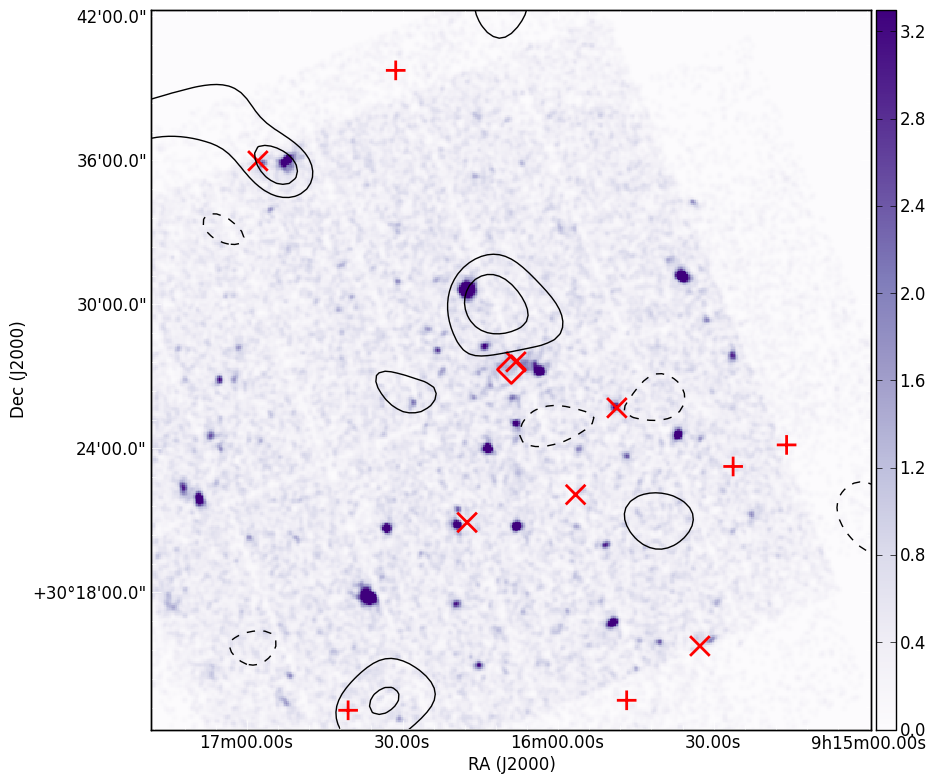}}

Left: XMJ0830+5241. Centre: XMJ0901+6006. Right: XMJ0916+3027.

\centerline{\includegraphics[width=6.0cm,,clip=,angle=0.]{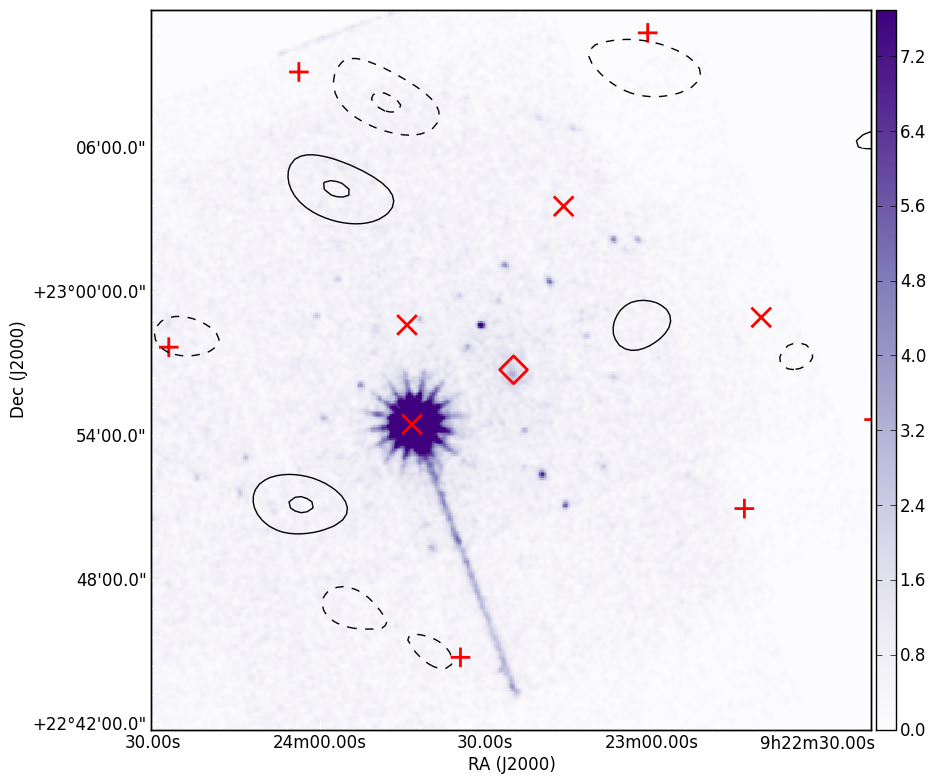}\qquad\includegraphics[width=6.0cm,,clip=,angle=0.]{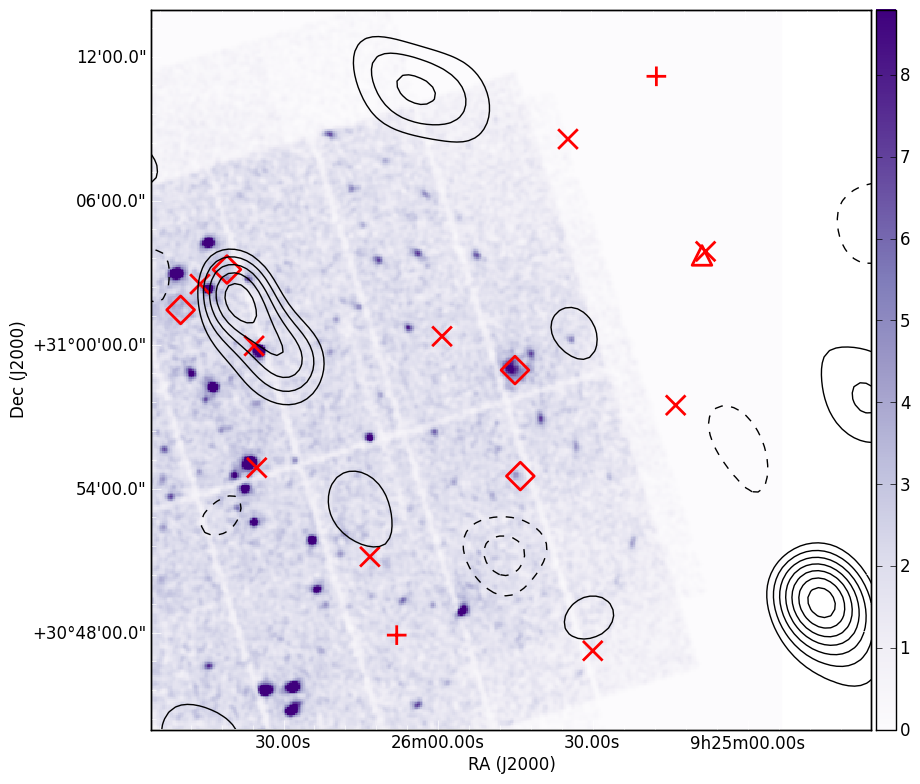}\includegraphics[width=6.0cm,,clip=,angle=0.]{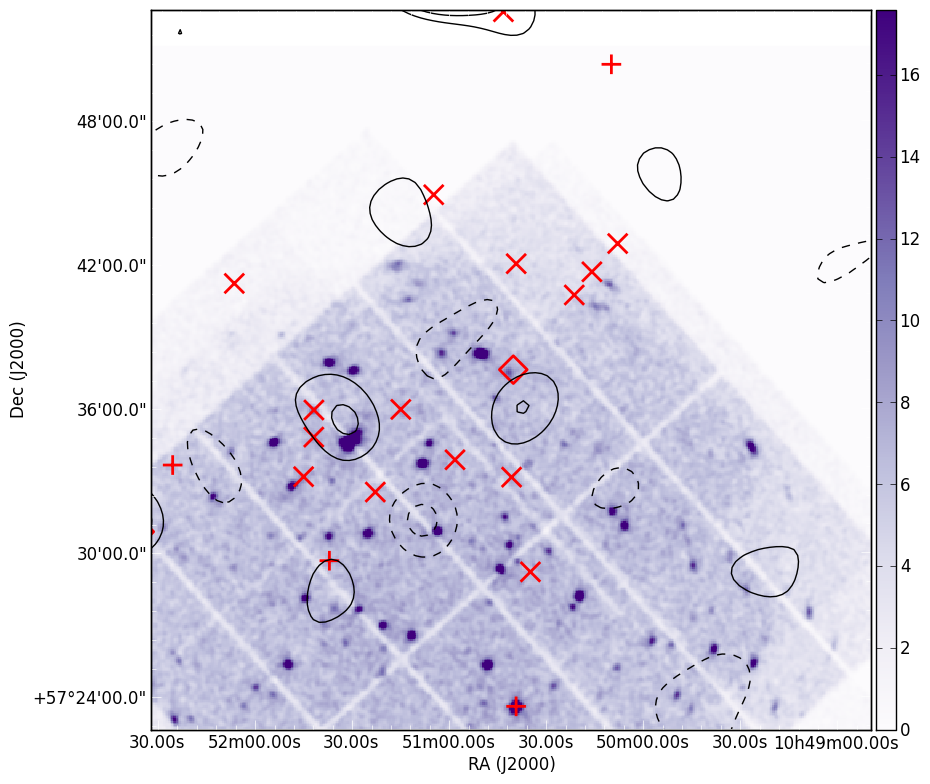}}

Left: XMJ0923+2256. Centre: XMJ0925+3059. Right: XMJ1050+5737.
\caption{XMM-Newton EPIC images of the 15 XCS galaxy clusters in our SZ sample overlaid with contours showing to the SA signal to noise ratio (see Table \ref{XMM_CLUSTERS_LIST} for the noise at the SA pointing centres). The X-ray images correspond to the observation IDs given in \citealt{Mehrtens_2011} and were taken from the XMM-Newton science archive. The colour bar shows the X-ray counts per pixel and the discontinuities in the X-ray images corresponds to the edge of the XMM-Newton detector. The SA contour levels are 2, 3, 4, 5, 6, 7, 8, 9, and 10 times the  SA receiver noise level.  Contours showing a positive SA signal to noise ratio are given by solid lines whilst dashed lines correspond to a negative signal to noise ratio. The $+$ symbols represent faint radio sources far from known clusters, the $\times$ symbols are brighter sources or faint sources closer to clusters, the $\Diamond$ symbols are XCS clusters (see Tables \ref{XMM_CLUSTERS_LIST} and \ref{ALL_CLUSTERS_CLOSE}) and the $\triangle$ symbols are other known clusters (see Table \ref{ALL_CLUSTERS_CLOSE}). Continued overleaf.}
\label{Fig:SA-obs1}
\end{figure*}

We present, in Figure \ref{Fig:SA-obs1},  maps of our 15 targets with contours of signal:noise, of which each is simply a  \textsc{clean}ed image of SA brightness divided by it \textsc{sourcefind} noise map.
%In this paper we present signal divided by noise maps which give an accurate indication of the significance of features on the SA images. These are created by simply dividing the \textsc{clean}ed SA images by \textsc{sourcefind} noise map. The noise levels presented in Table  \ref{XMM_CLUSTERS_LIST}, which are  for the centre of the SA images,  can be used to convert  signal divided by noise images to flux density maps.

\begin{comment}
\begin{figure}
\begin{center}
\includegraphics[width=8.0cm,clip=,angle=0.]{figures/XMJ1226+3332-SA-noise.png}
\caption{The noise distribution for the primary beam uncorrected pointed SA observation towards XMJ1226+3332 as calculated by \textsc{sourcefind}.}
\label{fig:XMJ1226+3332-SA-noise}
\end{center}
\end{figure}
\end{comment}

\addtocounter{figure}{0}

\begin{figure*}
\centerline{\includegraphics[width=6.0cm,,clip=,angle=0.]{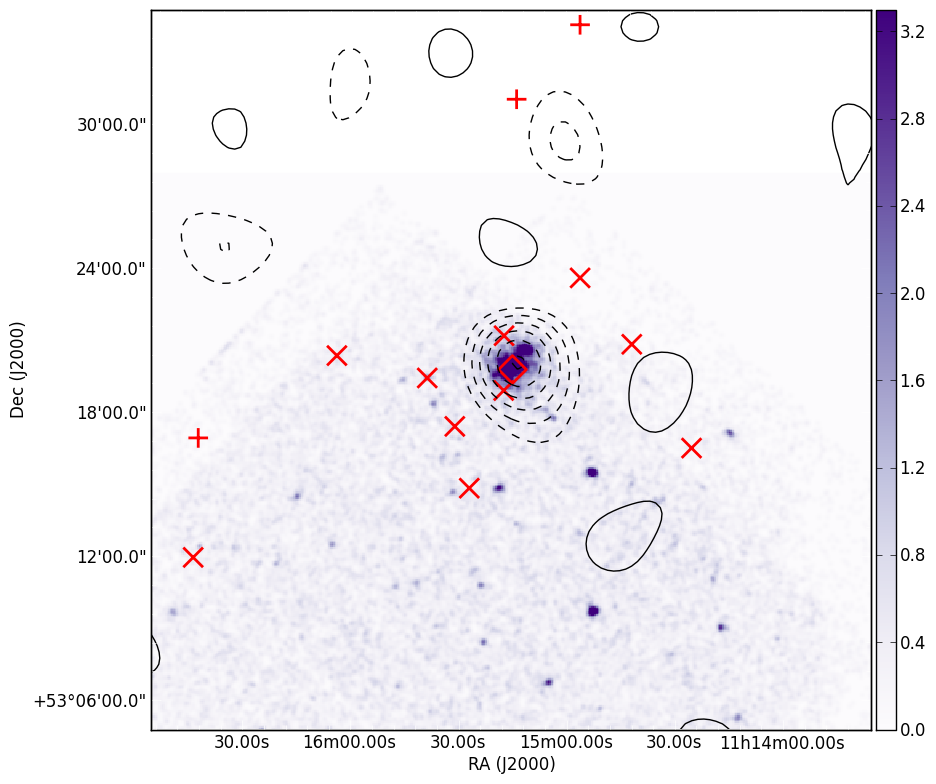}\qquad\includegraphics[width=6.0cm,,clip=,angle=0.]{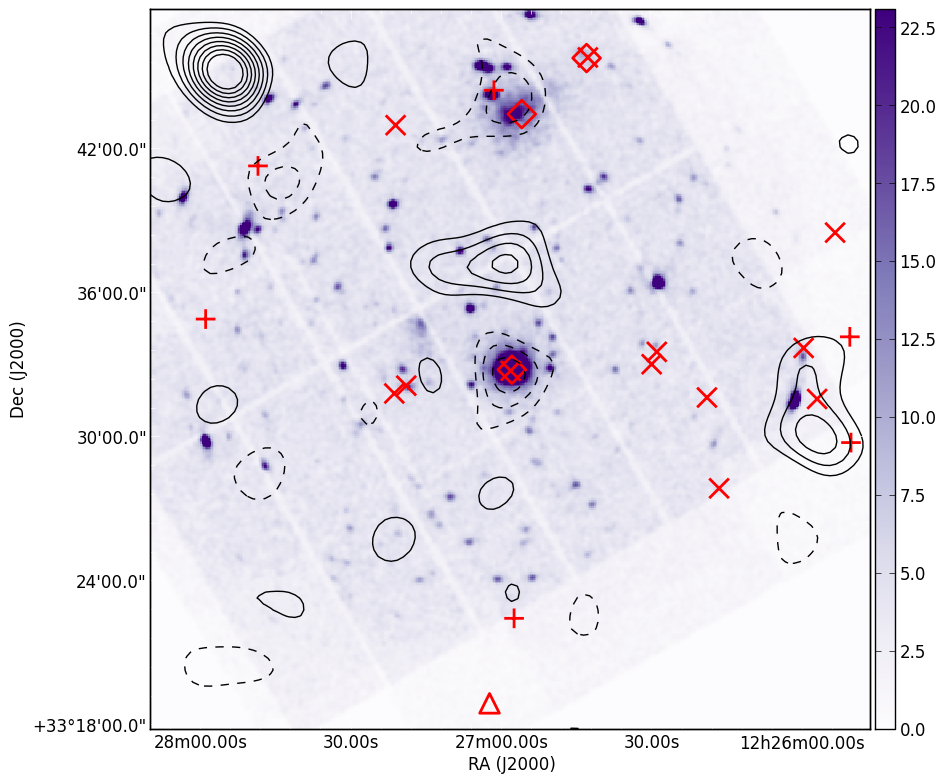}\includegraphics[width=6.0cm,,clip=,angle=0.]{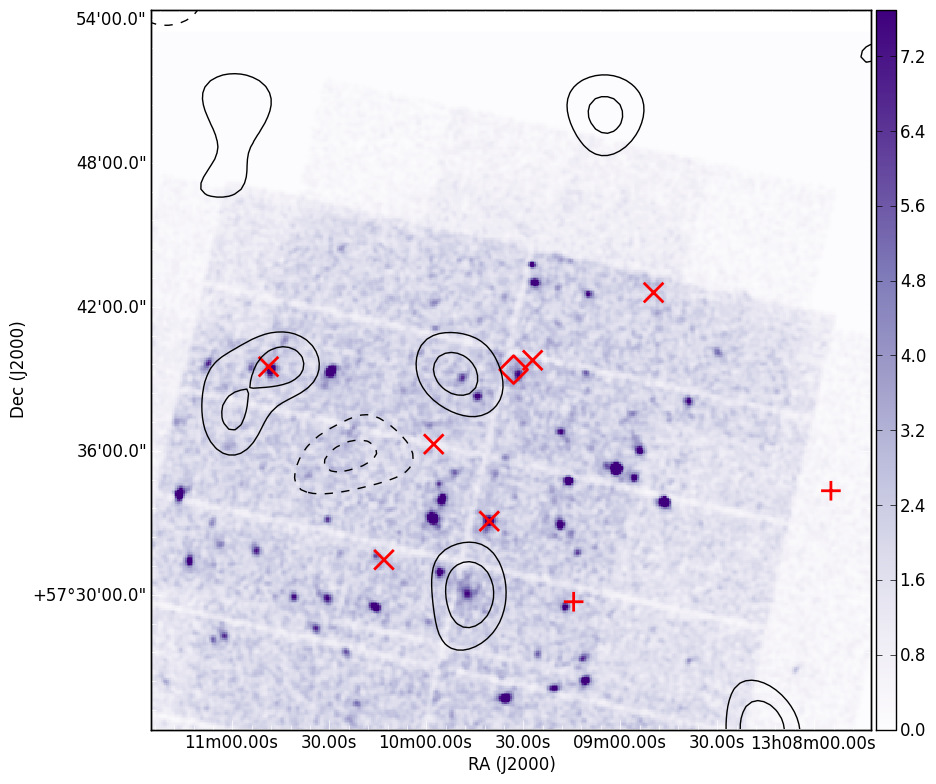}}

Left: XMJ1115+5319. Centre: XMJ1226+3332. Right: XMJ1309+5739.

\centerline{\includegraphics[width=6.0cm,,clip=,angle=0.]{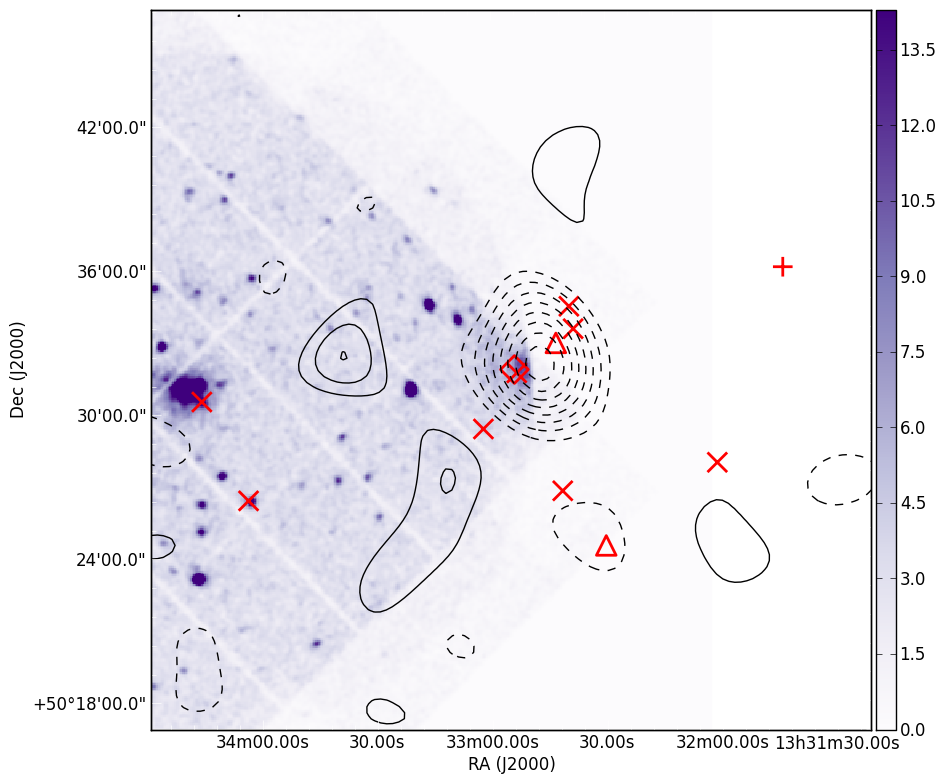}\qquad\includegraphics[width=6.0cm,,clip=,angle=0.]{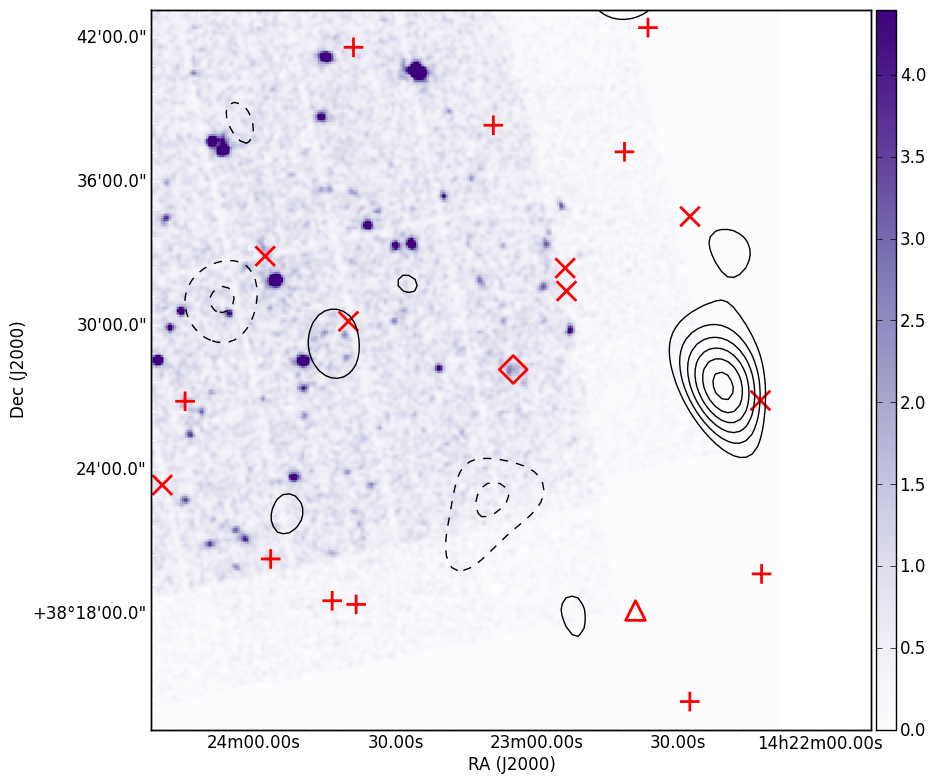}\includegraphics[width=6.0cm,,clip=,angle=0.]{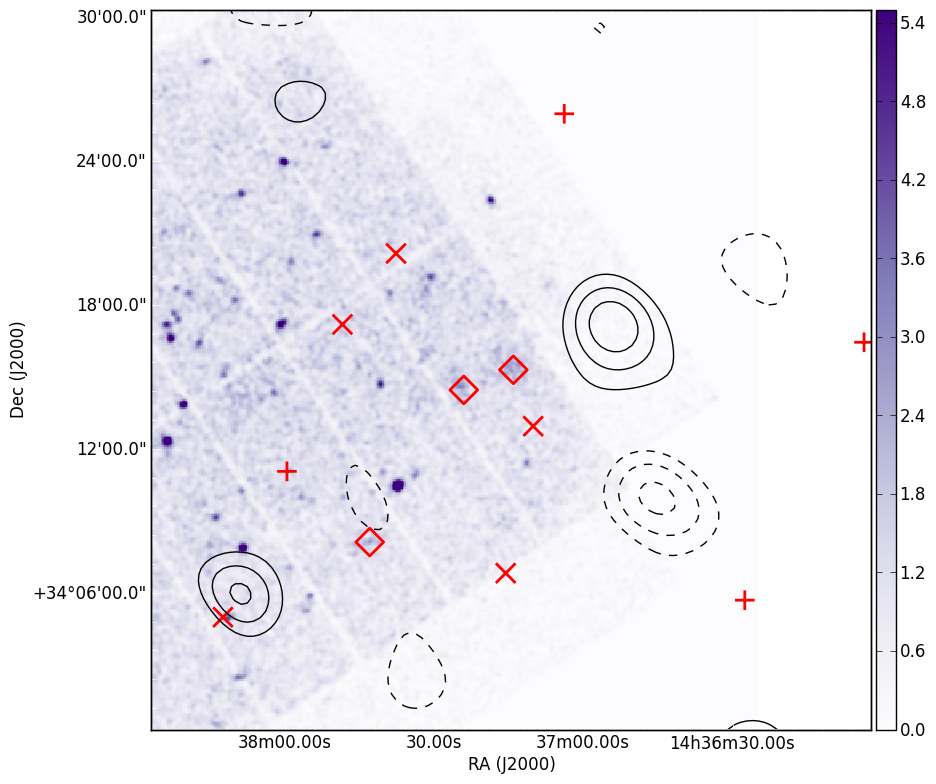}}

Left: XMJ1332+5031. Centre: XMJ1423+3828. Right: XMJ1437+3415.

\begin{minipage}[b]{1.0\linewidth}\centering
\vspace{0.2cm}
\bf{Figure 3 continued.}
\end{minipage}

\end{figure*}

\section{SZ Analysis}\label{SZ-analysis}

%{\bf{CHANGE TO BETA PROFILE FOR CONSISTENCY WITH XMM ANALYSIS???}}

The Bayesian analysis described in \cite{Feroz_2009} was used to model both galaxy clusters and radio sources. This applies the {\sc MultiNest} algorithm (\citealt{Feroz_2008a} and \citealt{Feroz_2008b}) to efficiently explore high-dimensional parameter spaces {\textcolor{black}{and calculate the Bayesian evidence}}. Here we use this software to estimate parameter probability distributions that describe the clusters and sources, and we use the Bayesian evidence for model comparison.

Within the framework of our analysis we account statistically for the contamination from the primary CMB anisotropies, we use our prior knowledge from high-resolution LA observations to constrain contaminating radio sources, and we model the SZ signal with a parameterised cluster model. %(see e.g. \citealt{Hurley-Walker_2011}, \citealt{Zwart_2011}, \citealt{Carmen_2011}, \citealt{Hurley-Walker_2012}, \citealt{Shimwell_2012} and \citealt{Carmen_2012}). 
We parameterise the SZ signal using the model described in \cite{Olamaie_2012}, which assumes the following: (a) an NFW (\citealt{Navarro_1997}) profile to model the density of the dark matter halo as a function of radius, $r$, %$\rho_{\rm{DR}}(r)$
\begin{align} 
  \rho_{\rm{DR}}(r) = \frac{\rho_{\rm{s}}}{\left(\frac{r}{R_{\rm{s}}}\right) \left( 1 + \frac{r}{R_{\rm{s}}} \right) ^2},
\end{align}
where $\rho_{\rm{s}}$ is a normalization density and $R_{\rm{s}}$ is the scale radius; (b) the electron pressure as a function of radius is described by a gNFW (\citealt{Nagai_2007}) profile,
\begin{align} 
 P_{\rm{e(r)}} =\frac{P_{\rm{e}i}}    {\left( \frac{r}{r_{\rm{p}}} \right)^c \left( 1 + \left( \frac{r}{r_{\rm{p}}}\right)^a \right)^{(b-c)/a}},
\end{align}
where $P_{\rm{e}i}$ is the normalised pressure, $r_{\rm{p}}$ is the scale radius, and a, b and c are 1.0620, 5.4807 and 0.3292 respectively as stated by \cite{Arnaud_2010}. (c) hydrostatic equilibrium relates the total cluster mass internal to radius r, $M_{\rm{SZ}}(r)$, to the gas pressure, $P_{\rm{gas}}(r)$,
\begin{align} 
 \frac{dP_{\rm{gas}}(r)}{dr} = -\rho_{\rm{gas}}(r) \frac{GM_{\rm{SZ}}(r)}{r^2},
\end{align}
where $\rho_{gas}$ is the gas density; (d) the gas mass fraction $f_g$ is small compared to unity ($\lesssim 0.15$). 

\begin{figure}
\begin{center}
\includegraphics[width=8.0cm,clip=,angle=0.]{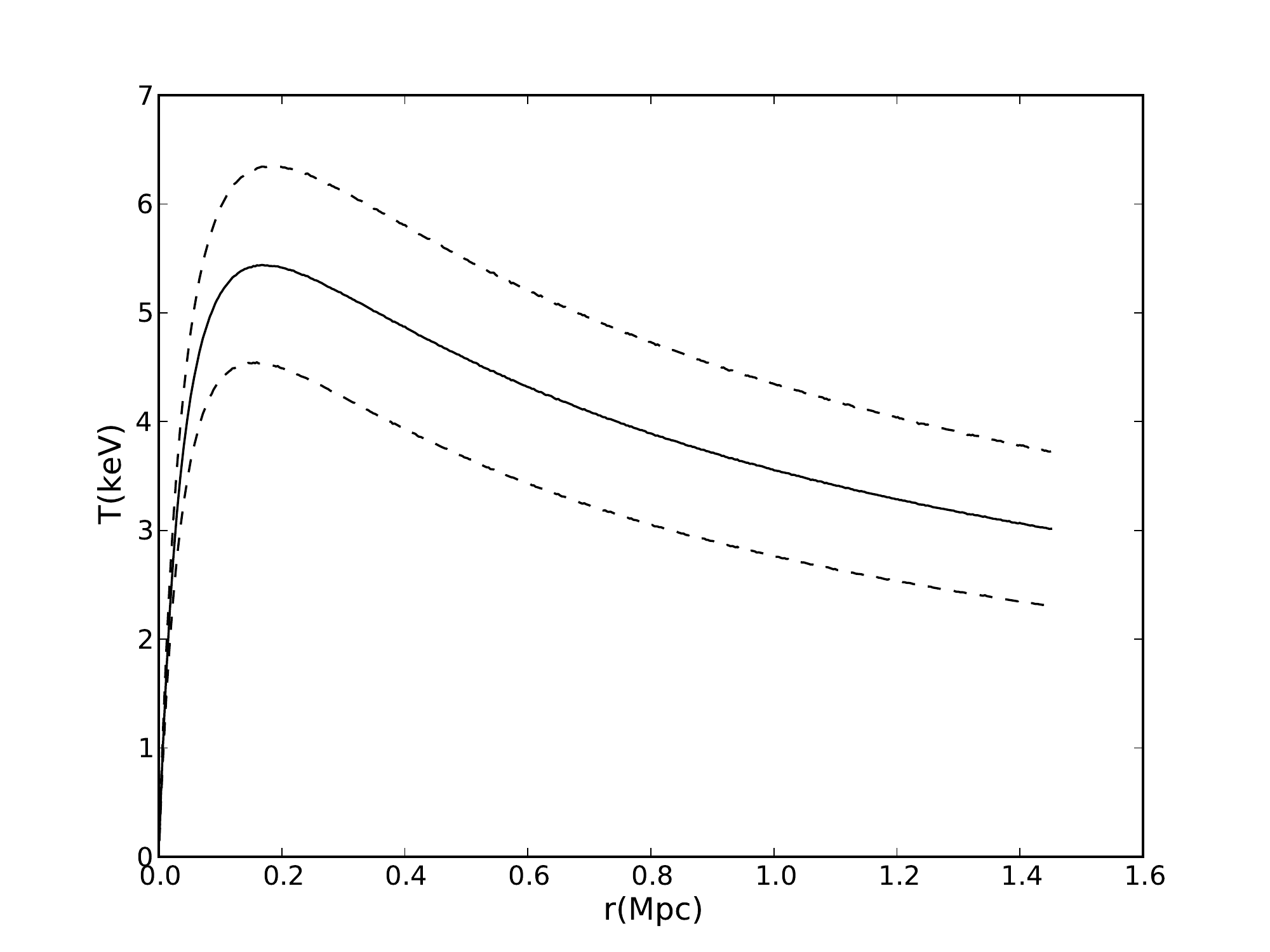}
\caption{An example of a cluster temperature profile used in our analysis. This profile was obtained using our data for XMJ1115+5319 with the assumption that the cluster density, pressure and temperature profiles can be described by the parameterisation of the \citealt{Olamaie_2012} model. The solid line shows our estimated temperature as a function of radius for XMJ1115+5319 and the dashed lines show similar profiles that correspond to the upper and lower 68\%  confidence limits of the parameter estimations for this cluster.}
\label{Fig:ami-example-temp}
\end{center}
\end{figure}

Using the  \cite{Olamaie_2012}  parameterisation we are able to fully describe the SZ signature of a cluster with the set $\Theta_{c}$ of parameters $x_{\rm{c}},y_{\rm{c}},\phi, \eta, M_{\rm{SZ},200}, f_{\rm{g},200}$ and  $z$.  Here, $x_{\rm{c}}$ and $y_{\rm{c}}$ give the cluster position, $\phi$ is the orientation angle measured from N through E, $\eta$ is the ratio of the lengths of the semi-minor to semi-major axes, $M_{\rm{SZ},200}$ is the SZ-derived cluster total mass within $r_{200}$, $f_{\rm{g},200}$ is the average baryonic gas fraction within $r_{200}$, and $z$ is the cluster redshift. The parameter $r_{200}$ is defined as the radius internal to which the mean total density is 200 times $\rm{\rho_{crit,z}}$. The \cite{Olamaie_2012} parameterisation is not isothermal, instead it assumes that the cluster temperature profile can be described by the formalism that is presented in that paper and this allows us to extract the temperature $T_{Y}$ at a given radius (an example of the temperature profile is given in Figure \ref{Fig:ami-example-temp}). Alternatively, from $M_{\rm{SZ,Y}}$, {\textcolor{black}{with the assumption that the cluster is virialized, isothermal, spherical and that all kinetic energy in it is plasma internal energy,}} we can derive the mean gas temperature, $T_{m,Y}$, within a specified radius $Y$:
\begin{align} 
 K_BT_{m,Y} = \frac{G\mu M_{SZ,Y}}{2r_{Y}} = \frac{G\mu}{2 \left( \frac{3}{4\pi(Y\rho_{crit,z})}\right)^{1/3}}M_{SZ,Y}^{2/3},
% K_BT = 8.2\,keV \left( \frac{M_{\rm{T_{Y}}}}{10^{15}h^{-1}M_{\odot}} \right)^{2/3} \left( \frac{H(z)}{H_0} \right)^{2/3}.
 \label{eqn:M-T-relation}
 \end{align}
where $\mu$ is the mass per particle, $\mu \approx 0.6m_{p}$, and $m_{p}$ is the proton mass. 

We model radio sources using the parameters $\Psi = (x_{s},y_{s},S_{0},\alpha)$, where $x_{s}$ and $y_{s}$ give the source position, $S_{0}$ is the flux-density and $\alpha$ is the spectral index. %The priors on cluster and source parameters are given in Table \ref{McAdam_Priors}.

%For the analysis of each of the 15 clusters in our SZ sample we additionally model all other known clusters within that field (see Tables  \ref{XMM_CLUSTERS_LIST}, \ref{XMM_CLUSTERS_CLOSE} and \ref{OTHER_CLUSTERS_CLOSE}). 

For each cluster, we perform two analyses: a) to quantify the significance of detection; and b) to obtain cluster parameter probability distributions. In previous analyses of AMI data it has not been necessary to separate analysis a) from b) and a single analysis has been used to both quantify the significance of detection and estimate parameters. However, unlike in all previous AMI analyses, here we use the X-ray-derived cluster mass estimate as a prior in our SZ analysis. This is problematic for quantifying the significance of detection, which we do with the Bayesian evidence ratio
\begin{align} 
R = \frac{Z_1}{Z_0},
 \label{eqn:evidence-ratio}
 \end{align}
where $Z_1$ is the Bayesian evidence for a model with a cluster of mass $\geq 2\times10^{14}M_{\odot}$ which also includes radio sources and the statistics of the primary CMB anisotropies, while $Z_0$ is the Bayesian evidence for the same model but without the cluster component ({\textcolor{black}{\citealt{Jeffreys_1961} provides an interpretive scale for
the $R$ value, as do revised scales  such as \citealt{Gordon_2007}}}). For some clusters in our SZ sample the X-ray mass estimates (see Section \ref{Sec:Xray-masses}) are  centred at low mass ($<2\times10^{14}M_{\odot}$) and the uncertainties can extend close to zero. If we use these low X-ray mass estimates as priors then the evidence ratio cannot be used to quantify the significance of detection. To circumvent this issue, analysis a) is performed with a log-uniform prior on the cluster mass to find $R$ and analysis b) is performed with a Gaussian prior on the cluster mass. If a cluster is not detected (i.e. $Z_0>Z_1$ in analysis a) then we provide only the mass and temperature upper limits output from analysis b). The priors used in our analyses are described further in Table \ref{McAdam_Priors}. 

Performing analysis b) can cause difficulties for our Bayesian analysis if the the XCS derived mass is significantly discrepant from the SA data. For example, XMJ0830+5241 has a well constrained but very low XCS mass estimate (1.04$\pm$0.17$\times10^{14}M_{\odot}$) and such a cluster would not produce a sufficiently strong SZ signal to be detectable by the SA. So, if we were to use the XCS mean mass with the corresponding error as a prior, our analysis will not correctly model the SZ signature that may be present in the data. To prevent this problem, we widen the priors and we use as the standard deviation on our Gaussian prior on cluster mass either the error from the XCS analysis or  $2\times10^{14}M_{\odot}$, whichever is larger. 

\begin{table*}
\caption{Priors used in our Bayesian analyses. The symbols used to describe priors on positions, fluxes and spectral indexes correspond to the annotations in Figure \ref{Fig:SA-obs1}. For analysis a), the detection analysis (see Section \ref{SZ-analysis}), the mass prior extends down to $1\times10^{14}M_{\odot}$ but the limiting cluster mass used for the evidence calculation (Equation \ref{eqn:evidence-ratio}) is $2\times10^{14}M_{\odot}$.}
 \label{McAdam_Priors}
\begin{tabular}{lcc}
\hline 
Parameter & Prior \\ \hline
Cluster position ($\bf x_{c}$) & Gaussian prior on the X-ray position, $\sigma$=60$\arcsec$. \\
                                                & $\Diamond$ XCS cluster (Tables \ref{XMM_CLUSTERS_LIST} and \ref{ALL_CLUSTERS_CLOSE}).  \\
                                                & $\triangle$ Other cluster (Table \ref{ALL_CLUSTERS_CLOSE}).  \\
Orientation angle ($\phi$) & Uniform between 0 and 180 \\ 
Ratio of semi-minor to semi-major axis ($\eta$) & Uniform between 0.5 and 1.0 \\ 
Mass ($M_{\rm{SZ},200}/M_{\odot}$) & Detection analysis: Uniform in log space over $(1.0-60.0)$ $\times$ $10^{14}$. \\
                      & Parameter analysis: Gaussian centred on the X-ray mean with a standard deviation \\ 
& equal to the largest uncertainty in the asymmetric  X-ray errors (Table \ref{scaled_mass}) \\ 
& or $2\times10^{14}M_{\odot}$, whichever is greater.\\

% Uniform in log space over, $(1.0-60.0)$ $\times$ $10^{14}$. \\
%
Redshift ($z$) & Delta-function on the value in Table \ref{XMM_CLUSTERS_LIST}. \\ 
Gas fraction ({\textcolor{black}{$f_{\rm{g},200}$}}) & Gaussian prior centred on 0.1, $\sigma$=0.02. \\ 
Source position ($\bf x_{s}$) & + :  Delta-function using the LA positions. \\ 
                              & $\times$: Gaussian centred on the LA positions with $\sigma$=5$''$. \\ 
Source flux densities ($S_{0}/\rm{Jy}$) & +: Delta-function on the LA value.\\
                                        & $\times$: Gaussian centred on the LA continuum value with a $\sigma$ of $0.4S_{0}$. \\ 
Source spectral index ($\alpha$) & +: Delta-function on the LA value.   \\ 
                                 & $\times$: Gaussian centred on the value calculated from the LA channel maps with $\sigma$ as the LA error.\\ \hline
 \end{tabular}
\end{table*}

\section{X-ray mass estimates} \label{Sec:Xray-masses}

\begin{figure}
%\begin{center}
\includegraphics[width=8.0cm,clip=,angle=0.]{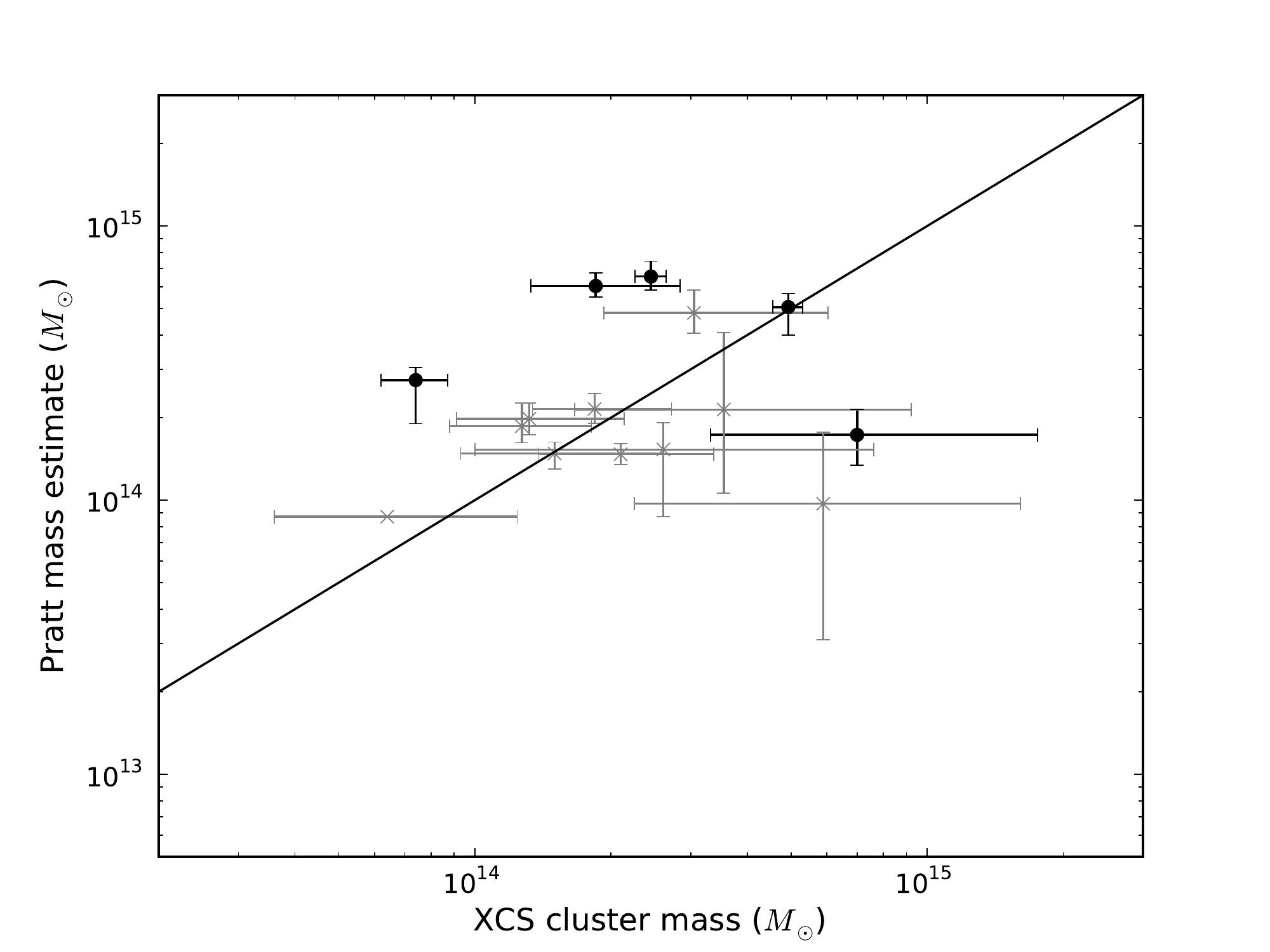}
\includegraphics[width=8.0cm,clip=,angle=0.]{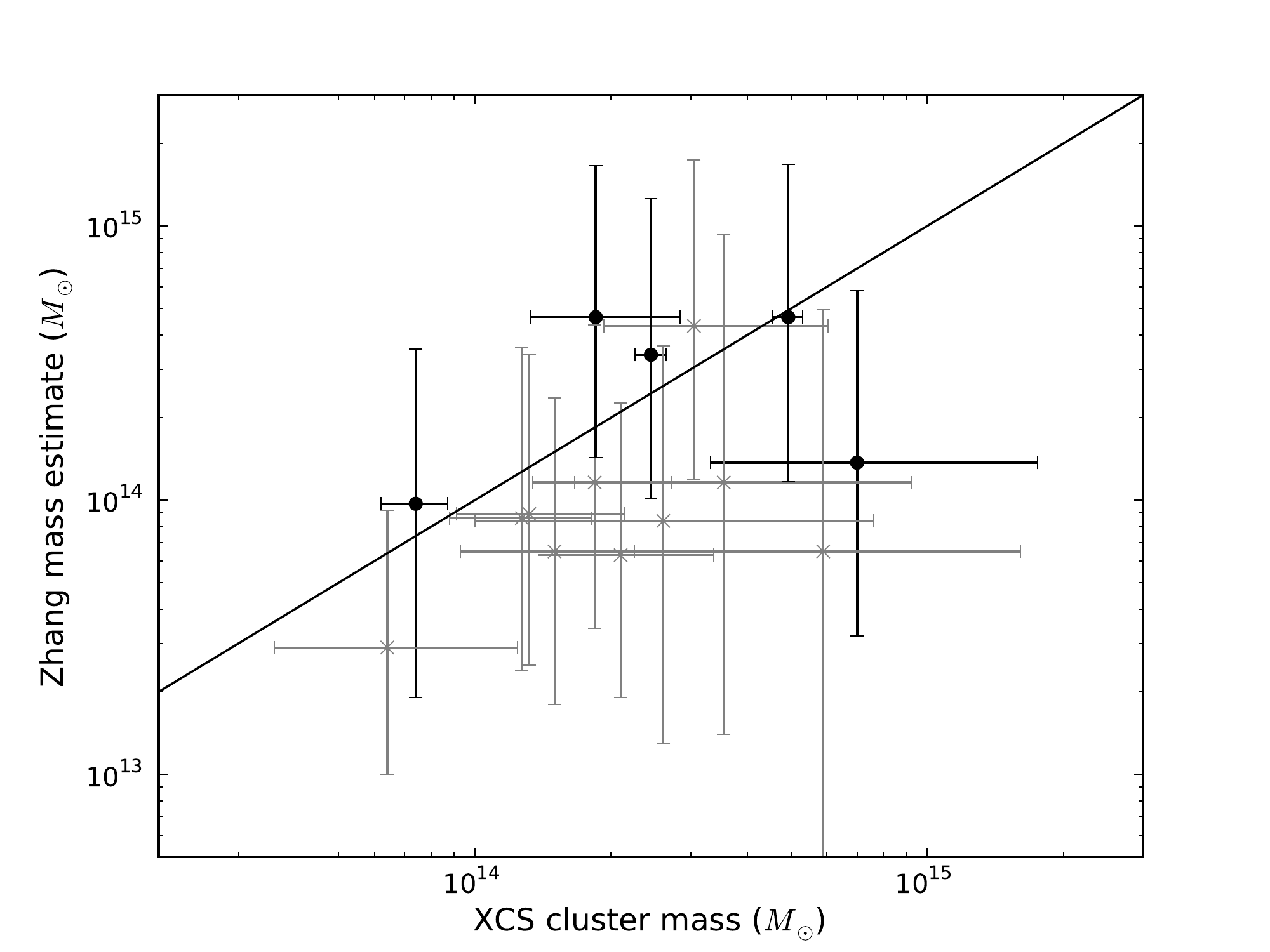}
\includegraphics[width=8.0cm,clip=,angle=0.]{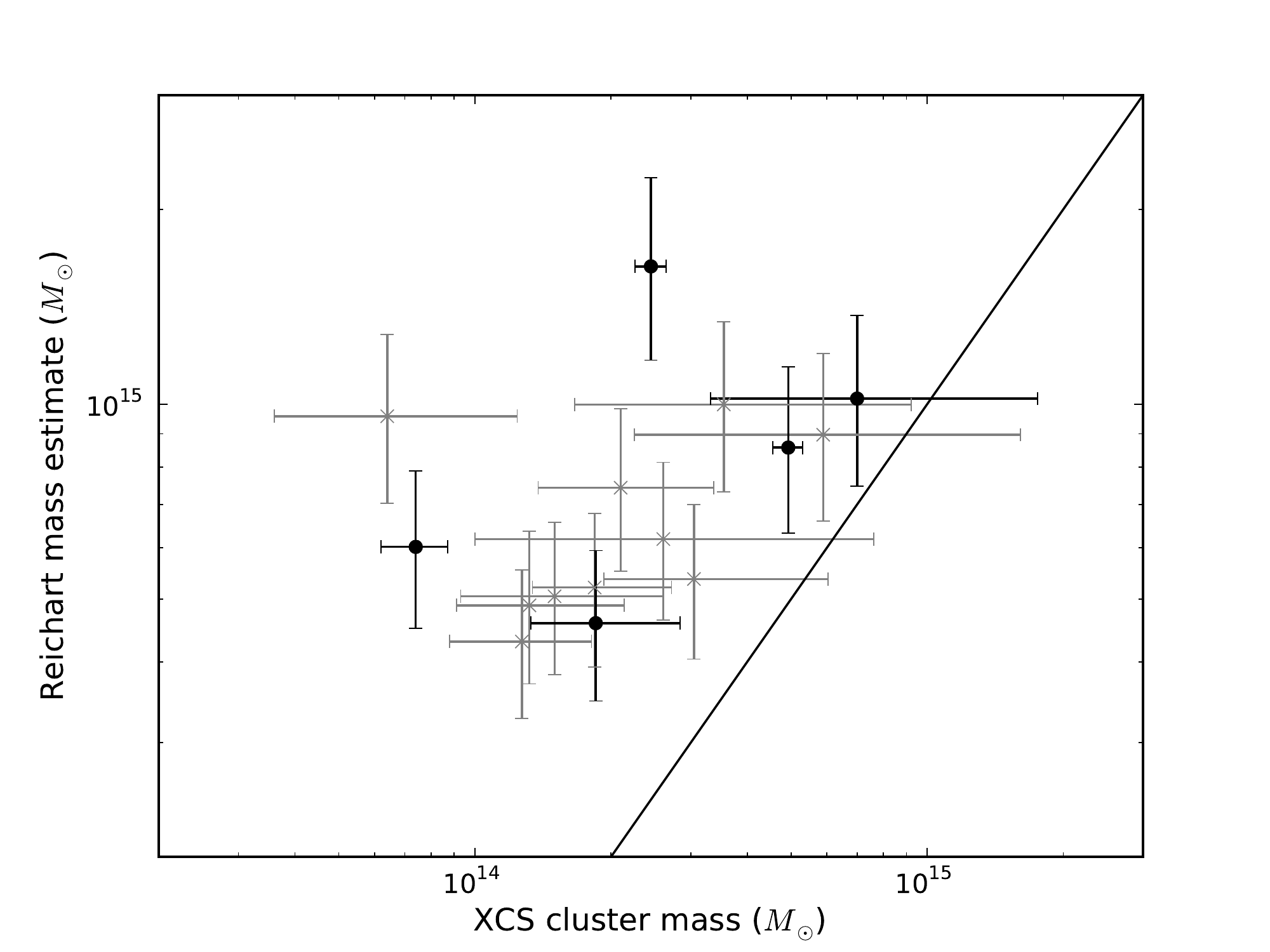}
\caption{Cluster masses derived from the relations given in Section \ref{Sec:Xray-masses}; as detailed  in Table \ref{scaled_mass}. The scaling relations used to estimate the cluster mass are  \citealt{Pratt_2009} (top), \citealt{Zhang_2011} (middle) and \citealt{Reichert_2011} (bottom). A straight line indicates equality between the scaling-relation derived cluster mass and the XCS cluster mass. Clusters that are detected by AMI are represented with black circles and clusters that are undetected by AMI are represented by $\times$ symbols.}
\label{Fig:scaling-relations-mass}
%\end{center}
\end{figure}

We use three approaches to estimate the mass from {\textcolor{black}{parameters presented in the XCS catalogue}}. \\
\\
1) Assuming the cluster is spherically symmetric, we can use $\rho_{\rm{crit,z}}$ and $r_{Y}$ to determine the cluster X-ray derived mass :
\begin{equation}
M_{X,Y} = \frac{4\pi}{3} Y r^3_{Y} \rho_{\rm{crit,z}}.
\end{equation}
For two of the clusters in our SZ sample (XMJ1050+5737 and XMJ1115+5319)  the XCS catalogue entries $r_{200}$ and $r_{500}$ are equal. For these clusters we assume $r_{200} \approx 1.5r_{500}$ (which is typical for the clusters in our sample) and that the errors  are a factor of 1.5 larger at $r_{200}$ than at $r_{500}$. \\
\\
2) %We use $M_{SZ,200}$ to derive the cluster temperature, similarly $M_{XT,200}$ can be estimate from the X-ray temperature using Equation \ref{eqn:M-T-relation}. 
We use the \cite{Reichert_2011}  mass-temperature relation,
\begin{equation}
M_{R,500} = (0.24 \pm 0.03)T_{X}(\rm{keV})^{1.76\pm0.08}10^{14}M_{\odot},
\end{equation}
to obtain a mass estimate from the X-ray temperature. \\
\\
3) We use masses from mass-luminosity scaling relations; there are many such relations in the literature, see e.g., \cite{Mantz_2010} and \cite{Rykoff_2008} amongst others. For this paper we make use of the \cite{Pratt_2009} and \cite{Zhang_2011} relations. The \cite{Pratt_2009} relation is given by
\begin{equation}
E(z)^{-7/3}L_{X,500} = C \left( \frac{M_{P,500}}{2\times10^{14}M_{\odot}} \right)^{\alpha},
\end{equation}
where C = $1.45\pm0.12$, $\alpha=1.90 \pm0.11$, E(z) is the Hubble parameter normalized to its present day value\footnote{$E^2(z) = \Omega_{m}(1+z)^3 + \Omega_{\Lambda}$} and $M_{P,500}$ is the \cite{Pratt_2009} estimated total mass within $r_{500}$. \cite{Zhang_2011} relates $L_X$ to the gas mass within $r_{500}$, $M_{Zg,500}$, by
\begin{equation}
{\rm{log}}_{10}\left( \frac{L_{X,500}}{E(z)\,{10^{-7}\rm{W}}}\right) = A + B{\rm{log}}_{10}\left( \frac{M_{Zg,500}E(z)}{10^{14}\,M_{\odot}} \right),
% L = L_{bol,500}^{co}
\end{equation}
where A = 45.06$\pm$0.68 and B = 1.29$\pm$0.05. \cite{Zhang_2011} convert their estimated gas mass to total mass, $M_{Z,500}$, via $E(z)^{1.5}{\rm{ln}}(M_{Zg,500}/M_{Z,500})=-2.37+0.21{\rm{ln}}(M_{Z,500}/2\times10^{14}M_{\odot})$.
\\
\\
Mass estimates obtained from the above relationships are presented in Table \ref{scaled_mass} and Figure \ref{Fig:scaling-relations-mass}. For the XCS clusters we find that the mass-luminosity scaling relationships (\citealt{Pratt_2009} and \citealt{Zhang_2011}) tend to predict a smaller mass than the mass-temperature scaling relationship (\citealt{Reichert_2011}).

\begin{comment}
The validity of these mass estimates is not clear. Firstly, the assumption of spherical symmetry is known to be over-simplistic in many cases, such as mergers, which, at expected to be more ubiquitous at high redshifts; our SZ sample comprises
 fairly high-$z$ systems and the nature of their selection, as serendipitous detections in XMM data, is likely to bias the selection of these clusters towards a more luminous and dynamical systems. {\bf{NOT SURE I UNDERSTAND THIS BIAS}}
\end{comment}

\begin{table*}
\caption{Cluster masses for clusters in our SZ sample derived from the relations given in Section \ref{Sec:Xray-masses}. For two of the clusters in our SZ sample (XMJ1050+5737 and XMJ1115+5319) the luminosity errors are not provided in the XCS catalogue and the XCS catalogue the entries $r_{200}$ and $r_{500}$ are equal; for these clusters we assume $r_{200} \approx 1.5r_{500}$ and that the errors also increase by a factor of 1.5. These cluster masses are plotted in Figure \ref{Fig:scaling-relations-mass}. On average we find that $M_{Z,500}/M_{X,500}=0.78$, $M_{P,500}/M_{X,500}=1.40$ and $M_{R,500}/M_{X,500}=4.04$.
} 
 \label{scaled_mass}
 \begin{tabular}{lcccccc}
 \hline
Cluster name & $M_{X,200}$  & $M_{X,500}$ & $M_{Z,500}$ & $M_{P,500}$ & $M_{R,500}$ \\
                       & $10\times^{14}M_{\odot}$ & $10\times^{14}M_{\odot}$ & $10\times^{14}M_{\odot}$ & $10\times^{14}M_{\odot}$ & $10\times^{14}M_{\odot}$ \\ \hline
XMJ0110+3305 & $1.77^{+0.77}_{-0.55}$ & $1.27^{+0.54}_{-0.39}$ & $0.86^{+2.74}_{-0.62}$ & $1.86^{+0.40}_{-0.24}$ & $4.30^{+1.25}_{-1.03}$ \\ [0.2em]
XMJ0116+3303 & $1.85^{+1.14}_{-0.57}$ & $1.32^{+0.82}_{-0.41}$ & $0.89^{+2.51}_{-0.64}$ & $1.98^{+0.28}_{-0.25}$ & $4.89^{+1.47}_{-1.19}$ \\[0.2em]
XMJ0515+7939 & $2.94^{+1.77}_{-1.02}$ & $2.1^{+1.27}_{-0.72}$ & $0.63^{+1.63}_{-0.44}$ & $1.47^{+0.14}_{-0.12}$ & $7.43^{+2.41}_{-1.91}$ \\[0.2em]
XMJ0830+5241 & $1.04^{+0.17}_{-0.17}$ & $0.74^{+0.13}_{-0.12}$ & $0.97^{+2.58}_{-0.78}$ & $2.74^{+0.31}_{-0.84}$ & $6.02^{+1.87}_{-1.51}$ \\[0.2em]
XMJ0901+6006 & $4.26^{+4.19}_{-1.55}$ & $3.05^{+2.99}_{-1.12}$ & $4.32^{+13.1}_{-3.13}$ & $4.82^{+1.01}_{-0.75}$ & $5.37^{+1.63}_{-1.33}$ \\[0.2em]
XMJ0916+3027 & $2.1^{+1.53}_{-0.8}$ & $1.5^{+1.1}_{-0.57}$ & $0.65^{+1.71}_{-0.47}$ & $1.48^{+0.15}_{-0.18}$ & $5.05^{+1.52}_{-1.23}$ \\[0.2em]
XMJ0923+2256 & $9.75^{+14.6}_{-5.13}$ & $7.0^{+10.53}_{-3.68}$ & $1.37^{+4.43}_{-1.05}$ & $1.73^{+0.41}_{-0.39}$ & $10.2^{+3.5}_{-2.73}$ \\[0.2em]
XMJ0925+3058 & $2.57^{+1.23}_{-0.71}$ & $1.84^{+0.88}_{-0.5}$ & $1.16^{+3.19}_{-0.82}$ & $2.15^{+0.3}_{-0.24}$ & $5.21^{+1.57}_{-1.28}$ \\[0.2em]
%Cant obtain estimates for XMJ1050+5737 as no model fit parameters
%Cant obtain estimates for XMJ1050+5737 as Luminosity errors too large 
%For cluster XMJ1050+5737 r200 = r500
XMJ1050+5737 & $0.86^{+0.82}_{-0.38}$ & $0.64^{+0.6}_{-0.28}$ & $0.29^{+0.63}_{-0.19}$ & $0.87$ & $9.58^{+3.24}_{-2.55}$ \\[0.2em]
%Cant obtain estimates for XMJ1115+5319 as no model fit parameters
%Cant obtain estimates for XMJ1115+5319 as Luminosity errors too large 
%For cluster XMJ1115+5319 r200 = r500
XMJ1115+5319 & $2.49^{+1.34}_{-0.70}$ & $1.85^{+0.99}_{-0.52}$ & $4.65^{+12.0}_{-3.22}$ & $6.04^{+0.7}_{-0.54}$ & $4.59^{+1.35}_{-1.11}$ \\[0.2em]
XMJ1226+3332 & $3.42^{+0.28}_{-0.26}$ & $2.45^{+0.2}_{-0.19}$ & $3.39^{+9.2}_{-2.38}$ & $6.54^{+0.9}_{-0.7}$ & $16.32^{+6.06}_{-4.63}$ \\[0.2em]
XMJ1309+5739 & $8.23^{+14.2}_{-5.09}$ & $5.89^{+10.2}_{-3.64}$ & $0.65^{+4.31}_{-0.61}$ & $0.97^{+0.80}_{-0.66}$ & $8.97^{+3.0}_{-2.37}$ \\[0.2em]
XMJ1332+5031 & $6.88^{+0.52}_{-0.53}$ & $4.93^{+0.37}_{-0.37}$ & $4.65^{+12.1}_{-3.48}$ & $5.06^{+0.61}_{-1.06}$ & $8.57^{+2.85}_{-2.25}$ \\[0.2em]
XMJ1423+3828 & $3.64^{+6.98}_{-2.24}$ & $2.61^{+5.0}_{-1.61}$ & $0.84^{+2.81}_{-0.71}$ & $1.53^{+0.39}_{-0.66}$ & $6.19^{+1.94}_{-1.55}$ \\[0.2em]
XMJ1437+3415 & $4.95^{+7.9}_{-2.63}$ & $3.55^{+5.66}_{-1.89}$ & $1.16^{+8.13}_{-1.02}$ & $2.14^{+1.95}_{-1.08}$ & $9.99^{+3.41}_{-2.67}$ \\ \hline
  \end{tabular}
 \end{table*}

\section{Results}

\subsection{SZ measurements}

In Figure \ref{Fig:SA-obs1}, we present the source-subtracted SA signal divided by noise maps with a 0.6\,k$\lambda$ $uv$-taper for all clusters except the high-redshift clusters XMJ0830+5241 and XMJ1226+3332 ($z$=0.99 (X-ray) and $z$=0.89 (phot) respectively) which are presented with no $uv$-taper. In the Bayesian analysis of the SA data, we modelled: high flux density sources ($S_0 \geq 4 \sigma_{SA}$); sources within 5$\arcmin$ of the hot XCS clusters; and sources within 5$\arcmin$ of other known clusters from Table \ref{ALL_CLUSTERS_CLOSE}. Faint sources ($S_0 \le 4 \sigma_{SA}$) further than 5$\arcmin$ from a cluster and sources outside the 10\% point of the SA power primary beam were not modelled. The subtraction of modelled sources uses the source flux, spectral index and position obtained from our Bayesian analysis. For sources that are not modelled we subtracted the LA measured values.  Our SA images of XMJ0925+3059 and  XMJ1226+3332 each contain a $>4\sigma$ source that has not been subtracted as it is outside the region covered by the LA observations. Nevertheless, the effects of these two unsubtracted sources are negligible in the regions of the known clusters.

Out of the 15 clusters comprising our SZ sample, we detect three clusters -- XMJ1115+5319, XMJ1332+5031 and XMJ0830+541 -- at high significance: log$\frac{Z_1}{Z_0}$ $\geq 5$, i.e. a model with a cluster is $\geq e^5$ more likely than one without. We also probably detect two other clusters  -- XMJ0923+2256 and XMJ1226+3332 -- at lower significance  ($5 \geq$ log$\frac{Z_1}{Z_0}$ $\geq 0$).  The remaining 10 clusters in our SZ sample were not detected by AMI.

\begin{figure}
\begin{center}
\includegraphics[width=8.0cm,clip=,angle=0.]{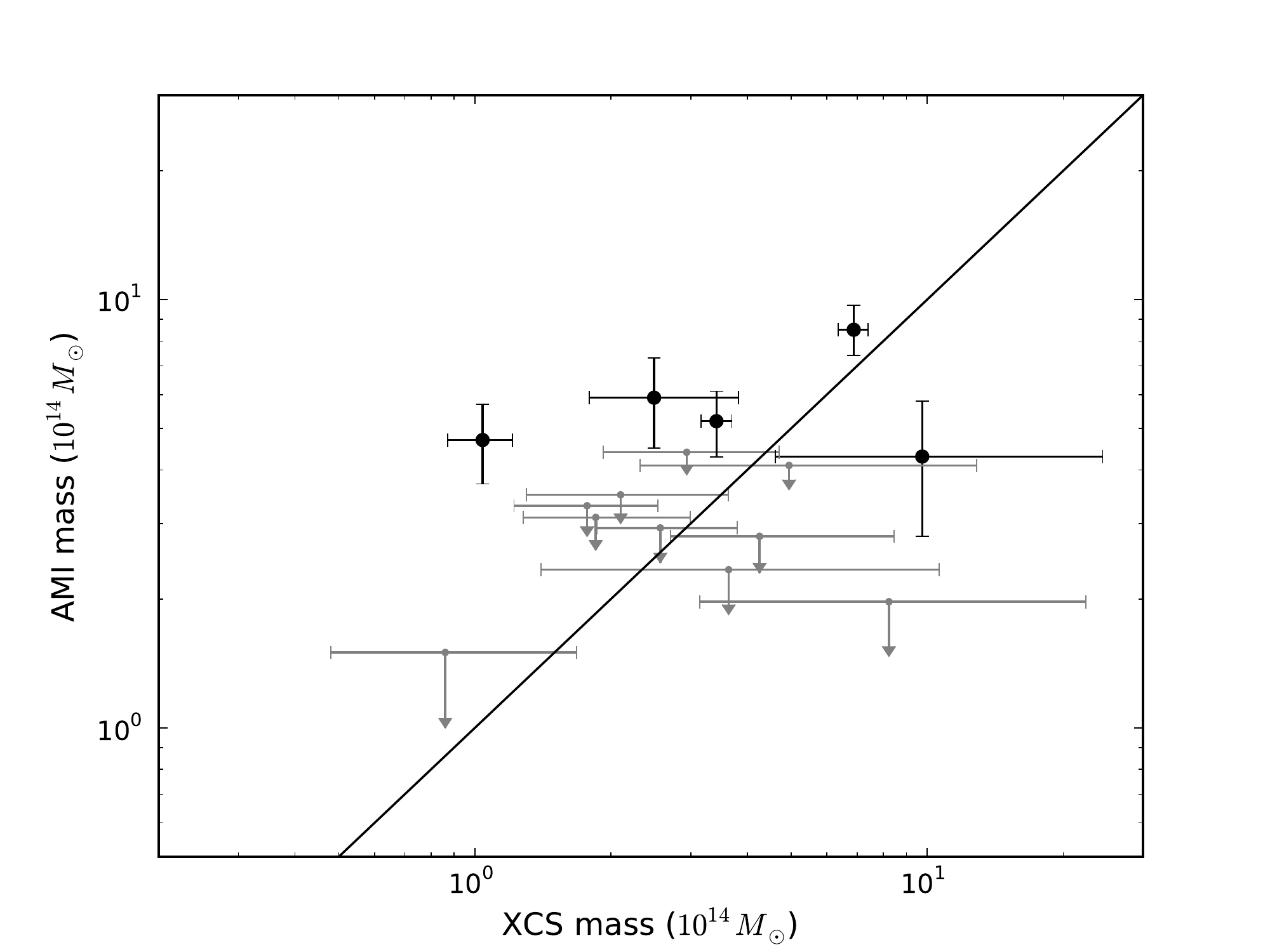}
\caption{AMI and XCS derived cluster masses. We present AMI masses for the 5 clusters we detect (black circles) and for the other 10 XCS clusters (grey dots) in our SZ sample we present AMI upper limits on the cluster mass. The AMI masses are derived from a Bayesian analysis using the XCS mass estimates as priors.}
\label{Fig:xcs-versus-ami-mass}
\end{center}
\end{figure}

\begin{figure}
\begin{center}
\includegraphics[width=8.0cm,clip=,angle=0.]{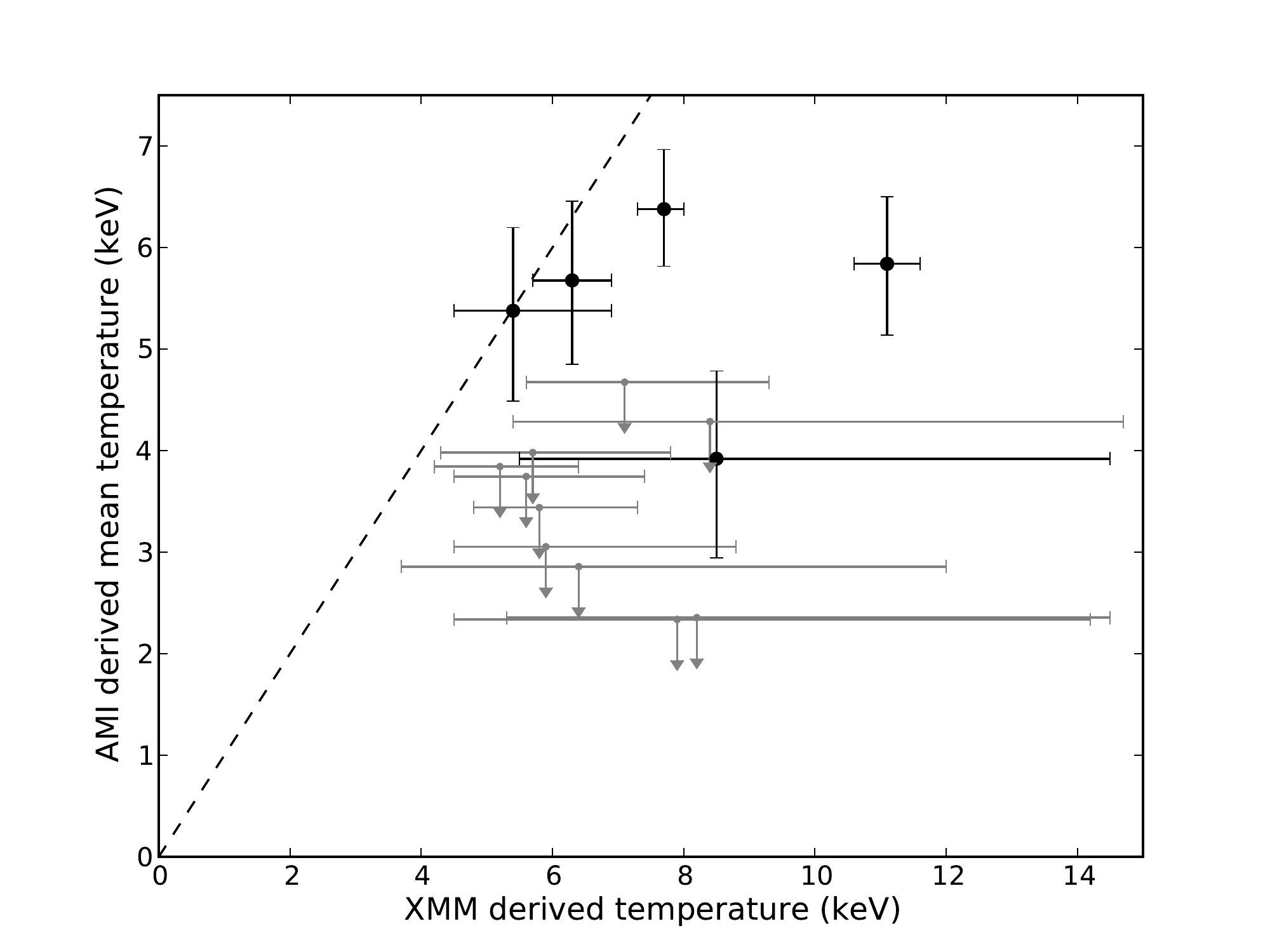}
\caption{AMI and XCS derived temperatures. The AMI temperature is derived from Equation \ref{eqn:M-T-relation} and represents the mean temperature within $r_{200}$. The XCS temperature is derived from the X-ray spectra.}
\label{Fig:xcs-versus-ami-temp}
\end{center}
\end{figure}

We note that XMJ0830+5241 (2XMM J083026.2+524133; \citealt{Culverhouse_2010} and \citealt{Schammel_2012}); XMJ1332+5031 (Abell 1758A; see e.g., \citealt{Carmen_2012}) and XMJ1226+3332 (CLJ1226.9+3332; see e.g., \citealt{Korngut_2011} and \citealt{Muchovej_2007}) have previously been observed in SZ. In the literature we have not found attempts of SZ observations towards the other XCS clusters studied in this paper. The evidence ratios for all 15 XCS clusters and the derived cluster mass (for detected clusters) or upper limit on the cluster mass (for undetected clusters) are given in Table \ref{Tab:mass-values} and plotted against the XCS mass estimates in Figure \ref{Fig:xcs-versus-ami-mass}. Similarly, Figure \ref{Fig:xcs-versus-ami-temp} shows a comparison of the XCS and SA derived temperatures.

In the following subsections we describe each of the clusters in our SZ sample. Unless otherwise specified, the cluster was discovered in the XCS and is not known to be present in any other cluster catalogue in the literature.

\begin{table*}
\caption{Log-evidence ratios from the analysis with a log-uniform mass prior for the 15 hot XCS clusters in our SZ sample, together with the derived mass parameters from runs with Gaussian mass priors. Three clusters -- XMJ1115+5319, XMJ1332+5031 and XMJ0830+541 -- are detected at high significance (log$\frac{Z_1}{Z_0}$ $\geq 5$) and a further two clusters  -- XMJ0923+2256 and XMJ1226+3332 -- are detected at lower significance  ($5 \geq$ log$\frac{Z_1}{Z_0}$ $\geq 0$). For these five detected clusters, we present the AMI-derived total cluster mass estimates within $r_{200}$ and $r_{500}$, the mean temperature within $r_{200}$ ($T_{m,200}$; Equation \ref{eqn:M-T-relation}), and the cluster temperature $T_{200}$ at $r_{200}$ (an example temperature profile is shown in Figure \ref{Fig:ami-example-temp}). For undetected clusters we present an upper limit on the mass. The errors and the upper limits correspond to 68\% confidence limits.}
 \label{Tab:mass-values}
\begin{tabular}{lcccccc}
\hline 
Cluster name &  Log Evidence Ratio          & $M_{SZ,200}$ & $M_{SZ,500}$ & $T_{200}$ & $T_{m,200}$ \\
                      &  log$(\frac{Z_1}{Z_0})$ & ($\times 10^{14} M_{\odot}$) &  ($\times 10^{14} M_{\odot}$) & keV & keV\\ \hline
XMJ0110+3305 & -1.50 & $<3.3$ & -- & -- & --\\ 
XMJ0116+3303 & -3.14  & $<3.1$& -- & -- & --\\ 
XMJ0515+7939 & -5.09  & $<4.4$& -- & -- & --\\ 
XMJ0830+5241 & 5.08  & $4.7_{-1.0}^{+1.0}$ & $3.5_{-0.7}^{+0.8}$ & $3.1_{-0.5}^{+0.5}$ &$5.7^{+0.8}_{-0.8}$ \\ 
XMJ0901+6006 & -1.61  & $<2.8 $& -- & -- & --\\ 
XMJ0916+3027 & -0.79  & $<3.5$& -- & -- & --\\ 
XMJ0923+2256 & 1.21  & $4.3_{-1.5}^{+1.5}$ & $3.2_{-1.1}^{+1.1}$ & $2.1_{-0.5}^{+0.5}$ & $3.9^{+0.9}_{-1.0}$ \\ 
XMJ0925+3059 & -4.34  & $<2.9$& -- & -- & --\\ 
XMJ1050+5737 & -1.58  & $<1.5$& -- & -- & --\\ 
XMJ1115+5319 & 7.50  & $5.9_{-1.4}^{+1.4}$ & $4.3_{-1.0}^{+1.0}$ & $3.0_{-0.5}^{+0.5}$ & $5.4^{+0.8}_{-0.9}$ \\ 
XMJ1226+3332 & 4.20  & $5.2_{-0.9}^{+0.9}$ & $3.8_{-0.7}^{+0.7}$ & $3.2_{-0.4}^{+0.4}$ & $5.8^{+0.7}_{-0.7}$ \\ 
XMJ1309+5739 & -1.68  & $<2.0$& -- & -- & --\\ 
XMJ1332+5031 & 6.77  & $8.5_{-1.1}^{+1.2}$ & $6.3_{-0.8}^{+0.9}$ & $3.5_{-0.3}^{+0.3}$ & $6.4^{+0.6}_{-0.7}$ \\
XMJ1423+3828 & -1.46  & $<2.3$& -- & -- & --\\ 
XMJ1437+3415 & -0.42 & $<4.1$& -- & -- & --\\ \hline
 \end{tabular}
\end{table*}

\subsubsection{SZ non-detections}

\paragraph*{XMJ0110+3305.}

We detect no SZ decrement towards this cluster although a $2\sigma$ negative feature lies at the pointing centre. Neither our data quality nor analysis is hindered by bright sources in the region of the cluster. The highest-flux-density source close to the cluster has a flux density of 2.6\,mJy and lies $\approx 5\arcmin$ to the South of the cluster. After the subtraction of this source, there is a 2$\sigma$ positive feature which is unlikely to effect our null detection. 

The XCS-derived mass, $M_{X,200}$, for this cluster is low ($1.8^{+0.8}_{-0.6}\times10^{14}M_{\odot}$). But the X-ray luminosity appears to be typical for a cluster of this mass and estimates of the mass from the mass-luminosity scaling relationships, $M_{P,500}$ and $M_{Z,500}$, are in reasonable agreement with $M_{X,500}$. However, the XCS temperature estimate is higher than expected for  a cluster with a mass $<2\times10^{14}M_{\odot}$ and as a consequence the mass estimated from the \cite{Reichert_2011} mass-temperature scaling relationship, $M_{R,500}$, is  higher ($4.3^{+1.3}_{-1.0}\times10^{14}M_{\odot}$). From our SZ data we derive a higher upper limit of 3.3$\times10^{14}M_{\odot}$. The mass estimates suggest that the cluster could be hot for its mass but the uncertainties are large.

The cluster NGC 0410 lies 7.9$\arcmin$ to the East of XMJ0110+3305. We see no decrement at the position of NGC 0410 but it: has a low mass (0.21$\times10^{14}M_{\odot}$; \citealt{Piffaretti_2011}); lies at 64\% of the SA power primary beam in our observation; is coincident with a radio source with an LA measured flux-density of 1.5\,mJy; and it will be affected by the brighter 4.2\,mJy and 5.1\,mJy sources $\approx 4\arcmin$ South-East (after subtraction these sources leave $3\sigma$ residuals on our SA image.)

\paragraph*{XMJ0116+3303}

Our LA observations show a point source of flux density 0.48\,mJy close to the cluster position but given the faintness of this source and the slight separation it is unlikely to be the cause of a null SZ detection.

The XCS-derived $M_{X,200}$ and the masses derived from the XCS luminosity ($M_{Z,200}$ and $M_{P,200}$) are below $2\times10^{14}M_{\odot}$. However, the mass derived from the XCS temperature ($M_{R,200}$) is significantly higher ($4.9^{+1.5}_{-1.2}\times10^{14}M_{\odot}$). From the SA SZ data, the upper limit on the cluster mass is 3.1$\times10^{14}M_{\odot}$. Although the uncertainties are large these mass estimates again suggest that the cluster may be hotter than expected.

Five arcmin South of the pointing centre we observe an extended 2-3$\sigma$ negative feature in source-subtracted tapered SA map with a peak flux density of -0.66\,mJy. The position of the decrement matches that of XMJ0116+3257, a low temperature XCS cluster separated from XMJ0116+3303 by 6.4$\arcmin$ which corresponds to 75\% of the SA power primary beam (see Table \ref{ALL_CLUSTERS_CLOSE}). We expect XMJ0116+3257 not to be a massive object, its $M_{X,200}$ is 0.28$\times10^{14}M_{\odot}$ and from the cluster scaling relationships we predict $M_{500} < 1.5 \times10^{14}M_{\odot}$. Further observations would be required to assess whether this extended decrement is caused by XMJ0116+3257 or is simply a noise fluctuation.

\paragraph*{XMJ0515+7939}

The LA flux density measurement of the source $\approx$3$\arcmin$ East of XMJ0515+7939 is 3.4\,mJy and the source $\approx$4$\arcmin$ South is measured to be 0.45\,mJy. These sources are not bright enough and are sufficiently separated from the XCS cluster position as to not significantly hinder an SZ detection. 

We do measure a 3$\sigma$ decrement slightly South (separated by 3$\arcmin$) of the XCS position but our Bayesian evidences favour a model without a cluster. Due to the weak decrement we derive a high upper limit of $M_{SZ,200}<4.4\times10^{14}M_\odot$ from our SZ analysis and this is greater than the XCS derived mass estimates ($M_{X,200} =2.9^{+1.8}_{-1.0}\times10^{14}M_{\odot}$ and $M_{X,500} =2.1^{+1.2}_{-0.7}\times10^{14}M_{\odot}$). From the mass-luminosity scaling relations we derive a lower mass estimate than XCS ($0.6-1.5\times10^{14}M_{\odot}$), yet the mass-temperature scaling yields a high mass ($7.4^{+2.4}_{-1.9}\times10^{14}M_{\odot}$). The mass derived from the X-ray temperature is greater than the SZ detection limit and inconsistent with $M_{X,500}$, apparently suggesting that this cluster is unusually hot for its mass.

\paragraph*{XMJ0901+6006}

This cluster was first discovered in the maxBCG survey (MaxBCG J135.25325+60.10133; \citealt{Koester_2007}). The `corrected' $N_{200}$ value (\citealt{Rykoff_2012})\footnote{corrected $N_{200}$ values can be downloaded from the official site http://risa.stanford.edu/maxbcg/.} is 41.1$\pm$3.4, where $N_{200}$  represents the number of red sequence galaxies observed to be in the cluster that are above a limiting brightness and within $r_{200}$. The \citealt{Rozo_2009_mass_richness} mass-richness scaling relationship which relates the richness at $r_{200}$ to the mass at that radius is
\begin{equation}
\frac{M_{N,200}}{10^{14}M_{\odot}} = e^{B_{M|N}}\left( \frac{N_{200}}{40} \right)^{\alpha_{M|N}},
\label{Eqn:Mass-richness}
\end{equation}
where the constants $B_{M|N}$ and $\alpha_{M|N}$ are 0.95 and 1.06 respectively. Using Equation \ref{Eqn:Mass-richness} and the `corrected' richness for this cluster we estimate that  $M_{N,200} = 2.66\pm0.24\times10^{14}M_{\odot}$.

The cluster richness derived mass is below our SZ upper limit ($2.8\times10^{14}M_{\odot}$) but all other mass estimates exceed this limit. The XCS mass, the mass-luminosity derived mass, and the mass-temperature derived mass consistently indicate that this is a massive cluster ($M_{200} > 4.2\times10^{14}M_{\odot}$). Our null SZ detection is surprising and suggests that the X-ray derived mass estimates are high. %Our SZ null detection and the $M_{N,200}$ derived from the maxBCG catalogue suggests that the X-ray mass estimates and those from the scaling relations are high. 

In our SA data we find a 2$\sigma$ positive feature close to the pointing centre and a 4$\sigma$ positive feature only $\approx$4$\arcmin$ from the pointing centre. In our LA data where we do not detect any sources close to the cluster above the $4\sigma_{LA}$ threshold nor are there any corresponding sources in the NVSS catalogue. There is a 3.2$\sigma$ feature on the LA image close to the pointing centre; the location of this faint positive feature corresponds to the 2$\sigma$ peak on the SA map. The flux of this positive feature is below the LA detection threshold and therefore it was not included in our model but if we were to subtract this LA measured flux from our SA image we would obtain only a very weak  ($<0.25$\,mJy) decrement.

Zwicky 2094 is separated from XMJ0901+6006 by $\approx 15\arcmin$, a separation coresponding to the 21\% level of the SA power primary beam. At the position of this cluster (to the North-East of our map) we observe a $4\sigma$ decrement. Given that this lies so far down the primary beam, further observations are required to determine if this is an SZ decrement.

\paragraph*{XMJ0916+3027}

\begin{figure}
\begin{center}
\includegraphics[width=8.0cm,clip=,angle=0.]{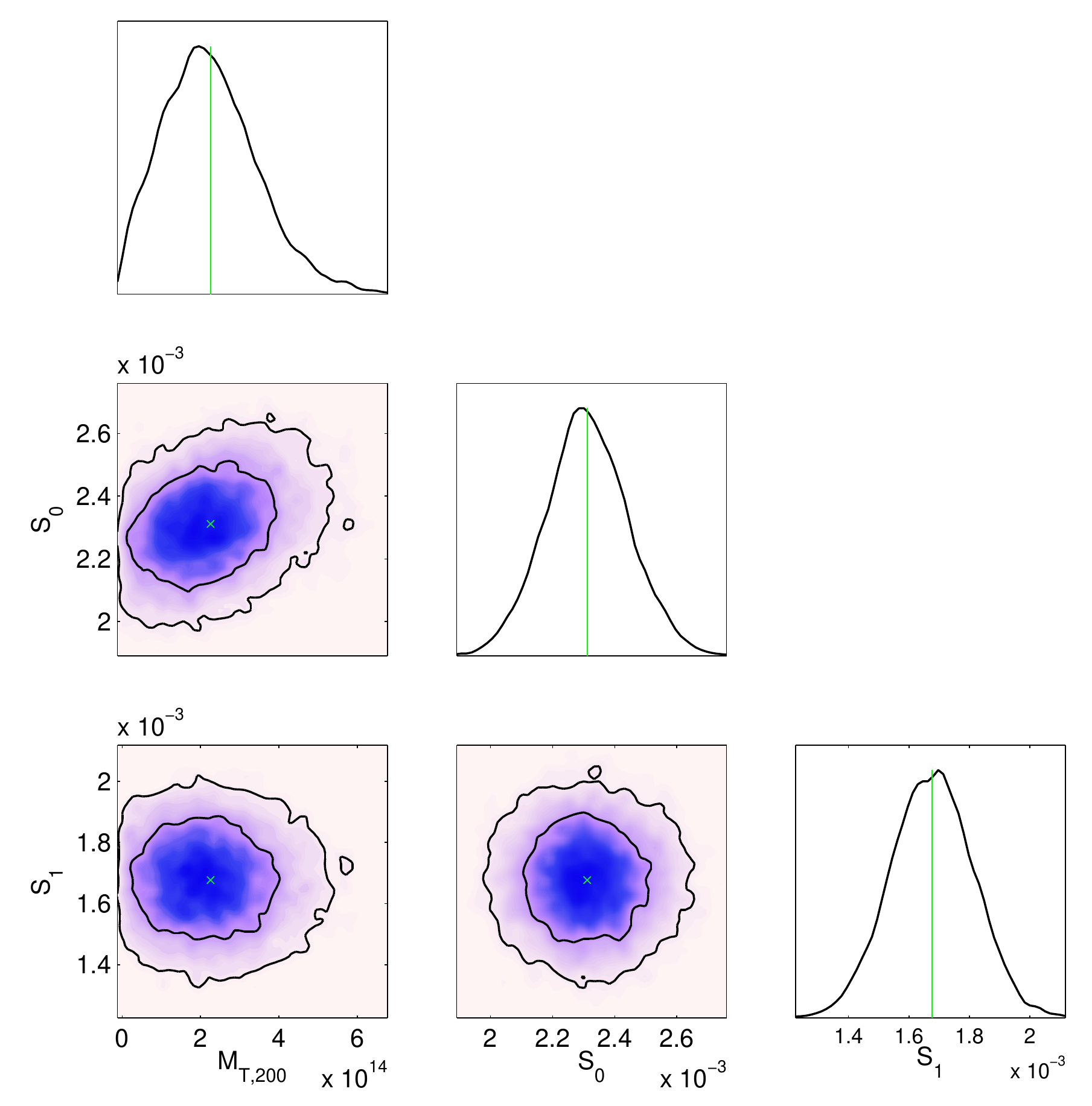}
\caption{The mass of XMJ0916+3027 and the derived flux densities for sources within 300$\arcsec$ of the cluster centre. $S_0$ and $S_1$ lie 14$\arcsec$ and 280$\arcsec$ from the cluster centre respectively. The green lines show the mean derived parameter value.}
\label{Fig:XMJ0916+3027ES_tri}
\end{center}
\end{figure}

The XCS derived angular extent of this cluster is small ($r_{200}=0.989$\,Mpc at $z$=0.59 corresponds to 2.5$\arcmin$), and in the SA data the cluster will be unresolved. With the AMI spectral coverage, separating the cluster signal from the coincident 2.1\,mJy source is difficult. In Figure \ref{Fig:XMJ0916+3027ES_tri} we demonstrate that the flux density of this source ($S_{0}$) is degenerate with the cluster mass but sources further away (e.g $S_{1}$ which is separated 4.7$\arcmin$) do not show this degeneracy. From our Bayesian analysis we derive a SA mean flux of 2.3\,mJy for the source coincident with the cluster, which is similar to that found in the LA data where the SZ signal is mostly resolved out.

Our upper limit on the cluster mass ($3.5\times10^{14}M_{\odot}$) exceeds both the XCS value ($M_{X,200}=2.1^{+1.5}_{0.8}\times10^{14}_{M\odot}$) and the values obtained from the mass-luminosity scaling relations ($0.65-1.5\times10^{14}M_{\odot}$). The mass derived from the mass-temperature scaling relation is high ($5^{+1.5}_{-1.2}\times10^{14}M_{\odot}$), apparently indicating that the cluster is hot for its mass.

\paragraph*{XMJ0925+3058}

Close to XMJ0925+3058 there are four known clusters: XMJ0925+3054 (4$\arcmin$ away), maxBCG141.2857+31.0615 (9$\arcmin$ away), XMJ0926+3103 (13$\arcmin$ away) and XMJ0926+3101 (14$\arcmin$ away). XMJ0926+3103 and XMJ0926+3101 are $\approx 1 \arcmin$ from a source with a LA measured flux density of 18.8\,mJy and are unlikely to be detectable even if they were not at $\approx 25\%$ of the SA power primary beam. MaxBCG141.2857+31.0615 is at 54\% of the SA power primary beam but is not a massive cluster, with an $N_{200}$ of 17.2$\pm$1.8 from which the estimated mass is $M_{N,200}=1.06\pm0.11 \times10^{14}M_{\odot}$ (Equation \ref{Eqn:Mass-richness}). At the location of this cluster we see no decrement in our map. XMJ0925+3054 is at 86\% of the SA power primary beam and for this XCS cluster $M_{X,200}= 0.25^{+0.09}_{-0.08} \times10^{14}M_{\odot}$. We observe no SZ decrement towards XMJ0925+3054 in agreement with XCS that this is not a massive cluster. 

For XMJ0925+3058 the XCS mass estimates are $M_{X,200} = 2.6^{+1.2}_{-0.7}\times10^{14}M_{\odot}$ and $M_{X,500} = 1.8^{+0.9}_{-0.5}\times10^{14}M_{\odot}$. $M_{X,500}$ is in reasonable agreement with masses derived from mass-luminosity scaling relations (1.2-2.2$\times10^{14}M_{\odot}$). The mass-temperature scaling relation predicts a more massive cluster ($5.2\times10^{14}M_{\odot}$) but our SZ upper limit, derived from data with very little source contamination, is $M_{200}<2.9\times10^{14}M_{\odot}$. Our limit combined with the other estimates suggests that the mass-temperature scaling relationship derived mass is high and this implies that the temperature of the cluster is hotter than expected for a cluster of this mass.

%The 3$\sigma$ decrement $\approx$10$\arcmin$ South of the pointing centre may a residual from the 6.5\,mJy source $\approx 5 \arcmin$ South-East of that.

\paragraph*{XMJ1050+5737}

This cluster is the second least luminous in our SZ sample yet XCS derived its temperature, at $8.2^{+6.3}_{-2.9}$\,keV, is relatively high.  The XCS-derived masses are low ($M_{X,200} = 0.9^{+0.8}_{-0.4}\times10^{14}M_{\odot}$ and $M_{X,500} = 0.6^{+0.6}_{-0.3}\times10^{14}M_{\odot}$) as are the masses derived from the X-ray luminosity (0.29-0.87$\times10^{14}M_{\odot}$) but the mass derived from the temperature is high ($9.6^{+3.2}_{-2.6}\times10^{14}M_{\odot}$). The SZ upper limit ($M_{SZ,200}<1.5\times10^{14}M_{\odot}$), the mass-luminosity derived masses, and the XCS masses all suggest that the mass-temperature derived mass is very high, apparently suggesting that this cluster has a very hot core given its mass.

We find a 3$\sigma$ positive feature at the pointing centre. There are no sources close to the cluster that are bright enough to contaminate the decrement but it is possible that undetected sources lie below our LA threshold. We find a 3$\sigma$ positive feature at the centre of the LA image which is below our detection threshold, but may be responsible for the observed positive feature at the centre of the SA image. Even if the LA flux of this low significance feature was subtracted from the SA data we would observe no SZ decrement at the SA pointing centre. 

\paragraph*{XMJ1309+5739}

This hot ($7.9^{+6.3}_{-3.4}$\,keV), XCS-derived, low-redshift cluster is the least luminous in the SZ sample. The cluster is also known as NSC J130931+574023 and was discovered in the Northern Sky Optical Cluster Survey (\citealt{Gal_2003}). The cluster richness is 33.6 which has a corresponding $M_{N,200} = 2.1\times10^{14}M_{\odot}$ (Equation \ref{Eqn:Mass-richness}). The XCS derived masses of $M_{X,200} = 8.2^{+14.2}_{-5.1}\times10^{14}M_{\odot}$ and $M_{X,500} = 5.9^{+10.2}_{-3.6}\times10^{14}M_{\odot}$ are higher than the mass derived from the optical richness but due to a low X-ray count the errors of the XCS parameters are very large. Given the low luminosity but high temperature of this cluster, the mass-temperature scaling relation predicts a high mass ($9.0^{+3.0}_{-2.4}\times10^{14}M_{\odot}$) in agreement with the XCS value, but the mass-luminosity scaling relations predict very low masses ($0.65-1.0\times10^{14}M_{\odot}$). The upper limit on mass from our SZ observations is $M_{SZ,200}<2.0\times10^{14}M_{\odot}$, which suggests that both the XCS and the mass-temperature scaling relation mass estimates are high.

At the pointing centre there is a 2.91\,mJy source but due to the extension of the cluster (the XCS derived $r_{200}$ of 1.81\,Mpc corresponds to an angular size of 9.1$\arcmin$) the degeneracy between the cluster mass and the source flux is not large. Rather than source flux-density contamination, there is a concern that we resolve out some of the signal from the cluster as the SA is optimised for smaller angular scale clusters. The natural resolution of the SA is $3\arcmin$ and with this resolution the SZ signal will be very extended compared to the beam, hence the total cluster mass within a beam is much smaller than the total cluster mass. Only the shortest SA baselines ($<6.7$m) have a resolution $>9\arcmin$ and only these baselines will capture all the mass within $r_{200}$ in a single beam.

\paragraph*{XMJ1423+3828}

With the lowest X-ray counts, this cluster is the least significant X-ray detection in our sample. As a consequence the errors in the XCS mass estimates are large: $M_{X,200} = 3.6^{+7.0}_{-2.2}\times10^{14}M_{\odot}$ and $M_{X,500} = 2.6^{+5.0}_{-1.6}\times10^{14}M_{\odot}$. The mass-luminosity scaling relations favour a lower mass cluster ($0.8-1.6\times10^{14}M_{\odot}$), but the mass-temperature relation favours a cluster of mass $6.2^{+1.9}_{-1.6}\times10^{14}M_{\odot}$. From our analysis we derive $M_{SZ,200}<2.3\times10^{14}M_{\odot}$ thus favouring a lower mass than obtained from the XCS or the mass-temperature relationship.

Our observations are not limited by point source contamination. There is a point source with an LA measured flux of 18\,mJy $\approx10\arcmin$ from the cluster position, the source lies $\approx50\%$ of the way down the SA power primary beam and its apparent flux on the SA map is 8.6\,mJy. After subtraction the source leaves a 7$\sigma$ positive residual but separated far enough from the cluster location to avoid contamination.

MaxBCG215.6632+38.301 is 11.3$\arcmin$ South-West of XMJ1423+3828. The  `corrected' richness for this cluster is 20.0$\pm$2.2 and from this we estimate that  $M_{N,200} = 1.24\pm0.15\times10^{14}M_{\odot}$. Given that this cluster lies at 40\% of the SA power primary beam and has a low mass, we would not expect a detection.

\paragraph*{XMJ1437+3415}

The XCS mass estimates ($M_{X,200}=5.0^{+7.9}_{-2.6}\times10^{14}M_{\odot}$ and $M_{X,500}=3.6^{+5.7}_{-1.9}\times10^{14}M_{\odot}$) are high but, due to few X-ray counts, the errors are large. The mass derived from the mass-temperature scaling relation suggests that the cluster mass could be as high as $1.0^{+0.3}_{-0.3}\times10^{15}M_{\odot}$. The mass-luminosity estimates are much smaller ($1.2-2.2\times10^{14}{M_\odot}$) and these coupled with our SZ upper limit ($4.1\times10^{14}M_{\odot}$) indicate that the cluster is less massive than the mass-temperature estimate.

There is no obvious contamination from point sources as the closest source is $4\arcmin$ South of the cluster and has a flux of only 1.6\,mJy. We have carefully flagged our data for interference and contaminated regions but we do detect a 4$\sigma$ positive feature on our SA image at a position where no corresponding $>4\sigma$ source is detected on the LA map. However, a 3.7$\sigma$ source is detected on the LA map. The SA and LA observations were not concurrent -- if this feature is a source then source variability may explain the different SA and LA fluxes. We detect a 4$\sigma$ negative feature about $8\arcmin$ South-West of the pointing centre at a position where there is no known cluster but, unfortunately, this lies outside the X-ray data. 

There are a further two XCS clusters within our SA field of view. XMJ1437+3414 lies close to the pointing centre at the 96\% level of the SA power primary beam. XMJ1437+3408 lies  South-East at a position corresponding to 53\% of the SA power primary beam. Neither system is thought to be massive: for  XMJ1437+3414 $M_{X,200} = 5.6^{+5.0}_{-2.3}\times10^{13}M_{\odot}$ and for XMJ1437+3408 $M_{X,200} = 1.5^{+2.1}_{-0.8}\times10^{13}M_{\odot}$. We detect a 2$\sigma$ decrement at the position of XMJ1437+3408  but further targeted observations would be needed to assess if this is a noise feature or a very low significance SZ structure.

\paragraph*{Stacked SZ non-detections}

\begin{figure}
\begin{center}
\includegraphics[width=8.0cm,clip=,angle=0.]{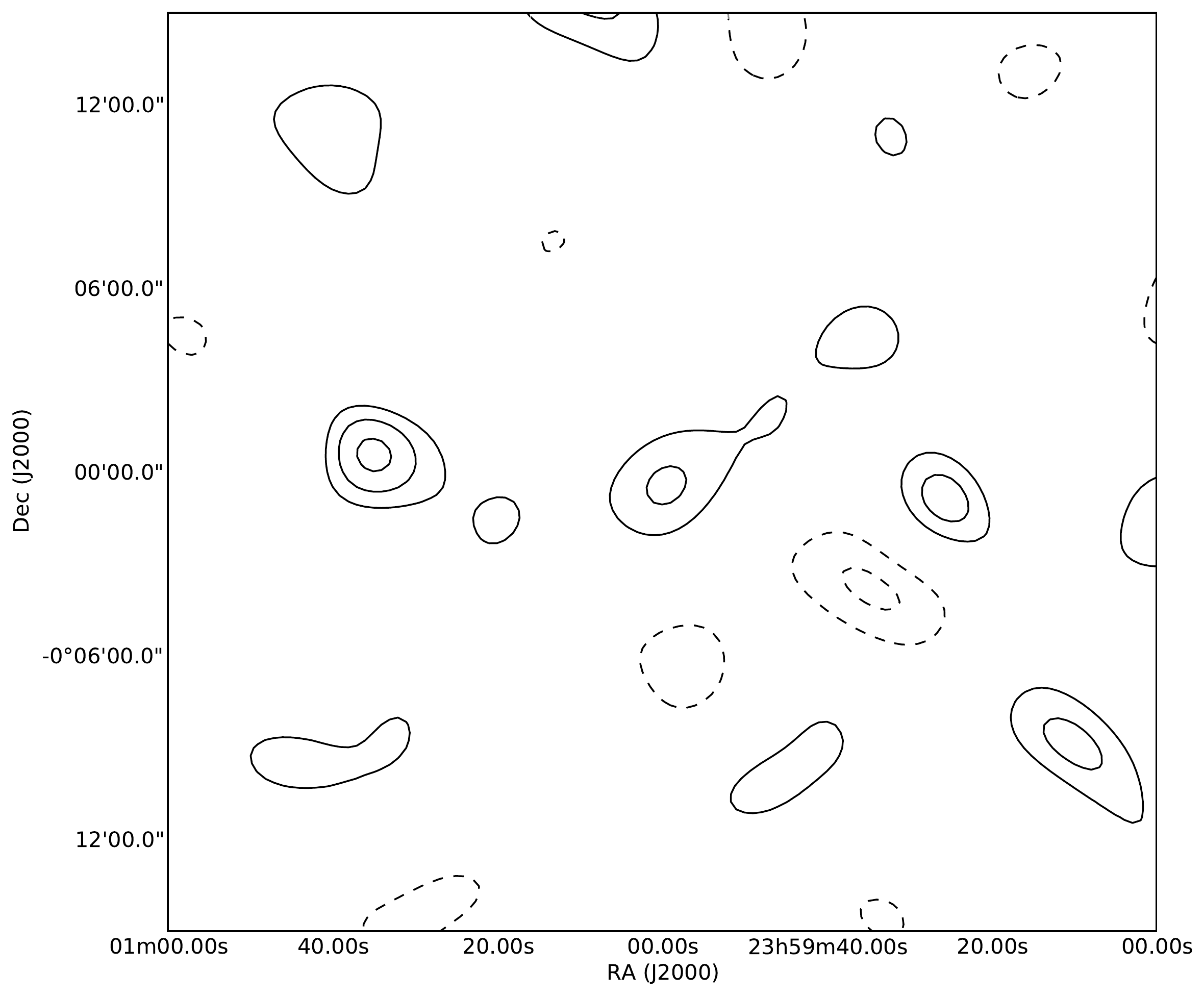}
\caption{A weighted stacked image of the 10 clusters where there was no SA detected SZ signal. The noise is 41$\mu$Jy and the contour levels 2, 3, 4, 5, 6, 7, 8, 9, and 10 times the noise level. }
\label{Fig:stacked-undetected}
\end{center}
\end{figure}

We have performed a simple stacking procedure on the SZ non-detections. The source-subtracted SA maps were stacked, placing all the undetected XCS clusters on top of each other. The stacked image is formed by taking the weighted sum ($w_{i}=1/\sigma_{i}^2$) of these collocated images. At the putative SA centre in the stacked image, we find 120$\pm$41$\mu$Jybeam$^{-1}$, where the synthesised beam is $3\arcmin \times 3\arcmin$, so we conclude that we find no SZ decrement (see Figure \ref{Fig:stacked-undetected}).

\subsubsection{SZ detections}

\paragraph*{XMJ0830+5241}

This high redshift, very luminous cluster was recorded in the literature (e.g., \citealt{Lamer_2008}) prior to the XCS and the SZ signature was observed by \cite{Culverhouse_2010} and \cite{Schammel_2012}. \cite{Lamer_2008} measure the bolometric luminosity to be $\approx 18 \times 10^{37}$\,W, a temperature of $8.2\pm0.9$\,keV and, assuming the cluster is isothermal, spherically symmetric and in hydrostatic equilibrium, they derived a total mass within $r_{500}$ of $5.6\times10^{14}M_{\odot}$. From SZ observations \cite{Culverhouse_2010} estimated a gas mass within $r_{2500}$ of $1.12^{+0.25}_{-0.25}\times10^{13}M_{\odot}$ and from X-ray observations this value to be $1.40^{+0.14}_{-0.20}\times10^{13}M_{\odot}$. From SA SZ observations \cite{Schammel_2012}) derived a total mass within $r_{200}$ of $3.56^{+1.10}_{-1.11}\times10^{14}M_{\odot}$ when modelling the decrement with a $\beta$-model density-profile, or $4.66^{+1.44}_{-1.41}\times10^{14}M_{\odot}$ when applying the same density-profile as used in this paper. Surprisingly, even with high X-ray counts, the XCS-fitted cluster parameters suggest that the cluster is not massive ($M_{X,200}= 1.0^{+0.2}_{-0.2}\times10^{14}M_{\odot}$). Furthermore both the mass-luminosity scaling relations imply that the mass is low ($1.0-2.7\times10^{14}M_{\odot}$). On the other hand the mass-temperature scaling relation estimates a higher mass ($6.0^{+1.9}_{-1.5}\times10^{14}M_{\odot}$) which is in better agreement with existing studies. This cluster is detected with a high Bayesian probability of detection and the SZ derived mass is $M_{SZ,200}=4.7^{+1.0}_{-1.0}\times10^{14}M_{\odot}$.

\begin{figure}
\begin{center}
\includegraphics[width=8.0cm,clip=,angle=0.]{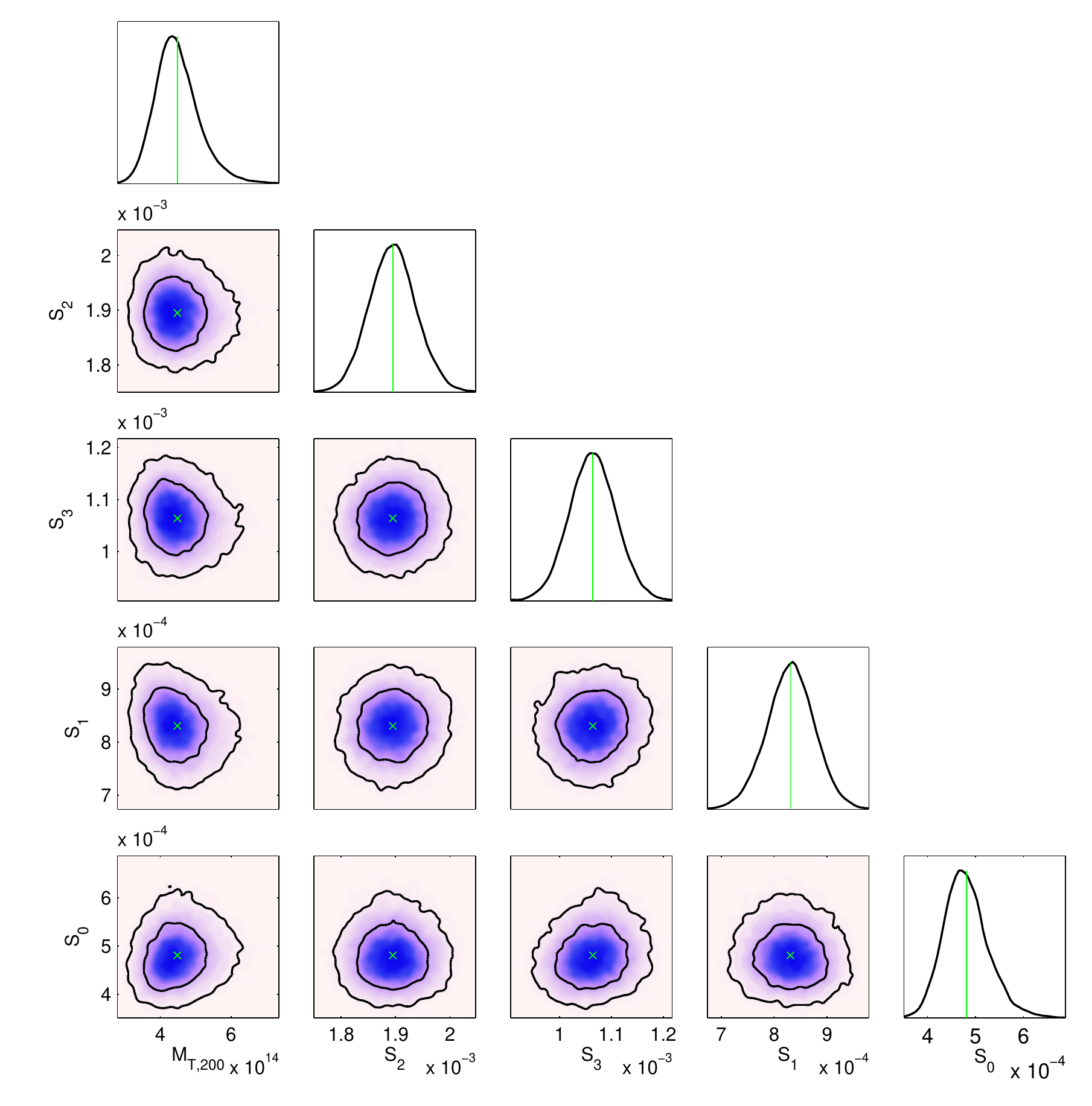}
\caption{The mass of XMJ0830+5241 and the derived flux densities for sources within 300$\arcsec$ of the cluster centre. $S_0$, $S_1$, $S_2$ and $S_3$ lie at 130$\arcsec$,  160$\arcsec$, 220$\arcsec$ and 240$\arcsec$ from the cluster centre respectively. The green lines show the mean derived parameter value.}
\label{Fig:XMJ0830+5241ES_tri}
\end{center}
\end{figure}

We expect high-redshift clusters to be unresolved with the SA and in these circumstances the SA has difficulty separating sources from the SZ by exploiting their different spectral indices. Hence, when studying small angular scale clusters it is vital that there is adequate knowledge of the source environment. For XMJ0830+5241 the source environment is good and the closest source ($\approx 2\arcmin$ from the cluster centre) has a flux of only 0.4\,mJy . In Figure \ref{Fig:XMJ0830+5241ES_tri} we plot the degeneracy between our derived mass and the closest sources and demonstrate that the sources do not significantly affect our SZ derived mass.

Two other clusters discovered in the XCS, XMJ0831+5234 and XMJ0831+5250, lie in the field of view and they are separated from the cluster centre by 10$\arcmin$ and 15$\arcmin$ (at 49\% and 18\% of the SA power primary beam respectively). For XMJ0831+5234 there are no parameter estimates in XCS and we are unable to estimate its mass, but XMJ0831+5250 is not massive with $M_{X,200}=2.32^{+0.68}_{-0.64}\times10^{14}M_{\odot}$. Given their separation from the pointing centre, it would be surprising if we were to have detected decrements corresponding to either cluster in the SA data.

\paragraph*{XMJ0923+2256}

This cluster, at a redshift of 0.19, is the closest cluster in our SZ sample and it was discovered in the MaxBCG survey (MaxBCG J140.85564+22.94378; \citealt{Koester_2007}).  The `corrected' $N_{200}$ value (\citealt{Rykoff_2012})  is 20.5$\pm$2.1, from which we calculate that $M_{N,200} = 1.27\pm0.14\times10^{14}M_{\odot}$ (Equation \ref{Eqn:Mass-richness}). This is in agreement with mass derived from the mass-luminosity relation but a factor of eight lower than the XCS derived mass of $M_{X,200} = 9.75^{+14.7}_{-5.1}\times10^{14}M_{\odot}$ (which has very large errors) or the mass derived from the mass-temperature relation ($1.0^{+0.4}_{-0.3}\times10^{15}M_{\odot}$). 

In our SA data, the ``peak'' decrement we obtain in our map is $\approx 2\sigma$ and we have found no close contaminating sources. Our Bayesian evidences demonstrate that our data supports a model with a cluster rather than one with no cluster. The ratio of the Bayesian evidences is $e^{1.3}$ which according to the \cite{Jeffreys_1961} scale corresponds to substantial support for the model with the cluster rather than the model without the cluster. From our low significance SZ detection we derive a mass of $M_{SZ,200} = 4.3^{+1.5}_{-1.5}\times10^{14}M_{\odot}$.

The XCS-derived $r_{200}$ for the cluster is 1922\,kpc which corresponds to an angular size of 10.1$\arcmin$. Only the shortest SA baselines ($<6$m) have sensitivity on angular scales around $10\arcmin$ and only these baselines will measure the entire mass within $r_{200}$.

\paragraph*{XMJ1115+5319}

This very luminous but quite cool cluster was known prior to the XCS and it is also known as SDSS J1115+5319. The cluster was discovered in the red-sequence cluster survey data (\citealt{Gladders_2005a} and \citealt{Gladders_2005b}) but the coordinates of the cluster were first published by \cite{Hennawi_2008} who identified strong gravitational lenses around the cluster.  \citealt{Bayliss_2011} performed further gravitational lensing observations and estimated the mass within the virial radius to be $\approx 6.4^{+3.7}_{-4.3}\times10^{14}M_{\odot}$.

The XCS fit to the cluster shape has not provided an estimate of $r_{200}$ so we use $r_{200}=1.5\times r_{500}$, and from XCS calculate that $M_{X,200} = 2.5^{+1.4}_{-0.7}\times10^{14}M_{\odot}$ and $M_{X,500} = 1.9^{+1.0}_{-0.5}\times10^{14}M_{\odot}$. Using the X-ray scaling relations, we find that the expected cluster mass is higher ($4.6-6.0\times10^{14}M_{\odot}$) than the XCS value. Our mass estimate from this 6$\sigma$ decrement ($M_{SZ,200}=5.9^{+1.4}_{-1.4}\times10^{14}M_{\odot}$) is significantly higher than the XCS estimate but comparable to estimates from scaling relations.

\begin{figure}
\begin{center}
\includegraphics[width=8.0cm,clip=,angle=0.]{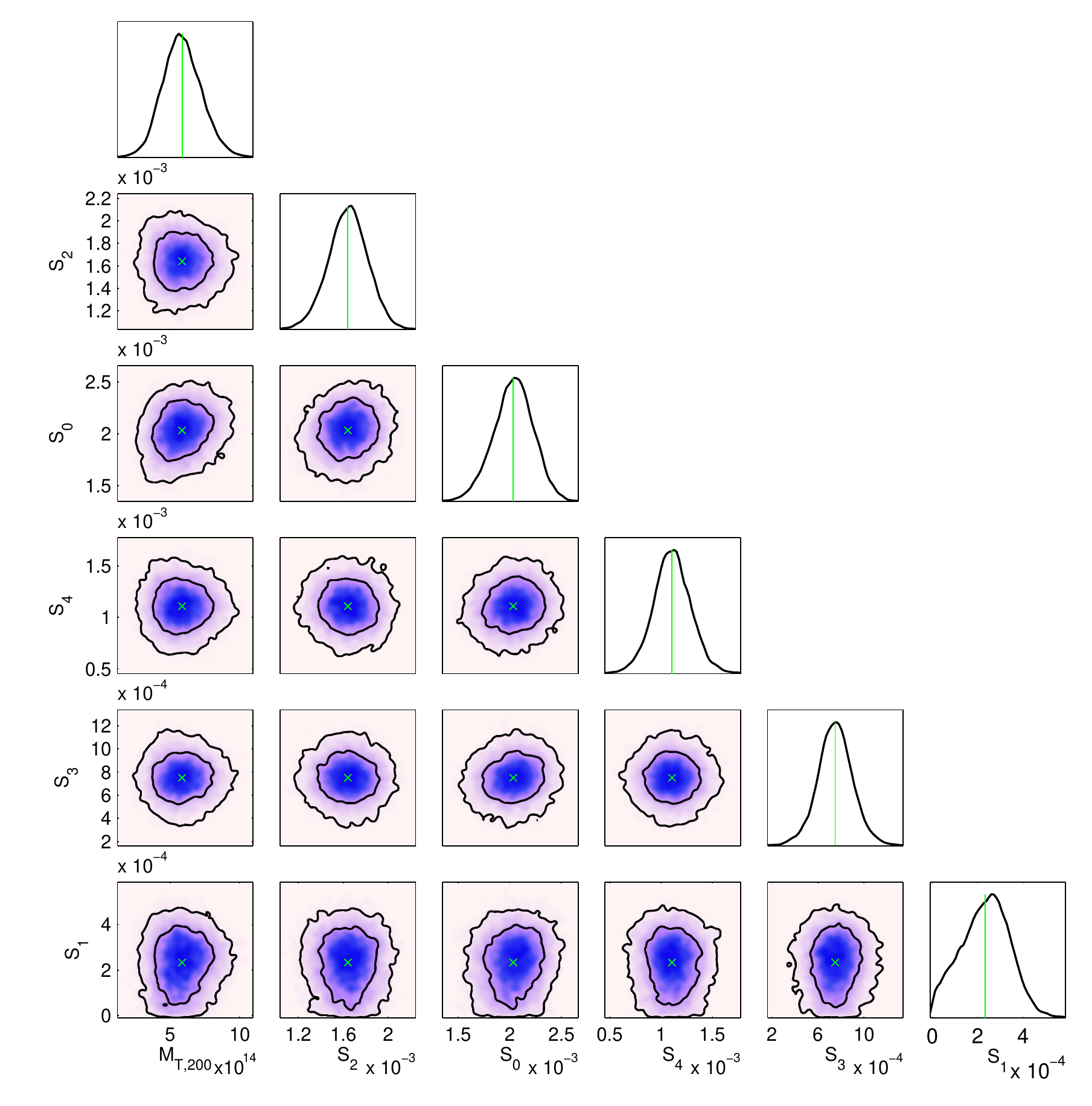}
\caption{The mass of XMJ1115+5319 and the derived flux densities for sources within 300$\arcsec$ of the cluster centre. $S_0$, $S_1$, $S_2$, $S_3$ and $S_4$ lie at 53$\arcsec$,  86$\arcsec$, 200$\arcsec$, 211$\arcsec$ and 287$\arcsec$ from the cluster centre respectively. The green lines show the mean derived parameter value.}
\label{Fig:XMJ1115+5319ES_tri}
\end{center}
\end{figure}

There are several radio sources within the SZ decrement, the source at the North of the decrement has a LA flux density of 0.55\,mJy and the source at the South of the decrement has a LA flux density of 2.52\,mJy. A further three sources lie within 5$\arcmin$ of the cluster centre. The degeneracy between these sources and the cluster mass is shown in Figure \ref{Fig:XMJ1115+5319ES_tri}. The sources within the decrement are the most degenerate with the brightest source ($S_{0}$) within the decrement causing the largest uncertainty in our mass estimate.

%The derived parameters from our SZ analysis are shown in Figure \ref{Fig:XMJ1115+5319-tri}.
%\begin{figure}
%\begin{center}
%\includegraphics[width=8.0cm,clip=,angle=0.]{./figures/XMJ1115+5319-tri.png}
%\caption{The SZ derived parameters for XMJ1115+5319.}
%\label{Fig:XMJ1115+5319-tri}
%\end{center}
%\end{figure}

\paragraph*{XMJ1226+3332}

This massive high redshift cluster is by far the most luminous in the entire sample and also the hottest in the SZ sample. There are many studies of the cluster present in the literature and it has been previously observed in both X-ray and SZ. \cite{Maughan_2007} used deep XMM-Newton and {\sc{Chandra}} observations to analyse the cluster, they derived a total mass within $r_{500}$ of $5.2^{+1.0}_{-0.8}\times10^{14}M_{\odot}$ and a temperature of 10.4$\pm$0.6\,keV.  \citealt{Muchovej_2007} studied the SZ signature of the cluster with observations from the Sunyaev-Zel'dovich Array (SZA) and derived a total mass estimate within $r_{200}$ of $7.2^{+1.3}_{-0.9}\times10^{14}M_{\odot}$. The XCS mass estimate is lower ($M_{X,200}=3.4^{+0.3}_{-0.3}\times10^{14}M_{\odot}$) but is consistent with the \citealt{Zhang_2011} mass-luminosity scaling relation estimate ($3.4^{+9.2}_{-2.4}\times10^{14}M_{\odot}$). The \cite{Pratt_2009}  mass-luminosity scaling relation estimate is higher ($6.5^{+0.9}_{-0.7}\times10^{14}M_{\odot}$) but the \cite{Reichert_2011} mass-temperature scaling relation value is very high ($16.3^{+6.1}_{-4.6}\times10^{14}M_{\odot}$).

\begin{figure}
\begin{center}
\includegraphics[width=8.0cm,clip=,angle=0.]{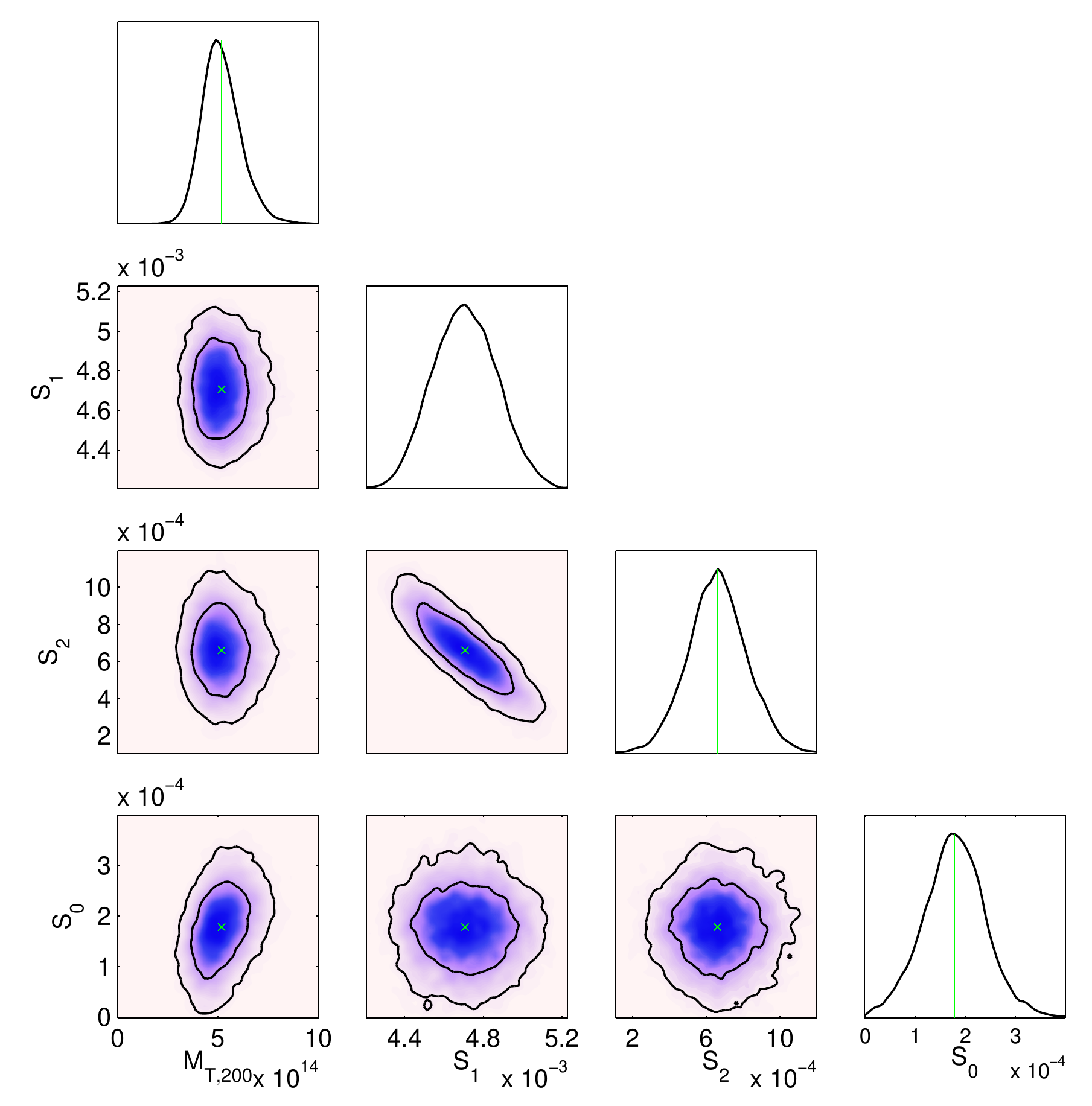}
\caption{The mass of XMJ1226+3332 and the derived flux densities for sources within 300$\arcsec$ of the cluster centre. $S_0$, $S_1$ and $S_2$ lie 4$\arcsec$, 261$\arcsec$ and 299$\arcsec$ from the cluster centre respectively. The green lines show the mean derived parameter value.}
\label{Fig:XMJ1226+3332ES_tri}
\end{center}
\end{figure}

We detect an SZ decrement corresponding to a cluster of mass $5.2^{+0.9}_{-0.9}\times10^{14}M_{\odot}$ which agrees well with previous studies but indicates that the \cite{Reichert_2011} value is high. The XCS derived $r_{200}$ is $1033\pm27$\,kpc which corresponds to an anglular size of 133$\arcsec$ and therefore we do not expect the cluster to be resolved by the SA, and as a consequence, the LA-measured 0.2\,mJy source at the pointing centre is degenerate with the cluster mass (this degeneracy is shown in Figure \ref{Fig:XMJ1226+3332ES_tri}).

Within 15$\arcmin$ of XMJ1226+3332 there are a further three clusters: XMJ1226+3343, XMJ1226+3345 and maxBCG 186.7603+33.3155. These lie at the 44\%, 28\% and 25\%  of the SA power primary beam respectively. It is unlikely that any of these systems is sufficiently massive for SZ detection: for XMJ1226+3345 there are no parameter estimates in XCS and we are unable to estimate its mass; for XMJ1226+3343 the $M_{X,200}=1.91^{+0.27}_{-0.23}\times10^{14}M_{\odot}$; and the maxBCG 186.7603+33.3155 `corrected' $N_{200}$ of  26.78$\pm$2.94 corresponds to a mass estimate of $M_{N,200} = 1.75\pm0.2\times10^{14}M_{\odot}$ (Equation \ref{Eqn:Mass-richness}). At the position of XMJ1226+3345 we find a 3$\sigma$ decrement in our SA image but we have subtracted a 0.5\,mJy source from this position directly from the LA flux estimates. Due to the SA sensitivity at this position in the power primary beam further targeted observations with the SA would be needed to determine if this is a genuine SZ effect.

\paragraph*{XMJ1332+5031}

This cluster is a well known Abell cluster, Abell 1758A. Not only is the cluster itself a complex merging system (e.g., \citealt{Durret_2011} and \citealt{David_2004}) but Abell 1758A also forms half of a spectacular double SZ system (see e.g., \citealt{Carmen_2012}). Abell 1758B, the second component, is nearly as massive as its partner and lies $8\arcmin$ to the South. 

Using XMM-Newton, \cite{Zhang_2008} derived a cluster mass within $r_{500}$ of 1.1$\pm$0.3$\times10^{15}M_{\odot}$, whereas a previous AMI study (\citealt{Carmen_2012}) found a mass within $r_{200}$ of $5.9\pm1.0\times10^{14}M_{\odot}$, and the XCS-estimates are $M_{X,500}=4.9\pm0.4\times10^{14}M_{\odot}$ and $M_{X,200}=6.9\pm0.5\times10^{14}M_{\odot}$. Furthermore, the scaling relations using either the X-ray luminosity or temperature produce comparable mass estimates ($4.7-8.6\times10^{14}M_{\odot}$). Our 9$\sigma$ detection is the most significant SZ signal detected in this study and the observed decrement corresponds to a cluster of mass $M_{SZ,200}=8.5^{+1.2}_{-1.1}\times10^{14}M_{\odot}$. The mass we have derived is comparable to previous estimates, to scaling-relationship masses, and to those derived in the XCS. 

\begin{figure}
\begin{center}
\includegraphics[width=8.0cm,clip=,angle=0.]{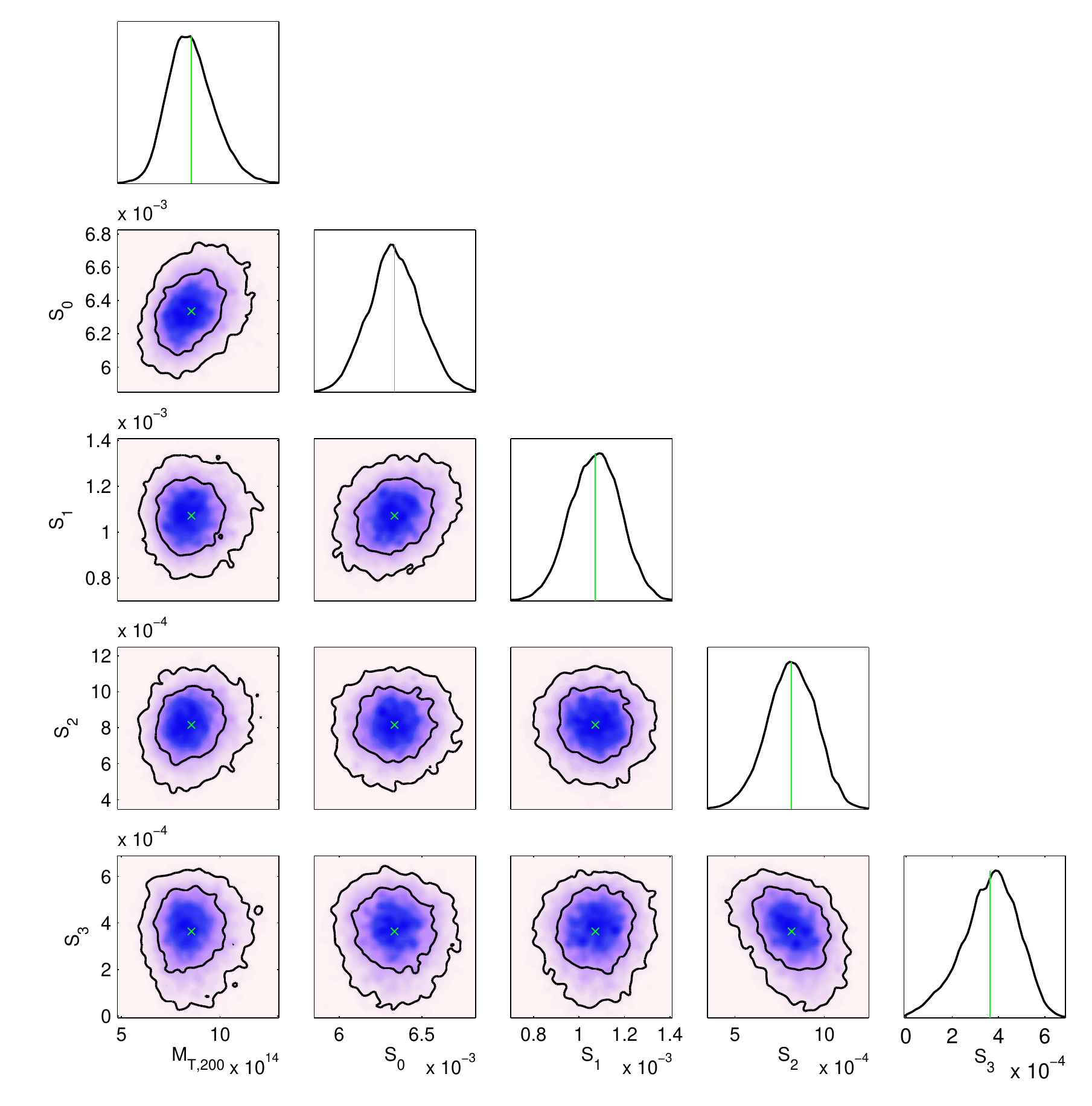}
\caption{The mass of XMJ1332+5031 and the derived flux densities for sources within 300$\arcsec$ of the cluster centre. $S_0$, $S_1$, $S_2$ and $S_3$ lie 19$\arcsec$, 185$\arcsec$ 163$\arcsec$ and 213$\arcsec$ from the cluster centre respectively. The green lines show the mean derived parameter value.}
\label{Fig:XMJ1332+5031ES_tri}
\end{center}
\end{figure}

There are three faint sources within the decrement itself but the cluster is quite extended and the degeneracy between source flux densities and the cluster mass is minimal (see Figure \ref{Fig:XMJ1332+5031ES_tri}).

\section{Discussion}

The cluster temperature is a good indicator of its mass and in this study we have obtained deep SZ observations of all observable XCS clusters that have an X-ray derived mean temperature greater than 5\,keV. If the observed clusters were isothermal, virialized, and spherical then their masses could simply be calculated from the XCS temperature and redshift with Equation \ref{eqn:M-T-relation} (for a 5\,keV cluster this corresponds to 6.16$\times10^{14}M_{\odot}$ ($r_{200}$=1.65Mpc) at redshift 0.2 or 3.86$\times10^{14}M_{\odot}$ ($r_{200}$=1.30Mpc) at redshift 1.0). These high X-ray temperatures suggest that the clusters in our sample are massive but the X-ray derived masses indicate that four of the 15 clusters in our SZ sample have a mass within $r_{200}$ of less than $2\times10^{14}M_{\odot}$. Additionally, only five of the clusters were detected in SZ yet to the sensitivities that we have reached in our SA cluster observations we would expect to detect clusters with mass $\gtrapprox 2\times10^{14}M_{\odot}$. % this limit depends upon many factors such as redshift, morphology, and temperature. 
Considering our careful selection in compiling our SZ sample and the previous successes of the SA in SZ detection from massive clusters at various redshifts, it is unlikely that either radio source contamination or instrumental effects are responsible for the low detection rate.

%These X-ray temperatures were obtained by fitting models to the X-ray spectra in the bright central region of the cluster. The errors of the derived temperatures can be large, and given that there are many more clusters with temperatures $>5$\,keV than $<$5\,keV, the \cite{Eddington_1913} bias implies that the true temperature of several of these clusters may be below 5\,keV. 

%The SZ effect is more sensitive to $T_{e}$ than the X-ray signal ($T_{e}$ dependence rather than $T_{e}^2$) so a lower temperature value would decrease the SZ magnitude. 

In our Bayesian analysis, we use wide Gaussian priors centred on the X-ray values to direct but not drive our SZ mass estimates. We find that out of the five detected clusters, four (XMJ0830+5241, XMJ1115+5319, XMJ1226+3332 and XMJ1332+5031) have higher SZ estimates of the mass within $r_{200}$ than the corresponding X-ray estimates. One system (XMJ0923+2256) has a lower mean SZ derived mass than X-ray derived mass but the X-ray error bars are large. Our derived upper limits on the cluster mass for six clusters are higher than the X-ray derived mean mass estimates. For the remaining four clusters our upper limits on the cluster mass are lower than, but not significantly discrepant from, the X-ray mean values. A simple stacking test on our undetected cluster images shows that even in this low noise (41$\mu$Jy) image no SZ decrement is observed, thus placing constraints on the cumulative SZ signal from these 10 undetected clusters. %The most significant outlier in our sample is XMJ0830+5241 for which the XCS derived mass, $1.04\pm0.17\times10^{14}M_{\odot}$, is well constrained but significantly lower than our SZ estimate. For this specific cluster our estimate agrees other measurements (e.g., from X-ray measurements \citealt{Lamer_2008} derived $M_{T,500}=5.6\times10^{14}M_{\odot}$ and from SZ measurements \citealt{Schammel_2012} derived $M_{T,200}=4.7\pm1.4\times10^{14}M_{\odot}$). 

\begin{figure}
\begin{center}
\includegraphics[width=8.0cm,clip=,angle=0.]{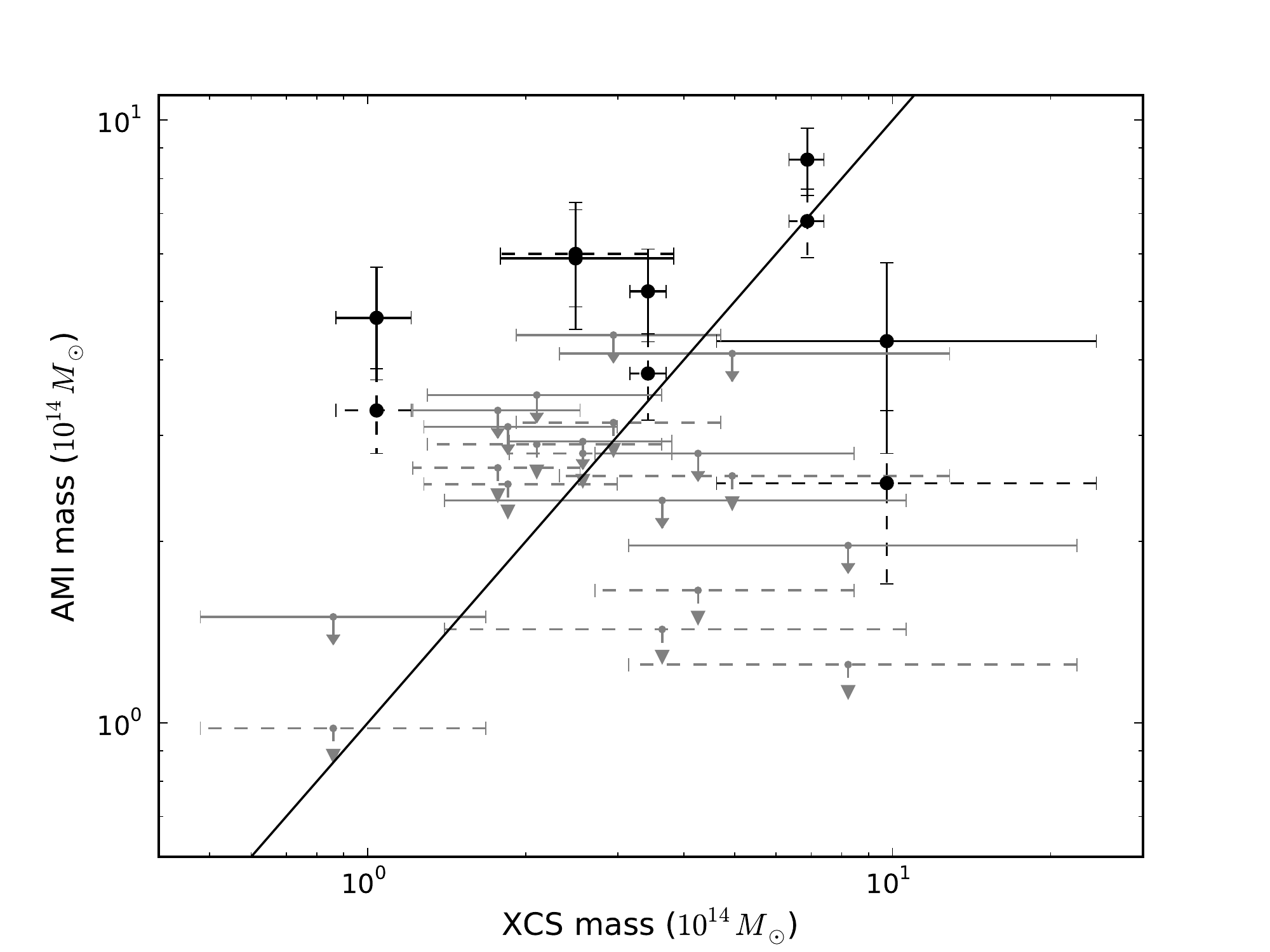}
\caption{{\textcolor{black} {AMI and XCS derived cluster masses for AMI analyses with two different $f_g$ priors. Black circles show the five clusters detected by AMI and grey dots show the 10 undetected XCS clusters. Solid lines give the results from the analyses with the $f_g$ prior centred on 0.10 and dashed lines give the results from the same analysis but with the $f_g$ prior centred on 0.18}}}
\label{Fig:fgas-mass-comparison} 
\end{center}
\end{figure}

{\textcolor{black} {Many studies have revealed that $f_{g}$ varies amongst similar clusters, furthermore, it is a function of radius and may also depend upon the cluster mass (see e.g. \citealt{Vikhlinin_2006}, \citealt{Arnaud_2007}, \citealt{Sun_2009}, \citealt{Landry_2012}  and \citealt{Sanderson_2013}). In our analysis we have used $f_{g}=0.1$ at $r_{200}$, which according to these same studies corresponds approximately to that expected at $r_{500}$ for a $M_{T,500}=2\times10^{14}M_{\odot}$ cluster. Our prior on the gas mass fraction is unlikely to be correct for all clusters in our sample and an incorrect $f_g$ prior will cause an error in the mass estimate. Due to our priors on cluster and source parameters our derived mass estimates do not change proportionally with the gas mass fraction prior. Instead, to assess the effects of an incorrect $f_g$ we analysed each cluster three times with different priors on this parameter (Gaussians of width 0.02 centred on 0.10, 0.14 and 0.18). We find that, on average, the derived mass for runs with $f_g=0.10$ were 1.2$\pm$0.13 times higher than runs with $f_g=0.14$ which in turn were 1.17$\pm$0.08 times higher than runs with $f_g=0.18$. The only exception to an increasing gas mass fraction corresponding to a decreasing $M_{SZ,200}$ is XMJ1115+5319, for which our mass estimates remain roughly constant throughout the $f_g$ range we have explored. In Figure \ref{Fig:fgas-mass-comparison} we present the derived mass estimates from the analyses with Gaussian priors on $f_g$ centred at 0.10 and 0.18. These analyses demonstrate that we can improve the agreement between the AMI and XCS mass estimates for two of the detected clusters (XMJ1226+3332 and XMJ1332+5031) by centring the $f_g$ prior on 0.18. However, for two of the five detected clusters (XMJ0830+5241 and XMJ1115+5319) a $f_g$ value far in excess of 0.18 would be required to bring the AMI and XCS values into agreement and for one of the five detected clusters a very low value of $f_g$ would be required ($\approx0.05$).}}
%The variation in mass is dependent upon the priors used in our analysis but to simplify the dependence we ran simulations ...}}

{\textcolor{black}{A further}} source of disagreement between X-ray and SZ mass estimates could be that the parametric profiles used to characterise the cluster shape differ for the SA and XCS analyses. We use a non-isothermal gNFW-profile whereas the XCS analysis uses an isothermal $\beta$-profile. For the analysis of SA data the derived mass does depend somewhat upon the density profile (see e.g., \citealt{Olamaie_2012}). Additionally, throughout our analysis we have assumed that the redshifts of the clusters are known without error but some of the redshifts are photometric. In our SZ analysis the redshift of a cluster is degenerate with the cluster mass. Thus, any errors in the redshifts would propagate through our analysis and influence the derived mass.

From the SA mass estimates, or upper limits, we calculate the mean cluster temperature. Besides XMJ1115+5319, which has an equal SZ and X-ray derived temperature, the SZ derived mean temperature (or upper limit on this value) for every other cluster in the SZ sample is lower than the X-ray core temperature (see Figure \ref{Fig:xcs-versus-ami-mass}). A possibility is that by selecting the hottest X-ray clusters we have chosen a sample of clusters that are far from idealised systems. Assuming such systems are isothermal is certainly not valid and it may be expected that the X-ray temperatures (which are measured in the bright X-ray core) are significantly larger than the mean cluster temperature. Certainly XMJ1332+5031 (Abell 1758A) is a well known merging system. Additionally, the errors of the X-ray derived temperatures can be large, and given that there are many more clusters with temperatures $>5$\,keV than $<$5\,keV, the \cite{Eddington_1913} bias (1913) suggests that the actual temperatures of some of these clusters may be lower.

Finally, we use scaling relationships to estimate cluster masses from the XCS derived temperatures and luminosities. We find that mass-luminosity scaled values are consistently lower than those obtained from mass-temperature estimates. Furthermore we find that the masses estimated from the mass-temperature relation frequently exceed the SZ values revealing that for this sample of clusters the mass-temperature relation is overestimating the cluster mass.

14/12/2012 Commented out

\begin{comment}
\subsection{Astrophysical}

The XCS temperatures are measured in the inner region of the cluster and unless the clusters are relaxed this temperature is likely to not be representative of the average cluster temperature. Errors in the temperature will propagate through the XCS analysis

Scaling relations not holding

underluminous in the SZ

over luminous in the X-ray -- merging systems

\end{comment}

\section{Conclusions}
We have presented an SZ study of a sample of the hottest galaxy clusters detected in the XCS catalogue. We find significant radio source contamination for 19 of the 34 clusters. High radio source contamination is likely to be a selection bias due to the XCS survey being conducted on the XMM-Newton archive which consists of targeted observations towards \emph{interesting} X-ray sources. For the 15 clusters with low point-source contamination, we have detected only five clusters, and from our analysis of these we have derived the cluster mass and temperature. For the remaining ten clusters we have provided upper limits on the cluster mass and temperature. From our study we find the following:
\begin{enumerate}
\item Many of the X-ray clusters are detected with low X-ray counts and we note that the five clusters that are detected in SZ all have X-ray counts $>700$ (a total of seven clusters in our SZ sample have counts $>700$).
\item Four out of the five detected clusters (XMJ0830+5241, XMJ1115+5319, XMJ1226+3332, and XMJ1332+5031) have higher SZ estimates of the mass within $r_{200}$ than the corresponding X-ray estimates. One system (XMJ0923+2256) has a lower mean SZ-derived mass than its X-ray-derived mass but the X-ray error bars are large.
\item Our upper limits on the cluster mass of undetected systems are in four cases lower than the mean XCS derived mass. For the remaining six undetected systems our upper limits exceed the mean XCS derived mass.
\item For this sample of hot XCS clusters, the mass estimates obtained from a mass-temperature scaling relation and the XCS derived temperature consistently overestimate the cluster mass.
\item The mean temperatures that we derive within $r_{200}$ are significantly lower than the core X-ray temperatures.
%\item A possibility is that by selecting our clusters on their high temperature we have biased our sample to merging clusters. For such clusters we would not expect scaling relations to hold and nor would be expect the core temperatures to be representative of the mean temperature.
\end{enumerate}

\section{Acknowledgements}

We thank the anonymous referee for constructive feedback. We thank Cambridge University and STFC for their support of AMI and its operations. We are grateful to the staff of the Cavendish Laboratory and the Mullard Radio Astronomy Observatory for the maintenance and operation of AMI. We thanks Julian Mayers for useful discussions regarding the XCS temperature estimates. CR and MPS acknowledge PPARC/STFC studentships. YCP acknowledges the support of a Rutherford Foundation/CCT/Cavendish Laboratory studentship. This work was performed using the Darwin Supercomputer of the University of Cambridge High Performance Computing Service (http://www.hpc.cam.ac.uk/), provided by Dell Inc. using Strate- gic Research Infrastructure Funding from the Higher Education Funding Council for England, and the Altix 3700 Supercomputer at DAMTP, University of Cambridge supported by HEFCE and STFC. We are grateful to Stuart Rankin and Andrey Kaliazin for their computing assistance.

\bsp
\label{lastpage}

\end{document}